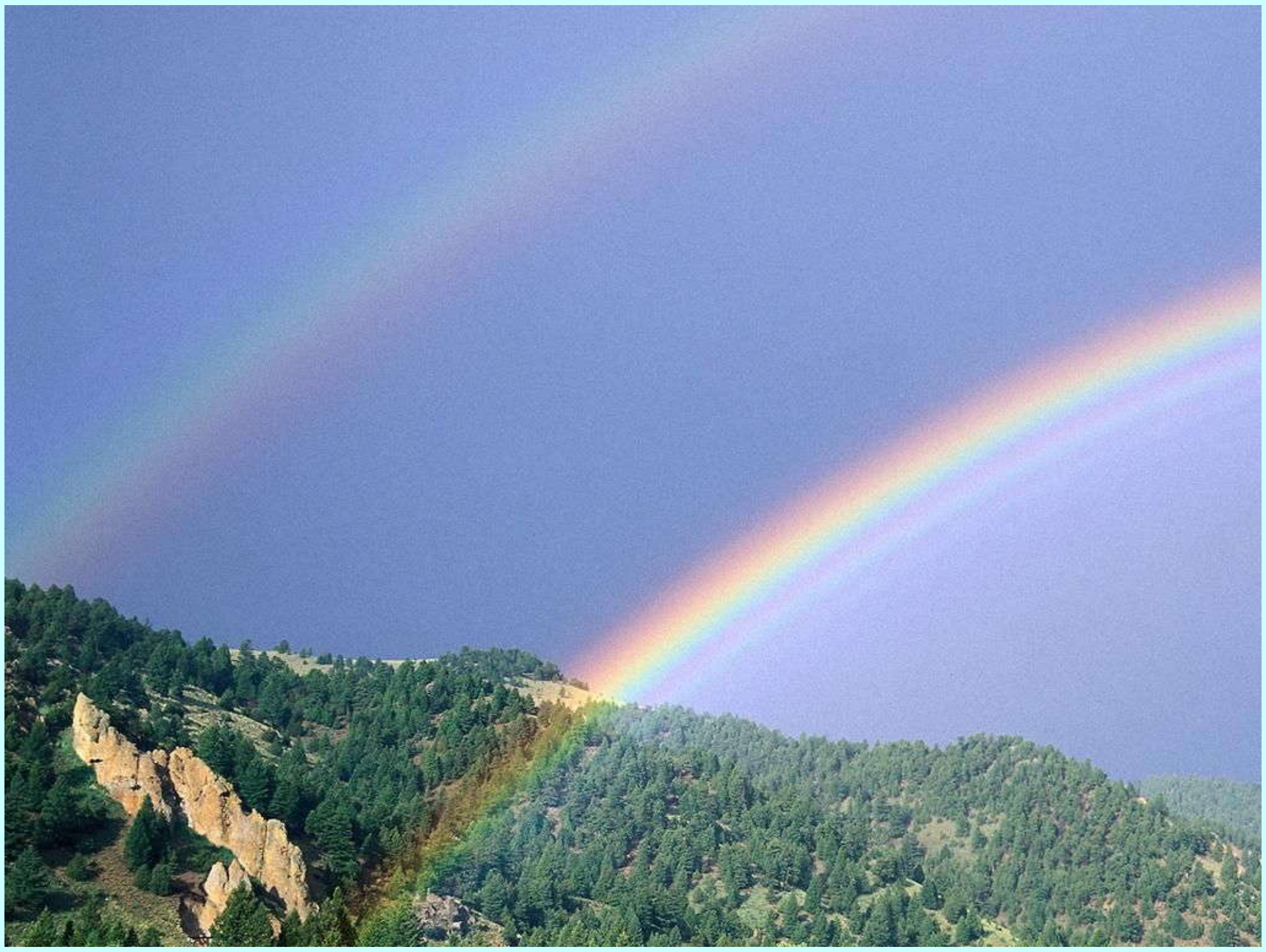

# Model of pathogenesis of psoriasis

## Part 2. Local processes

Edition e1.3

**Mikhail Peslyak**

---

---

Web: www.psorias.info, E-mail: info@psorias.info

---

---

The given monograph is the authorized translation of the book, published in Russian. Translation has been carried out with active support of Translation agency "LingLab".

The author thanks Vladimir Turbin for the help with the proofreading.

Periodic reprinting at occurrence of new materials or detection of serious errors is supposed. Further, updated texts (Russian and English) of the monograph will be regularly placed in the Internet.

Details and links on www.psorias.info.

## Abstract

**Observational and analytical research of results of experimental and theoretical studies on etiology and pathogenesis of psoriatic disease is carried out. The new model of pathogenesis of psoriasis – skin reaction to systemic psoriatic process SPP is formulated.**

**Because of PAMP-nemia in blood flow fractions of tolerized monocytes Mo-T and dendritic cells DC-T are formed. Tolerized Mo-T and DC-T also are kPAMP-carriers. Within the limits of PAMP-nemia also (PG-Y)-nemia takes place. Therefore in tolerized fractions subfractions of monocytes Mo-R and dendritic cells DC-R are formed. These Mo-R and DC-R also are (PG-Y)-carriers. Tolerized Mo-T and DC-T (incl. Mo-R and DC-R) have chemostatuses similar to nonactivated ones. Therefore these cells participate in homeostatic and-or inflammatory renewal of pool of dermal macrophages and dendritic cells of non-resident origin.**

**Psoriatic inflammation is regarded as a reaction of the skin immune system to activity of Mo-R and DC-R involved in derma from blood flow. They contain Y-antigen and, getting to derma, can be transformed in mature maDC-Y and present this antigen to Y-specific T-lymphocytes as well as activate them. Y-antigen is a part of the interpeptide bridge IB-Y. Therefore, the skin immune system can incorrectly interpret Y-antigen presentation as a sign of external PsB-infection and switch one of mechanisms of protection against bacterial infection – epidermal hyperproliferation.**

**Psoriatic plaque can be initiated only during action of local inflammatory process LP2 in derma causing not only innate, but also adaptive response. In particular, it is possible at LP2(IN) - open trauma of derma or at LP2(HPV) - HPV-carriage of keratinocytes. The level of Y-priming (presence and concentration of Y-specific T-lymphocytes in prepsoriatic derma and in lymph nodes) also determines possibility of psoriatic plaque initiation.**

**Existence and severity of psoriatic plaque is determined by intensity of Y-antigen income into derma (inside Mo-R and DC-R). Severity of plaque is aggavated by intensity of kPAMP income into derma (inside Mo-T and DC-T). Intensity of both incomes depends on SPP severity.**

**Severity of plaque is aggravated by LP2-inflammation if it persists after this plaque initiation. New Mo-T, DC-T (incl. Mo-R, DC-R) and Y-specific T-lymphocytes are constantly attracted into plaques from blood flow, and so support vicious cycles. Only at decrease of SPP severity, these vicious cycles weaken and natural remission of plaques takes place, up to their complete disappearance.**

**The detailed analysis comparing the new model of pathogenesis with five other previously published models is carried out.**

**The given monograph is the authorized translation of the book, published in Russian (ISBN 978-5-905504-03-7; edition r1.3).**

# Content

---

## *List of figures*

*«Is it true that firemen put fires out instead of going to start them?»*
Ray Bradbury, «Fahrenheit 451»

# Introduction

This book represents the second part of the monograph «Model of pathogenesis of psoriasis».

The first part of the monograph "Model of pathogenesis of psoriasis. Part 1. Systemic psoriatic process" has been published earlier (Peslyak 2011). For possible reader it is highly desirable to complete reading of Part 1, before starting Part 2. References to concrete sections of Part 1 are used, if necessary.

Part 2 is devoted to description and analysis of local processes determining skin manifestations of psoriatic disease - consequences of SPP – systemic psoriatic process. Their detailed description is given in the section "Y-model of pathogenesis".

### Research objective

The present work represents observational and analytical research of results of experimental and theoretical studies on etiology and pathogenesis of psoriatic disease. The main objective of the monograph is construction of a new model of pathogenesis of psoriasis (further called Y-model), based on results of latest researches (Fig. 2-6).

### Research methods

At creation of the second part of the book, search and analysis of works where well-based models of pathogenesis of psoriasis are offered have been carried out. Experimental works supporting these models have also been studied.

Besides, results of important experimental researches, those not yet employed in the creation of models of pathogenesis of psoriasis, were studied.

Also, publications investigating the condition of prepsoriatic skin and events initiating psoriatic plaque (Koebner's effect) have been analysed. Search of publications was carried out using Medline and Embase, publications in Russian – Central Scientific Medical Library and Scientific Electronic Library. Review and-or analysis of results of these researches are given at the description of (sub)processes. Real or prospective interaction of (sub)processes is given in the end of the description of dependent (sub)process.

### Abbreviations and terms

Appendices contains decoding of conventional and new abbreviations (App. 2-1 and App. 2-1a). Where possible, links to Wikipedia are given in these tables. App. 2-2 contains the list of new terms (sample from App. 2, Part 1).

### Novelty and hypotheses

A model cannot be called "new" if it does not contain new fragments that are absent in other models, including hypothetical (sub)processes and-or dependencies. All new fragments of Y-model do not contradict to commonly accepted facts. And some of new fragments are based on recently discovered facts. In the section "Discussion" and App. 2-9, the comparative analysis of Y-model with five other models of pathogenesis published earlier is carried out (Nestle 2009a and Nestle 2009b, Guttman-Yassky 2011, Tonel 2009, Gilliet 2008, Sabat 2011). The results of this comparison are collected in the block «Y-model novelty», presented in the same section. Similarly to Part 1, in App. 2-4 references to publications explaining some local processes are given. If such publications are not present, a brief suggested hypothesis has been formulated.

### Some facts and assumptions

The top layer of skin, epidermis, provides the first barrier of protection against external influence. Derma is separated from epidermis by the basal membrane, and contains vascular network for supplying with nutrients epidermis, which is lacking its own blood vessels.

Apparently healthy skin of patients with psoriasis (further psoriatics) without plaques (uninvolved, nonlesional, symptomless) is also called "prepsoriatic" or the abbreviation NLS is used (NonLesional Skin). Psoriatic skin is designated by the abbreviation PLS (Psoriatic Lesional Skin). Further, these terms and abbreviations will be used everywhere.

NLS is characterized by mild dermal inflammation (not exceeding basal layers of epidermis), increased vascularity, itch, dryness, decrease of barrier function of the cornual layer, larger vulnerability. After psoriatic plaque (further PLS-plaque or plaque) appearance, active priming of surrounding skin occurs (increase of Tem concentration). The width of the priming ring depends on the severity, size and growth rate of plaque and can reach 5-10 centimeters. The internal border of the ring (directly near to plaque) by its characteristics is very close to condition of skin inside the focus of plaque. At relapses and aggravations the width of the ring increases, at natural remission - decreases (Cameron 2002, van de Kerkhof 1996, Vissers 2004).

Immunocytes are present in both derma and in epidermis. These are epidermal dendritic LC (Langerhans cells) and dermal DC (DDC), macrophages (MF), mast cells and T-lymphocytes (TL). In the absence of inflammation, insufficient number of B-lymphocytes and NK cells is present in derma, and plasmacytoid DC (PDC) and neutrophils (Neu) are practically absent. All epidermal LC, as well as a big part of DDC (about 60%) at the homeostatis state are originated from the precursors constantly presented in skin. However, a smaller part (about 40%) of DDC and practically all MF are originated from attracted blood DC and Mo. At inflammation, LC pool starts to be replenished by blood CCR2+CD14+Mo attracted in skin and are gradually transformed in LC. At inflammation, the most part of DDC pool is replenished by blood DC and, probably, from blood Mo attracted in skin that can be transformed in DDC (Bogunovic 2006, Ginhoux 2007, Ginhoux 2006, Koch 2002). DDC are more active in presentation of antigens and make most part of traffic in regional lymph nodes. LC play the leading role in formation of tolerance of the skin immune system in relation to skin commensals. DDC provide adaptive response against bacteria and viruses during trauma or dermal infection (Baker 2006c).

Similar processes of renewal of DDC and MF pools occur in NLS and PLS. The critical role of dermal DC and MF in initiation and development of PLS-plaques became doubtless last years (Clark 2006b, Jariwala 2007, Marble 2007, Wang 2006, Zaba 2009b). First of all, these are proinflammatory tissue CD11c+DDC, actively secreting TNF-alpha and iNOS, so-called TipDC. In PLS, quantity of TipDC increases in derma and sublayers of epidermis as much as 4-6 times, and the most part of these cells is attracted from blood flow (Lowes 2005, Zaba 2009a). The reason of activity of TipDC is unknown. The cause of activity of part of dermal CD68+Mo and MF, intensively secreting TNF-alpha, is also unknown (Fig. 2-4, Fig. 2-5).

For the first time it has become possible to prove presence of PG in PLS with the help of 2E9 monoclonal antibodies in the study (Baker 2006a). PG has been found in CD68+MF and (possibly) in CD1a(-)DDC and has not been found in CD207+DC and CD83+DC. These results correlate with presviously obtained (Zaba 2009a) interms of CD11c+BDCA-1(-)TipDC expressing CD14 (and, therefore, almost not expressing CD1a (Angel 2006) and weakly expressing CD83).

In spite of the fact that BDCA-1(low)CD14+DDC are less mature than BDCA-1(high)CD1a+DDC, they may carry out presentation and activate TL (Angel 2006).

Precursors of BDCA-1(-)TipDC can be blood BDCA-1(-)CD14(-)CD16+slanDC (Hansel 2011) and blood CD14+CD16+Mo (Piccioli 2007). Among all fractions of blood DC and Mo, these cells secrete TNF-alpha in reply to PAMP-influence most actively.

For the first time the assumption that antigen presented by TipDC, is a fragment of the interpeptide bridges of BSPG made in (Baker 2006a). However, PG has not been found in CD83+maDC. Probably, during transformation of DC-R and Mo-R into CD83+maDC-Y, PG fragments degraded in such a way that epitopes corresponding to 2E9 monoclonal antibodies were not preserved.

Presence of DC-R and Mo-R in NLS is necessary, but not enough for transformation of NLS in PLS. Transition of DC-R and Mo-R from passive condition in the active one can occur only during adaptive response against local initiating and aggravating process (further process LP2). Existence and development of PLS-plaque is determined by constant attraction into derma of new DC-R and Mo-R from blood flow (more precisely – by the quantity of PG-Y fragments brought by them). At decrease of the level or full termination of such attraction remission of PLS-plaque up to its full disappearance in connection with natural extinction of maDC-Y pool occurs.

# Y-model of pathogenesis

For the detailed description of Y-model, all events, both previous and concurrent with development of one plaque, should be divided on processes and phases. Y-model in the form of the scheme of interaction of local processes is represented in Fig. 2-6. Plaque development by phases is shown in Fig. 2-8 (colors as well as in Tab. 2.1). Initiating and aggravating process LP2 is not specified in these schemes.

The most important causal dependencies are represented by color arrows. The color of an arrow depends on the color of a causative process. For simplification of schemes, some dependencies are not designated. In App.2-3, the table including all causal dependencies between processes and subprocesses is given.

The more detailed interaction scheme is developed for two cases of LP2: LP2(IN) - open trauma of derma (Fig. 2-9) and at LP2(HPV) - HPV-carriage KC (Fig. 2-18).

For each cytokine (active in LP2- and-or at PLS-inflammation), secreting cells and effected cells are specified (App. 2-7).

Table 2-1 contains processes and phases of plaque development during constant action of SPP. Each phase is characterized by occurring (sub)processes (+ or *) during its duration, and also by their intensity. For each phase, one or two marker characterizing processes are is chosen.

It is supposed that Y-priming of skin has already occurred and, consequently, phase 0 is absent. More information about phase 0 and Y-priming is given in the section «Phases of psoriatic plaque development».

***Table 2-1. Processes and phases of psoriatic plaque development***

| | | Prepsoriasis | | | | Psoriasis | |
|---|---|---|---|---|---|---|---|
| | **Phases** | **Phase 1** | **Phase 2** | **Phase 3** | **Phase 4** | **Phase 5** | **Phase 6** |
| | The phase start determines: | - | LP2 start | LP4 start | Start of LP5 and/or LP7.1 | LP8 start | End of LP2, LP3 and LP5 |
| | Marker processes | Homeostatic LP1 | LP2 and LP3 | LP4 | LP5 | LP5 and LP8 | Self-sufficient LP8 |
| **(Sub)processes** | Comment | LP2 is absent. Common for any LP2. | | | | | LP2 ended. Common for any LP2. |
| **SPP** | Systemic psoriatic process. | * | * | * | * | * | * |
| **LP1** | Attraction of immunocytes from blood flow. | + | ++ | +++ | +++ | +++ | +++ |
| **LP2** | Initiating and aggravating process. | | ++ | +++ | +++ | +++ | |
| **LP3** | Innate response against LP2. | | ++ | +++ | +++ | +++ | |
| **LP4** | Trigger of adaptive response against LP2. | | | +++ | ++ | | |
| **LP5** | Adaptive response against LP2. | | | | +++ | +++ | |
| **LP6** | Mo and DC transformations. | | | | | | |
| **LP6.1** | Loss of tolerance to kPAMP. | | | | ++ | +++ | +++ |
| **LP6.2** | MF and MoDC formation. | + | ++ | ++ | +++ | +++ | +++ |
| **LP6.3** | maDC-Z formation. | | | ++ | +++ | +++ | |
| **LP6.4** | maDC-Y formation. | | | | ++ | +++ | +++ |
| **LP7** | Lymph nodes. Clonal proliferation. | | | | | | |
| **LP7.1** | TL-Z formation. | | | | ++ | +++ | |
| **LP7.2** | TL-Y formation. | | | | | ++ | +++ |
| **LP8** | False adaptive response to imaginary PsB-infection. | | | | | | |
| **LP8.1** | Y-antigen presentation by maDC-Y to effector ThN-Y. | | | | | +++ | +++ |
| **LP8.2** | KC hyperproliferation. Change of skin architecture. Growth of basal membrane area and vascularity increase. | | | | | +++ | +++ |

**Notes to Tab. 2-1**.

Color in the first column (here and everywhere further) is used for highlighting of particular process and its subprocesses. An empty table cell means that (sub)process does not occur. SPP activity mark with the symbol *. SPP occurs constantly with intensity not depending on local processes. However, initiation and support of any PLS-plaque depends on SPP intensity. Local (sub)process: + (white) - occurs with mild intensity; ++ (beige) - inflammatory, average intensity; +++ (pink) - inflammatory, high intensity;

Illustrations of interacting processes for each phase (under condition of Y-priming preceding PLS-plaque initiation) are presented in the series of figures. The figures corresponding to phase 1 and phase 6, do not depend from LP2 (Tab.2-2).

***Table 2-2. Processes and phases in figures***

<table>
<tr><th rowspan="2">Initial and aggravating process</th><th rowspan="2">Model of patho-genesis</th><th rowspan="2">Graphs of antigenic presentation and Tem level</th><th colspan="4">Prepsoriasis</th><th colspan="2">Psoriasis</th></tr>
<tr><th>Phase 1</th><th>Phase 2</th><th>Phase 3</th><th>Phase 4</th><th>Phase 5</th><th>Phase 6</th></tr>
<tr><td>LP2(IN) – open trauma of derma</td><td>Fig. 2-9</td><td>Fig. 2-10</td><td rowspan="2">Fig. 2-11</td><td>No</td><td>Fig. 2-14</td><td>Fig. 2-15</td><td>Fig. 2-16</td><td rowspan="2">Fig. 2-17</td></tr>
<tr><td>LP2(HPV) – Human Papilloma Virus – HPV-carriage KC</td><td>Fig. 2-18</td><td>Fig. 2-19</td><td>Fig. 2-20</td><td>Fig. 2-21</td><td>Fig. 2-22</td><td>Fig. 2-23</td></tr>
</table>

In illustrations, symbols from Part 1 (Fig. 2-1, Fig. 2-2) and additional symbols offered here (Fig. 2-3) are used. In following section the detailed description of each local process is given.

# Local processes

## *Process LP1. Attraction of immunocytes from blood flow.*

Renewal of pool of dermal immunocytes of non-resident origin is constant process occurring in the absence of local inflammation as well (Fig. 2-11). Process LP1 occurs during any of phases (Tab.2-1).

After LP2 start, process LP1 is intensified.

Renewal of this pool takes place due to action of chemokines, as well as AMP (antimicrobial proteins) and some cytokines possessing chemokine properties. All of them are produced by various skin cells, and their assortment and quantity depend on skin condition. Ligands of chemokines are certain chemokine receptors expressed by blood immunocytes. The list and description of all chemokines and their corresponding receptors, as well as parameters of their interaction, can be found in (Commins 2010) and on the resource IUPHAR (Murphy 2010). The observational information on chemokine receptors, chemokines and AMP, involved in the given process, can be found in two mutually supplementing Appendices:

Appendix 2-5. Chemokine receptors, chemokines and AMP, certainly or possibly involved in LP1 process.

Appendix 2-6. Chemokine receptors and blood immunocytes.

### *Subprocess LP1.1. Attraction of Mo and DC, Mo-T and DC-T (incl. Mo-R and DC-R) from blood flow.*

The nomenclature of blood monocytes and dendritic cells has been recently accepted (Ziegler-Heitbrock 2010). It is based on results of many previous studies including (Piccioli 2007, Serbina 2008, Tacke 2006).

According nomenclature population of blood Mo consists of three fractions: Classical CD14++CD16(-)Mo (in the normal state - to 90%, everywhere further they are designated as CD14++Mo), intermediate CD14+CD16+Mo (in the norm - to 10%) and nonclassical CD14(low)CD16+Mo (in the norm to 5%). So far monocytes from two last fractions together taken are considered, they can be is designated as CD16+Mo.

According to this nomenclature, the population of blood myeloid DC consists of two fractions: BDCA-1+DC and BDCA-3+DC. However, the fraction of CD14(-)CD16+BDCA-1(-)slanDC does not match this nomenclature completely, as they express CD16, and expression of BDCA-3 is not obligatory for cells of this fraction (Hansel 2011, Schakel 2006). Therefore, authors of reports devoted to slanDC divide the population of blood myeloid dendritic cells in two fractions: BDCA-1+DC (smaller) and BDCA-1(-)DC (bigger). They include slanDC subfraction in BDCA-1(-)DC fraction. Such divergence in classification can be explained by affinity of characteristics of slanDC to characteristics of CD14(-)CD16+Mo (Schakel 2006). Some authors believe that slanDC should be considered as monocytes, namely a part of the fraction of nonclassical CD14(low)CD16+Mo (Teunissen 2011).

In Fig. 2-4 variants of origin of dermal macrophages MF (further DMF) and dermal dendritic cells DC (further DDC) are represented. This figure also represents the subsequent possible transformation of DDC and MoDC in mature dendritic cells maDC (Angel 2006, Haniffa 2009, Zaba 2009a).

The role of chemokines and AMP, determining traffic of Mo and DC is analysed in some works (Bachmann 2006, Diamond 2009, Gautier 2009, Sozzani 2005). Attraction of Mo and DC from blood flow into derma takes place due to interaction of their chemokine receptors and chemokines expressed by endothelial cells and secreted in derma. This attraction occurrs at homeostasis and at inflammation due to various combinations of chemokine receptors and chemokines. This attraction is analysed for fraction of medium-classical monocytes CD14+CD16+Mo (Part 1, SP8, fig. 9).

In PLS-derma abrupt (more than 30 times) growth of BCDA-1(-)TipDC number can be seen (Zaba 2009a).

They can originate from blood tolerized BDCA-1(-)DC, in particular from the subfraction CD14(-)CD16+slanDC well expressing CX3CR1 (Hansel 2011). However, they can originate from blood tolerized CD14+CD16+Mo those well expressing CCR2 and CX3CR1.

Attraction of these cells from blood flow occurs first due to chemokine CX3CL1 (fractalkine) secreted by keratinocytes KC (Sugaya 2003) and DDC (Raychaudhuri 2001) in PLS, and, additionally, CCL2, HBD-2 and HBD-3, abundantly secreted in PLS-derma.

Recently it has been proved that CD163+MF in PLS-derma are classically activated (Fuentes-Duculan 2010), it is probable that their part originates from CD14++Mo (Fig. 2-4).

As it is known, at intensive inflammation the part of LC is formed from Mo and-or DC attracted from blood (Angel 2006, Ginhoux 2007, Ginhoux 2010). In PLS-epidermis the quantity of LC is not increased in comparison with the norm (Jariwala 2007, Sabat 2007). Within Y-model limits, LC may play the most essential role during LP5, while it still depends on LP2.

At inflammation, CCL2 secretion (ligand of CCR2) essentially increases, therefore attraction of CD14++Mo, CD14+CD16+Mo, as well as parts of DC and PDC increases too.

It has been established recently that HBD-2 and HBD-3 are ligands of CCR2 and support attraction of immunocytes as well as CCL2 (Rohrl 2010). At inflammation, secretion of CX3CL1 (ligand of CX3CR1) increases, which promotes CD16+Mo attraction. At inflammation, secretion of CCL20 (ligand of CCR6) increases, which strengthens attraction of CCR6+DC. At inflammation, chemerin secretion (ligand of ChemR23) increases, which strengthens attraction of PDC, as well as parts of DC and CD16+Mo.

All tolerized Mo-T and DC-T (incl. Mo-R and DC-R) participate in renewal of pool of dermal macrophages MF and dendritic cells DC of non-resident origin. It occurs because their chemostatuses are similar to chemostatuses of nonactivated Mo and DC (Part 1, App. 4; App. 2-2).

### *Subprocess LP1.2. Attraction of other immunocytes from blood flow.*

Homeostatic attraction of other blood immunocytes - plasmacytoid dendritic cells PDC, natural killers NK, neutrophils Neu, and also T-lymphocytes (TL) occurs in the absence of local inflammation as well. This attraction is intensified during local inflammation. It is necessary during immune responses LP3, LP5 or LP8.

PDC play key role in LP4. Mass secretion of IFN-alpha by PDC precedes plaque development (Nestle 2005a). Attraction of PDC in NLS-derma takes place due to the receptor ChemR23 - ligand of chemerin (Albanesi 2010, Nakajima 2010, Skrzeczynska-Moncznik 2009b). In PLS, chemerin is secreted both in epidermis and in dermal fibroblasts, while in NLS and in the norm - mainly in epidermis. NLS surrounding active and early PLS, is characterized by strong expression of chemerin in derma and presence of CD15+Neu and CD123+BDCA-2+ChemR23+pDC (Albanesi 2009). PDC traffic at homeostasis and inflammation in skin, besides receptor ChemR23, is also provided by receptors CXCR3 (ligand of CXCL10) and CXCR4 (ligand of CXCL12) (Sozzani 2010).

The role of natural natural killers NK and NKT (TL with properties of killers) at psoriasis has been carefully studied for the last decade. Their distribution in NLS and PLS studied in (Cameron 2002). In one study (Gilhar 2002) it is shown that at injection of NK in NLS-transplant replaced to SCID-mice, PLS-plaque develops, which proves NK role in psoriasis pathogenesis. Later, the same researchers have made an attempt to define this role more precisely (Gilhar 2006). The observational study (Peternel 2009) is devoted to NKT role at psoriasis. Attraction of NK is carried out due to receptors CCR5, CXCR1, CXCR3, CX3CR1, ChemR23.

The role of neutrophils Neu in SPP has been suggested in Part 1. Experimental decrease of Neu level in PLS does not render a positive influence and, hence, skin Neu are not obligatory for maintenance of already existing plaques (Numerof 2006). However, Neu play essential role in LP4, actively secreting LL37. Attraction of Neu to inflammation locus is carried out with the help of CXCR1 and CXCR2 (ligands of CXCL8 and CXCL1), as well as CCR2.

Attraction of TL from blood flow is carried out with the help of CCR4, CCR6, CCR10 and CXCR3 expressed by various TL fractions in various combinations (Teraki 2004, Kagami 2010). CCR6 is considered as a key receptor on attraction of Th1 and Th17 in PLS (Hedrick 2010).

More information about expression of chemokine receptors in immunocytes, and their interaction with ligands (chemokines and AMP) can be found in App. 2-5 and 2-6.

<u>LP1.1 depends on SPP.</u> A share of Mo-T and DC-T (incl. Mo-R and DC-R) among attracted blood Mo and DC depends on SPP intensity.

<u>LP1 depends on LP3 and LP5</u>. From phase 2 and to phase 4 (i.e. at the prepsoriasis stage), the spectrum of secreted chemokines and antimicrobial proteins with chemokine properties is determined exclusively by LP2.

CCL2 and CX3CL1 are actively secreted at LP2(IN), as well as antimicrobial proteins LL37, HBD-2 and HBD-3 (Fig. 2-9). The detailed information about it is given at description of LP3(IN) and LP5(IN) processes. At LP2(HPV), CCL2, HBD-2 and HBD-3 are actively secreted (Fig. 2-18). The detailed information about it is given at description of LP3(HPV) and LP5(HPV).

<u>LP1 depends on LP8.</u> After phase 5 start (i.e. after LP8 start) the spectrum of dermal chemokines and AMP extends, and their secretion increases. For example, HBD-2 and HBD-3 are secreted in large amounts (De Jongh 2005, Gambichler 2008). Thereby, the rate of attraction of all blood immunocytes (in comparison with previous phases) is amplified.

## *Process LP2. Initiating and aggravating process.*

LP2 is specific local inflammatory process in skin, at which conditions for initiation (and, maybe, support) of process LP8 - false adaptive response against PsB-infection can be created. It can occur, if LP3 (innate response against LP2) will be insufficient for suppression and elimination of LP2. In this case, LP4 switches (phase 3 begins) and then LP5 begins (phase 4 begins).

During phase 4, conditions for LP6.1 start, and then for LP6.4 start can appear. Without action of these subprocesses the start and support of LP8 is impossible.

With LP2 start - phase 2 begins, with LP2 end - phase 5 terminates (Tab.2-1, Fig. 2-8).

If LP2 exists for long time after LP8 initiation (phase 5), it aggravates LP8 through LP3 and LP5 as intensive functioning of vicious cycles B and C is maintained (Fig. 2-6):

B = {LP1.1 > LP6 > LP8 > LP1};

C = {LP6.4 > LP7.2 > LP1.2 > LP8 > LP6}.

If the vicious cycle B has begun acting, LP2 end may not render an influence on functioning of this cycle. If the vicious cycles B and C (with SPP support) appear to be self-sufficient, phase 6 begins.

LP2 role may be played by any of the following influences and-or processes (Fry 2007b):

- LP2(IN) = open trauma of derma (Fig. 2-9; Fig. 2-10; Fig. 2-11; Fig. 2-14; Fig. 2-15; Fig. 2-16; Fig. 2-17)
- LP2(HPV) = HPV-carriage KC (Fig. 2-18; Fig. 2-19; Fig. 2-11; Fig. 2-21; Fig. 2-22; Fig. 2-23; Fig. 2-17)
- LP2(PsB) = skin PsB-infection
- Combustion, contact dermatitis
- S. aureus (further SA)
- Malassezia Species (further MS)
- Candida Albicans (further CA).

The trauma and any PsB-infection of skin are only initiating processes since, as a rule, at the very initial stage of PLS-plaque they are usually eliminated (rapid transition to phase 6 takes place).

Bacterial or virus LP2 can persist in PLS-epidermis (phase 5). Diffusion of LP2 outside PLS-plaque on surrounding NLS can promote expansion of this plaque. Diffusion of LP2 on remote NLS provokes appearance of new PLS-plaques.

Microflora in normal skin and PLS-microflora was investigated and compared several times. It is well-known that SA, MS and CA in PLS-plaques are found more often than in the norm and, as a rule, aggravate such plaques (Kozlova 2007, Fomina 2009, Balci 2009, Chiller 2001, Doern 1999, Gao 2008). The suggestion about their triggerring role is based also on results of studeis (Ashbee 2002, Lober 1982, Tomi 2005).

However, at psoriasis the most widely spread is HPV-carriage of KC, and its rate essentially exceeds such one in healthy patients (Cronin 2008, Mahe 2003, Majewski 2003, Prignano 2005). It is interesting that prevailing HPV types depend on a region, in which researches were conducted. It is probable that HPV-carriage of KC is the most widespread process initiating and aggravating PLS-plaques. Therefore, further in processes and subprocesses connected with LP2, Y-model is specified and illustrated at LP2(IN) - trauma and at LP2(HPV) - HPV-carriage of KC.

LP2 is suppressed by LP3 and LP5.

LP2 is suppressed by LP8.2. It is not represented in schemes.

At LP2(IN), hyperproliferation of KC promotes the prompt healing of trauma area.

At LP2(HPV), hyperproliferation of KC reduces the time of stay of KC (including KC-v) in basal layers of epidermis. As a result, HPV-virions concentration goes down in intercellular space of these layers, and, as a consequence, the probability of HPV diffusion among basal KC decreases.

### *LP2(IN). Open trauma of derma*

Trauma of NLS does not always cause the subsequent PLS-plaque initiation (it is known, the Koebner effect appears in 25-30% of cases). The Koebner effect is impossible, when derma remains intact and if trauma is closed (Kalayciyan 2007, Weiss 2002). Wound repair is always accompanied by inflammation, but only invasion of bacterial and virus epidermal commensals into derma can cause dermal LP3 and LP5 against them. Further everywhere we consider only open trauma of derma, at which the Koebner effect, i.e. PLS-plaque initiation, is possible. During trauma activation of resident and attraction of blood Mo, DC and PDC occurs. While being destructed, damaged skin cells release intracellular proteins into the intercellular medium, including fragments of self-DNA and self-RNA.

### *LP2(HPV). HPV-carriage KC*

HPV - Human Papilloma Virus (Hebner 2006, Stanley 2006, Stanley 2009). Chronic asymptomatic HPV-carriage of KC (including stem basal) not oncogenous or of weak-oncogenous HPV types can last for years without any external manifestations. Simultaneous carriage of several various HPV types is possible. Extraneous contamination may be developed by trauma, infection or from adjacent KC-v.

For the first time, connection of psoriasis with HPV-carriage of KC of prepsoriatic and psoriatic skin was noted in (Favre 1998). It has been shown that epidermal flakes in 92% of psoriatics (37 people) contain DNA of EV-HPV (HPV types connected with epidermodysplasia verruciformis). There it has been shown that in blood of 24,5% of psoriatics (38 from 155) antibodies to HPV5 are detected. In the control group, they have been found in 5% of healthy subjects (3 of 60). The similar result has been observed for antibodies to HPV8 (43% of psoriatics against 7% controls) (Pfister 2003).

In one study (Pfister 2003) it is shown that epidermal flakes (54 people) contain DNA of HPV5 in 74% of cases if the biological material is taken from PLS, and in 33%, if samples are received from NLS. At the same time, in the control group (20 people) DNA of HPV5 was not found.

Similar results were received in one study (Weissenborn 1999). DNA of HPV was found in 83% of samples of skin taken from PLS (54 people), while in control group DNA of HPV was found in 19% (42 people). Types of HPV were the most common 36 (62%), 5 (38%), 38 (24%). Later, model of pathogenesis of psoriasis based only on EV-HPV-influence was suggested (Majewski 2003).

Many types of HPV (including EV-HPV) are skin commensals. Skin of 74% of healthy patients (23 people) contains constant set of beta-HPV without any external manifestations (more than 6 months) (De Koning 2005).

This is proved by extended research work (lasting more than 6 years), which showed that 70% of healthy patients (42 people) have constant skin HPV-carriage without any external manifestations (samples undertook from forehead). During the term of observation, the spectrum of HPV types found in one patient, changed, but not completely. In 48% of patients, at least one, but being of the same type of HPV was detected (Hazard 2007).

However, in (Cronin 2008) it has been shown that EV-HPV-spectrum of carriage in psoriatics is similar to the spectrum of carriage in control groups, and any specific HPV types do not prevail. Prevalent spread of EV-HPV-carriage in psoriatics is connected by many authors with earlier treatment (first of all, with PUVA-therapy). For this reason, psoriatics (20 people), who never received UV-therapy, external or systemic immunodepressants were selected and enrolled in the study. Samples of NLS (forearm) and follicles of hair from brows were taken and investigated. The general HPV prevalence in psoriatics appeared to exceed the control group of 23 people (83% against 46%). EV-HPV (HPV5, HPV36) were found not more often than other types.

Commensal HPV-carriage of KC is proved in on study (Viviano 2005), when HPV was found in PLS in 71% of psoriatics (38 people) and only in 58% of the control group (36 people).

DNA of HPV is found in 87% of samples of PLS (60 of 69), and only in 44% of samples of normal skin of the control group (20 of 46) ($p <0,001$). Virus load HPVL = lg ((amount of HPV DNA) / ($10^5$ human cells)) was determined.

It has appeared that HPVL < 2 for 40% of HPV+psoriatic samples and for 83% of HPV+norm, 2 < HPVL < 3 for 45% of HPV+psoriatic samples and for 17% of HPV+norm and HPVL >3 for 15% of HPV+psoriatic samples and for 0% of HPV+norm. So, it has appeared that excess of HPVL in psoriatics compared with healthy subjects from the control group is a reliable fact. Also, it has been shown that HPVL is significantly lower in psoriatics in the remission stage in comparison with the aggravation stage. HPVL increase in PLS correlates with disturbances of skin microflora, in particular, the number of S.aureus increases (Kozlova 2007, Fomina 2009).

It correlates with the average value (n - number of carriers) on virus load for PLS (HPV5, n=19, HPVL=2,3); (HPV20, n=11, HPVL=3); (HPV36, n=41, HPVL=2). These results are presented in (Pfister 2003, Tab. 1) and for comparison are recalculated using the logarithmic formula for HPVL (Fomina 2009).

The answer to the question on primacy or secondariness of increased HPV-carriage in psoriatics is in one research (Mahe 2003), where it is shown that the frequency of EV-HPV detection in psoriatic flakes in children coincides with its detection frequency in adult patients. It means that HPV-carriage is of primary character in relation to psoriasis initiation. In the same research it is shown that connection between EV-HPV-carriage and severity or duration of psoriasis is absent.

Examination of 26 psoriatics (average age 44, average PASI 10,2) has shown 100% HPV-carriage in hair from brows (including hair bulb). On the average, 4,8 various types of HPV were found. Antibodies to HPV were present in blood of 56% of psoriatics. It was not revealed any correlation between age and quantity of found types HPV and-or presence of HPV-antibodies at psoriatics while such correlations it is observed at the healthy population (De Koning 2011).

The results of researches listed above corresponds to the trigger role of HPV-carriage in Y-model. High % of HPV-carriage in psoriatics means that LP2(HPV) can be the basic trigger of PLS-plaques.

## *Process LP3. Innate response against LP2.*

During LP3, process LP1 is always intensified. If LP3 is insufficient for LP2 elimination (restraint), LP4 works and LP5 begins against LP2. Within LP3 limits PDC attraction from blood flow takes place, without which LP4 is impossible. Other immunocytes (which assortment depends on specific LP2) are also attracted from blood flow.

LP3 operates simultaneously with LP2, i.e. begins at phase 2 start, and terminates at phase 5 end (Tab.2-1).

LP3 depends on LP1.2. LP3 intensity is defined not only by resident, but also by non-resident immunocytes, such as Neu, NK, PDC and others.

LP3 is initiated and supported by LP2.

LP3 depends on LP5. It well-known that innate and adaptive responses strengthen each other.

This dependence is represented only in schemes at LP2(IN) and at LP2(HPV).

LP3 depends on LP6.2. LP3 intensity is defined by MF, formed from Mo (both resident and non-resident origin), as well as MF-T and MF-R, formed from Mo-T and Mo-R. This dependence is not shown in schemes.

### *LP3(IN). Innate response after trauma.*

Owing to trauma, release of AMP (antimicrobial proteins) and chemokines takes place (Fig. 2-9, Fig. 2-14). In particular, neutrophils Neu are the first cells that start to secrete LL37 saved in advance, and KC secrete HBD-2, HBD-3, LL37, CCL2, CX3CL1, etc. (Gillitzer 2001, Ishida 2008).

CCL2, HBD-2 and HBD-3 actively support chemotaxis of blood CCR2+Mo and CCR2+DC.

CX3CL1 actively supports chemotaxis of blood CX3CR1+Mo. Thus, intensification of LP1.1 occurs. In the course of healing, all antibacterial and anti-virus systems of skin protection are activated.

### *LP3(HPV). Innate response against HPV.*

KC-v (KC, infected by HPV) secrete cytokines, chemokines and AMP (in particular, IL-8, CCL2, CCL5, CXCL10, HBD-2, HBD-3, etc.), initiating LP3 against HPV (Chong 2006, De Andrea 2007, Kreuter 2009, Woodworth 2002). As a consequence, in addition to the resident, attraction of CXCR2+Neu, CCR2+Mo, CCR2+DC and CCR5+NK from blood flow into derma takes place. Also, attraction of CXCR3+PDC and TL takes place (Fig. 2-18, Fig. 2-20).

Let's notice that at HPV-affection of epithelial tissues, mRNA HBD-2 exceeds the normal range more than 1000 times (Chong 2006), which correlates with its high level of presence in PLS (Harder 2005) and excess of mRNA HBD-2 in PLS in 5000 (Gambichler 2008) or 20000 (De Jongh 2005) times.

Excess mRNA of HBD-3 in PLS in comparison with the norm is less considerable - 50 times (Gambichler 2008).

Mo (MF) and DC interact with HPV-antigenes, secrete TNF-alpha, IL-1beta, IL-6, IL-12, IL-23. Under the influence of these cytokines, initial events of angiogenesis and increase of endothelium permeability appear.

LP3 can be insufficient for complete release from HPV, but it is sufficient for restraint of its expansion and external HPV-manifestations. The balance between LP2(HPV) and LP3(HPV) processes leads to latent HPV-carriage.

Intensity of LP3(HPV) can be sufficient for epidermal HPV-expansion restraint (increase of KC-v number in basal layers). But some types of HPV influence invasiveness of KC-v, i.e. their ability to get through the basal membrane into derma. For example such ability is known for HPV8, HPV16 and HPV18 (Akgul 2005, Barbaresi 2010, Yoshida 2008).

Invasiveness of KC-v is raised by HPV-proteins, such as E2, E7, etc. In particular, they force KC-v to secrete metalloproteinases MMP (MMP1, MMP9, MMP14, etc.) actively. MMP influence permeability of basal membrane for KC-v and, thereby, let the beginning of dermal HPV-expansion.

MF, NK can distinguish KC-v and destroy them (Fig. 2-20, green arrows **kill**) both in epidermis, and in derma (Woodworth 2002). At disintegration of destroyed KC-v, the whole virions and fragments of HPV-DNA containing CpG get to the extracellular space.

At dermal HPV-expansion, the quantity of such virions and fragments of HPV-DNA in derma increases. On the other hand, in NLS-derma the most of PDC are located. It inevitably cause operation of LP4(HPV) - trigger of adaptive response against HPV.

It is possible to assume that at LP2(HPV), KC-v invasiveness (depending on specific HPV types) determines possibility of transition from latent HPV-carriage (phase 2) to the subsequent phases up to LP8 initiation (phase 5). I.e. determines ability of LP2(HPV) to be initiating and aggravating process. Let's notice that the increased level of various MMP in NLS and PLS is well-known (Lee 2009).

KC-v can process viral antigens and present them on MHC of class I (Black 2007). Thus, KC-v are transformed in KC-Z1 and become capable of interaction with TcN-Z1 (LP5.2). And, if TcN-Z1 is already in NLS, LP5.2(HPV) will begin earlier than LP5.1(HPV).

## *Subprocess LP3.1. Formation of LL37-complexes.*

### *LP3.1(IN). Formation of self-DNA-LL37 and self-RNA-LL37 complexes.*

At LP2(IN), KC and FB are damaged and self-DNA and self-RNA get into the extracellular dermal space. During LP3(IN) KC, Neu and FB actively secrete LL37 (Dorschner 2001).

At interaction of LL37 with self-DNA and self-RNA, complexes self-DNA-LL37 and self-RNA-LL37 are formed (Ganguly 2009, Gregorio 2010, Nograles 2010, Tang 2010). Only in a form of such complexes, PDC are able to both endocytose and deliver self-DNA and self-RNA to endosomal TLR9 for interaction.

Process LP4 (phase 3) begins when PDC endocytose complexes self-DNA-LL37 (cooperating with endosomal TLR9), and when PDC start to secrete IFN-alpha actively.

Complexes self-RNA-LL37 promote DC maturation (LP6.3 and LP6.4) (Ganguly 2009, Nograles 2010).

### *LP3.1(HPV). Formation of complexes self-DNA-LL37 and self-RNA-LL37 as well as CpG-LL37.*

At LP2(HPV), occurrence of self-DNA, self-RNA and HPV-DNA in the extracellular dermal space happens at destruction of KC-v getting from epidermis into derma. Such destruction occurs during LP3 and LP5 against HPV.

Immune response against HPV includes increased secretion of LL37 (Conner 2002).

At interaction of LL37 with self-DNA and self-RNA complexes self-DNA-LL37 and self-RNA-LL37 are formed.

At interaction of LL37 with CpG-motives complexes CpG-LL37 are formed. These complexes are able to be endocytosed by PDC much better (than free CpG), and then be delivered to endosomal TLR9 (Hurtado 2010).

Process LP4 (phase 3) begins when PDC endocytose free CpG-motives, complexes CpG-LL37 and self-DNA-LL37 (interacting with endosomal TLR9) and start to secrete IFN-alpha actively.

Complexes self-RNA-LL37 promote DC maturation (LP6.3 and LP6.4) (Ganguly 2009, Nograles 2010).

## *Process LP4. Trigger of adaptive response against LP2.*

Active peak-like secretion of IFN-alpha by PDC, is a signal formed only during innate response LP3 against LP2. This trigger signal provides active start of adaptive response LP5 against LP2 (Lande 2010, McKenna 2005, Seo 2010, Tang 2010, Zhang 2005).

This trigger starts working if LP2 intensity is combined with LP3 insufficiency.

LP4 start determines phase 2 end and short-term phase 3 start. This short-term phase 3 terminates at phase 4 start, namely with the launch of adaptive response LP5 against LP2.

During phase 4, process LP4 goes down (being suppressed) and by the time of phase 5 start, practically stops (Tab.2-1, Fig. 2-7, Fig. 2-8, Fig. 2-10).

LP4 has limited duration, occuring and initiating other processes and subprocesses (LP5, LP6.2 and LP6.3) one time (Fig. 2-6, dot-dash line).

PDC produce and secrete considerable amount of IFN-alpha approximately 7 days prior to appearance of pinpoint psoriatic plaque. For the first time it has been proved at carrying out experiments with NLS-transplants transferred to AGR129-mice (Boyman 2004, Nestle 2005a). More information about these experiments:

Appendix 2-8. Psoriasis on NLS-transplants transferred to AGR129-mice.

The majority of models of pathogenesis of psoriasis, published in recent years (App. 2-9), include dermal PDC as obligatory components. Those dermal PDC transitively and actively secrete IFN-alpha before plaque development (Albanesi 2010, Gilliet 2008, Nestle 2009a, Nestle 2009b, Tonel 2009).

When plaque has already started to develop, the level of IFN-alpha gradually decreases to the norm (LP4 calms down). Therefore, only in acute (but not chronic) plaques high level of MxA protein (which is a marker of previously high level of IFN-alpha) is detected (Fah 1995, Nestle 2005a).

PDC continue to be involved into PLS-derma and appear there in increased amount, however they lose their ability to intensive secretion of IFN-alpha (despite increased level of CpG-motives or self-DNA-LL37 complexes). It occurs under the influence of suppressive activity of high level of TNF-alpha (being secreted during LP5 and-or LP8) on PDC (Palucka 2005).

According to other data, PDC are represented also in PLS-derma, but only at early stages of plaque development, while at later stages they are absent (Albanesi 2010).

Decrease of IFN-alpha level by a single intravenous injection of MEDI-545 (monoclonal antibodies to IFN-alpha) does not influence psoriasis condition. 36 psoriatics participated in this study (Bissonnette 2010). This negative result confirms that IFN-alpha plays only a transit role during plaque initialization.

PDC attraction is carried out through their receptors CXCR3 (ligand of CXCL9, CXCL10, CXCL11) and ChemR23 (ligand of chemerin) (App. 2-5 and 2-6). In NLS, PDC level is essentially higher than in norm.

Levels of mRNA CXCR3 and its ligands CXCL9, CXCL10, CXCL11 are considerably increased in PLS in comparison with NLS, and in NLS they are significantly higher than the norm. The majority of CXCR3+ cells are localized in PLS-derma and almost all BDCA-2+PDC in PLS were CXCR3+ (Chen 2010).

Chemerin providing attraction of PDC, is mostly secreted by fibroblasts, as well as by mast and endothelial cells (Albanesi 2009, Albanesi 2010).

Increased HPV-carriage in NLS (Favre 1998, Prignano 2005) can cause increased secretion of both chemerin and CXCL10. Attraction of PDC during LP3 plays important preparatory role.

Condition for LP4 launch is an increased level of CpG-motives (or self-DNA-LL37 complexes) in the amount sufficient for activation of PDC, which massive secrete IFN-alpha. At psoriasis, LL37 are secreted by KC in all layers of epidermis (initially KC have its stocks in granules). LL37 is also represented in derma near to PDC accumulation. Fibroblasts and-or Neu can produce it (Lande 2007, Gilliet 2008).

At LP2 (PsB) Neu secrete LL37 in large amounts as they contain its stocks in granules (Dorschner 2001). Blood Neu are involved through CXCL8 (IL-8), secreted by Mo and MF in reaction to PsBP (in particular, PG-Y).

LP4 depends on LP1.2. Increased level of PDC arrived from blood flow in the area of future plaque development provides intensity of subsequent LP4.

LP4 depends on LP2. Intensity of LP2 at insufficiency of LP3 makes LP4 inevitable. At LP2(IN) this dependence is defined by the amount of self-DNA and self-RNA getting to the extracellular space. At LP2(HPV) this dependence is defined by the amount of CpG-motives.

LP4 depends on LP3. Insufficiency of LP3 at intensity of LP2 makes LP4 inevitable. At LP3(IN), the amount of secreted LL37 determines the amount of formed self-DNA-LL37 and self-RNA-LL37 complexes. At LP3 (HPV) the amount of HPV-DNA in the extracellular dermal space is determined by cytotoxic activity of NK, Mo and MF.

LP4 is suppressed by LP5. It occurs because of TNF-alpha, increased level of which is formed during LP5. If this suppression appears insufficient, additional suppression takes place from the moment of LP8 start (it is not shown in schemes and figures).

### *LP4(IN). Trigger of adaptive response after trauma.*

PDC endocytose self-DNA-LL37 and self-RNA-LL37 complexes, formed during LP3.1(IN). Inside PDC they are delivered to endosomal TLR7 and TLR9 and interact with them (self-DNA-LL37 with TLR9, self-RNA-LL37 with TLR7). Thereof PDC in large quantities secrete IFN-alpha (Ganguly 2009, Gregorio 2010, Nograles 2010, Tang 2010).

Such mechanism of activation PDC also works in case of combustion or contact dermatitis when the commensal (pathogenic) invasion almost or absolutely is not present. Owing to trauma irrespective of commensal (or pathogenic) invasions PDC intensively secretes IFN-alpha (Fig. 2-9, Fig. 2-14). It occurs right after traumas, therefore phase 2 which take place at HPV-carriage, is absent.

### *LP4(HPV). Trigger of adaptive response against HPV.*

During HPV-expansion, PDC react through TRL9 to increased concentration of HPV-DNA - fragments of dsDNA (two-chained DNA), containing CpG (Fig. 2-18, Fig. 2-21).

Such reaction develops at active epidermal HPV-expansion, and is almost inevitable at attempt of dermal HPV-expansion, i.e. at carriage of invasive HPV types.

PDC endocytose HPV-DNA, CpG-LL37 complexes (LP3.1(HPV)), so that CpG optimally gets to endosomal TLR9. As a result of interaction of CpG with TLR9, PDC secrete IFN-alpha intensively. At that, TNF-alpha and small amount of IL-6 are also being secreted.

Intensive secretion of IFN-alpha is a signal to LP5(HPV) start - adaptive response against HPV. During LP5(HPV), effector Tem-Z should at least promote suppression of HPV-expansion and, as maximum, - complete destruction of KC-v, i.e. HPV-carriage termination.

Increase of amount of dermal PDC at HPV-carriage was studied only for oncogenous types of virus (HPV16, HPV33 and HPV18). Attraction of PDC in derma also takes place through CXCR3, being a ligand of CXCL10, secretion of which at HPV-carriage is essentially increased (Santegoets 2008, van Seters 2008). It is possible to assume that the increase of PDC amount observed in NLS is connected with latent HPV-carriage of other types (Favre 1998).

## *Process LP5. Adaptive response against LP2.*

Trigger process LP4 provides active start of LP5 – adaptive response against LP2 (Lande 2010, McKenna 2005, Seo 2010, Tang 2010, Zhang 2005).

Processing and presentation of Z-antigens (Z1 or Z2) can be carried out by both maDC (LP5.1), and KC (LP5.2) (Nestle 2009a). This possibility for KC is considered only at LP2(HPV), since at LP2(IN) it is supposed to be insignificant.

At LP2(HPV), Z1 is a dominant HPV-antigen presented through MHC class I and distinguished by TcN-Z1, and Z2 is a dominant HPV-antigen, presented through MHC class II and distinguished by ThN-Z2. For example, at LP2(HPV16) Z1 is mainly E6, and at Z2 is mainly E2 (Stanley 2009).

As a result of interaction with maDC-Z or with KC-Z1 effector Tem-Z are activated, proliferate and secrete cytokines (Clark 2010).

At LP2(IN), Tem-Z mainly consist from ThN-Z, and at LP2(HPV), Tem-Z mainly consist from helper ThN-Z2 and cytotoxic TcN-Z1. Interaction of TcN-Z1 and KC-Z1 leads to destruction of KC-Z1.

Cytokines secreted by Tem-Z, influence FB and KC.

KC and FB secrete chemokines and AMP with chemokine properties.

These are CCL2, CCL20, CX3CL1, LL37, HBD-2, HBD-3 and others, assortment of which depends on concrete LP2. Thereby, LP1 is intensified.

Process LP5 exists during phases 4 and 5 (Tab.2-1). LP5 start determines transition of NLS in phase 4. If during phase 4 dermal secretion of cytokines-deprogrammers IFN-gamma and GM-CSF is sufficient to start LP6.1, launch of LP6.4 becomes possible.

If action of LP6.4 provides enough maDC-Y, preconditions for LP8.1 start , i.e. phase 5, will appear.

During phase 5, active coexistence and interaction of processes LP2, LP3, LP5 and LP8 take place. Process LP4 becomes suppressed while level of TNF-alpha increases.

To provide certain duration of phase 5, it is necessary to achieve the state when processes do not suppress but support each other (for example, at LP2(HPV)). Otherwise, phase 5 will not be able to last long (for example, at LP2(IN)) and, either plaque will be able to pass into phase 6, or return to phase 4.

<u>LP5 depends on LP1.2.</u> First of all, it depends on income from blood flow of Tem-Z already presented in it or are recently formed during subprocess LP7.1.

<u>LP5 depends on LP2.</u> For LP5 begins because of LP2, operates against LP2 and comes to the end at LP2 completion.

<u>LP5 depends on LP3.</u> Innate and adaptive responses strengthen each other. This dependence is represented only in schemes at LP2(IN) and LP2(HPV).

<u>LP5 depends on LP4.</u> IFN-alpha also influences activation of Tem-Z, especially, if LP2 is a virus (in particular HPV) (Seo 2010, Zhang 2005). IFN-alpha promotes activation of Th1 and increase their secretion of IFN-gamma (Eriksen 2005). Besides, LP6 is intensified during LP4. And since LP5 depends on LP6.2 and LP6.3, LP5 (indirectly through LP6) depends on LP4.

<u>LP5 depends on LP6.</u> First - the more maDC-Z is formed, the more intensity of LP5.1 is. And, secondly, all Mo and DC (irrespective of their origin) transformed during LP6 secrete TNF-alpha, IL-1beta, IL-12, IL-23, etc. and thus actively influence LP5.

### *Subprocess LP5.1. Z-antigen presentation by maDC-Z to effector Tem-Z.*

This process interacts with subprocess LP6.3 as intensity of presentation depends on maDC-Z amount, and maDC-Z continue maturation during presentation.

#### *LP5.1(IN).*

Tem-Z, resident and involved from blood flow are mostly ThN-Z.

maDC-Z process and present Z-antigens to ThN-Z (Fig. 2-9; Fig. 2-15; Fig. 2-16). At that, ThN-Z are activated, proliferate and secrete TNF-alpha, IFN-gamma, IL-17 and IL-22 (Eyerich 2009, Sasaki 2011), and maDC-Z secrete IL-12 and IL-23.

#### *LP5.1(HPV).*

Tem-Z, resident and coming from blood flow are mainly TcN-Z1 and ThN-Z2.

Cytotoxic TcN-Z1 are formed as a result of interaction with maDC-Z1, carrying out cross-presentation of virus Z1-antigen on MHC class I.

TcN-Z1 are intended for KC-v destruction.

Helper ThN-Z2 are formed as a result of interaction with maDC-Z2 that present virus Z2-antigen on MHC class II.

ThN-Z2 are intended for coordination and intensification of immune response against HPV, as a whole.

maDC-Z (maDC-Z1 and maDC-Z2) process and present endocytosed Z-antigens (Z1 and Z2) to Tem-Z (TcN-Z1 and ThN-Z2, accordingly) (Fig. 2-18; Fig. 2-22; Fig. 2-23). At that, Tem-Z are activated, proliferate and secrete TNF-alpha, IFN-gamma, IL-17, IL-22, and maDC-Z secrete IL-12 and IL-23.

### *Subprocess LP5.2(HPV). Z1-antigen presentation by KC-Z1 to effector TcN-Z1.*

KC-v can process virus antigens and present them on MHC class I. At that, KC-v are transformed into KC-Z1 (LP3(HPV)). Interaction of TcN-Z1 and KC-Z1 leads to destruction of KC-Z1. IFN-gamma intensifies this subprocess (Black 2007).

At that, TcN-Z1 are activated, proliferate and secrete TNF-alpha, IFN-gamma, IL-17, IL-22.

If HPV-infection in the given site is not a primary one, skin is Z-primed, i.e. already contains TcN-Z1 and ThN-Z2.

I.e. in Z-primed skin subprocess LP5.2(HPV) can begin right after KC-Z1 formation. In particular, before LP5.1(HPV) start.

In the norm, the most part of Tem is located in derma, and the smaller one - in epidermis (Clark 2006a). Epidermal Tem are located mainly in basal and spinous layers. The beginning of pinpoint plaque is mutually connected with active attraction of dermal CD8+TL (i.e. TcN) in epidermis (Nestle 2009a, Nestle 2009b, Tonel 2009).

It is known that long latent HPV-carriage is possible, in particular, because LC practically do not react to HPV existence in epidermis. And only dermal DC appear to be capable of it (Stanley 2009).

It is possible to assume that at LP2(HPV) subprocess LP5.2 is the very first link of adaptive response against HPV (phase 4 start).

Processes LP5.2 and LP5.1 strengthen each other.

At destruction of KC-Z1, great amount of virus antigens (in particular, Z1 and Z2) enter the extracellular space, then being absorbed by phagocytes, including dermal Mo, MoDC and DC. Hence, maDC-Z amount is enlarged that intensifies LP5.1.

On the other hand, dermal interaction of maDC-Z1 with TcN-Z1 leads to their proliferation and activation and, hence, the larger amount of dermal TcN-Z1 can get to epidermis for participation in LP5.2.

And, if phase 4 appears intensive enough, LP6.1 and LP6.4 can begin during it, and LP6.2 can be activated.

In plaque the most part of epidermal TL is CD8+ (i.e. TcN), and the most part of dermal TL is CD4+ (i.e. ThN) (Cameron 2002).

If processes LP3 and LP5 fail to stop LP2(HPV) because of accelerated desquamation of KC-v, the number of virions produced per one unit of time multiplies, and that leads to increase of probability of infection of other skin areas.

## Process LP6. Mo and DC transformations.

Process LP6 consists of four subprocesses (Tab. 2-3). All dermal Mo and DC have possibility to be transformed. How this transformation will occur depends on their type and origin (resident or non-resident). And also from phase of psoriatic plaque development.

During phase 1 (homeostatic condition) only LP6.2 partly develops (Mo are transformed in MF, Mo-T in MF-T and Mo-R in MF-R), and subprocesses LP6.1, LP6.3 and LP6.4 do not take place at all. With the beginning of LP2-inflammation (phase 2) other subprocesses LP6 begin and will be intensified during the subsequent phases (Fig. 2-8).

Loss of tolerance of PG-Y(-)Mo-T, PG-Y(-)DC-T and PG-Y(-)MoDC-T (LP6.1) gives the chance of their participations in LP6.2 and LP6.3. In the right part of Tab. 2-3 all possible transformations of these cells are specified. Thus it is meant that they are PG-Y(-).

Loss of tolerance of Mo-R, DC-R and MoDC-R (LP6.1) is necessary for their subsequent transformations (LP6.2 and LP6.4). The table-cells of Tab. 2-3 containing PG-Y(+) cells, are allocated by beige color.

Possibly, subprocess LP6.1 (loss of tolerance) and subprocesses LP6.2, LP6.3 and LP6.4 occur simultaneously strengthening each other. Process LP6 is included into the vicious cycle B.

***Table 2-3. Dermal Mo and DC transformations***

| Origin | | | **LP6.1.** Loss of tolerance | Formation subprocesses **LP6.2.** MF и MoDC | **LP6.3.** maDC-Z | **LP6.4.** maDC-Y |
|---|---|---|---|---|---|---|
| Non-resident and resident | | Mo | | Mo —> MF | | |
| | | | | Mo —> MoDC | MoDC —> maDC-Z | |
| | | DC | | | DC —> maDC-Z | |
| Non-resident only | Mo-T | PG-Y(-)Mo-T | * | Mo-T —> MF-T | | |
| | | | * | Mo-T —> MoDC-T | MoDC-T —> maDC-Z | |
| | | Mo-R | * | Mo-R —> MF-R | | |
| | | | * | Mo-R —> MoDC-R | | MoDC-R —> maDC-Y |
| | DC-T | PG-Y(-)DC-T | * | | DC-T —> maDC-Z | |
| | | DC-R | * | | | DC-R —> maDC-Y |

These transformations occur at trauma (Fig. 2-9; Fig. 2-14; Fig. 2-15; Fig. 2-16) and at HPV-expansion (Fig. 2-18; Fig. 2-21; Fig. 2-22; Fig. 2-23).

LP6.2 and LP6.3 begin due to LP4 - under influence (IFN-alpha + TNF-alpha) and are activated during LP5 (phase 4), and then cooperative action of LP5 and LP8 (phase 5) - under cumulative influence of IFN-gamma, TNF-alpha, IL-12 and IL-1beta.

If both LP3 and LP5 stop expansion or completely eliminate LP2, then LP6.3 also stops.

And, if for maintenance of subprocesses LP6.1, LP6.2 and LP6.4, LP8 appears to be enough, self-sufficient phase 6 begins.

In Fig. 2-4 variants of origin of dermal macrophages DMF and dermal dendritic cells DDC are represented. In the absence of inflammation, population of dermal macrophages FXIIIa+CD14+MF by 60% consists of cells, derived from dermal stem cells-precursors i.e. have resident origin. The rest 40% of DMF have non-resident origin, as they transform from CD14++Mo and CD14+CD16+Mo involved from blood flow. At local inflammation, total DMF number per unit of dermal volume is enlarged, first of all at the expense of attraction and subsequent transformation of blood Mo, at that share of DMF of resident origin decreases to 30%, and share of DMF of non-resident origin increases to 70%.

Dermal dendritic cells DDC have unique receptor CD11c and consist of two basic fractions CD14+DC (mainly BDCA-1(-)) and CD1a+DC (mainly BDCA-1+) (Haniffa 2009).

In the absence of inflammation, CD1a+DC are located mainly in the top part of derma, and CD14+DC are distributed everywhere, including vascular areas (Angel 2006).

Even in the absence of inflammation, the most part (80-90%) of both DDC fractions has non-resident origin, and only 10-20% originate from dermal stem cells-precursors.

At local inflammation, a share of DDC of non-resident origin becomes maximal (up to 100%). In the absence of inflammation, non-resident DDC mainly originate from DC, those involved from blood flow (Haniffa 2009). At inflammation, DDC pool of non-resident origin (first of all, BDCA-1(-)DC fraction) is actively renewed. It occurs due to attraction of BDCA-1(-)DC from blood flow and also due to the intradermal transformation of some of involved blood monocytes Mo in MoDC. The most capable of such transformation are CD14+CD16+Mo, i.e. the fraction of blood monocytes which can be under chronic kPAMP-load tolerized much easier, and this fraction becomes CD14+CD16+Mo-T (incl. CD14+CD16+Mo-R) (Part 1, SP8).

### *Subprocess LP6.1. Loss of tolerance to kPAMP.*

Loss of tolerance (deprogramming) of tolerized Mo-T, DC-T and MoDC-T (incl. Mo-R, DC-R and MoDC-R) develops under the influence of cytokines-deprogrammers GM-CSF and IFN-gamma (Adib-Conquy 2002, Cavaillon 2008, Kylanpaa 2005, Randow 1997).

DC secrete IL-12 during their maturation (LP6.3 and LP6.4). Th1 and Tc1 secrete IFN-gamma under the influence of IFN-alpha, and also at interaction with maDC (LP5 and LP8). IL-12 actively influences Th1, Tc1 and NK all secreting IFN-gamma. Thus, IL-12 also indirectly promotes loss of tolerance (Randow 1997).

KC, FB, etc. secrete GM-CSF, it occurs under the effect on EGFR receptor, and also under the influence of IL-17 (Koga 2008, Mascia 2010). Besides, during activation and differentiation, TL also secrete GM-CSF (Male 2006, Shi 2006).

Subprocess LP6.1 covers all Mo-T, DC-T and MoDC-T (incl. Mo-R, DC-R and MoDC-R). These cells involved in skin prior to the beginning of inflammation (or during inflammation) lose tolerance. This subprocess begins during phase 4 and reaches its maximum during phases 5 and 6 (Tab.2-1).

As soon as tolirized phogocytes (incl. R-phagocytes) get to non-inflamed tissues (for example, in NLS-derma), kPAMP-load having influenced them in blood flow disappears. Tolerized phagocytes gradually degrade earlier endocytosed F-content, as kPAMP-load is absent in non-inflamed tissues. Simultaneously, spontaneous gradual loss of tolerance takes place - the level of blocking proteins decreases. In non-inflamed tissues, in tolerized phagocytes, degradation of previously endocytosed F-content advances decrease of the level of blocking proteins and, as a consequence, tolerized phagocytes are transformed into normal phagocytes. Such transformation of a specific tolerized phagocyte on the average takes several days, with duration being longer in case of stronger and prolonged previous (influencing blood flow) kPAMP-load (depth of reprogramming) (Fitting 2004).

Depth of reprogramming of Mo-T and DC-T (incl. Mo-R and DC-R) involved into derma determines the level of intensity of local inflammation, sufficient for their loss of tolerance (Adib-Conquy 2002). If tissues are inflamed Mo-T and DC-T involved from blood flow appear to be under the influence of cytokines-deprogrammers IFN-gamma, GM-CSF and quickly lose tolerance. It is observed, in particular, in connection with rapid decrease of the level of blocking proteins (IRAK-M, etc.). And, if this loss occurs before complete degradation of previously endocytosed F-content, R-phagocytes are activated. Activation of Mo-T (Mo-R) promotes their transformation in MF-T and MoDC-T (in MF-R and MoDC-R) and active secretion of proinflammatory cytokines (LP6.2).

PG-Y(-)MoDC-T and PG-Y(-)DC-T are activated and mature, however cannot present Y-antigen. Their subsequent transformation is similar to transformation of usual Mo and DC (Tab. 2-3).

MoDC-R and DC-R are activated and become mature (LP6.4), which, in the presence of previously endocytosed F-content, leads to secretion of proinflammatory cytokines (IL-12, IL-23 etc.) and presentation of antigens contained in F-content (including Y-antigen). And, if the local immune system considers Y-antigen pathogenic along with adaptive response against Z-antigen (LP5), false adaptive response against Y-antigen (LP8) can develop.

### *Subprocess LP6.2. MF and MoDC formation.*

During this subprocess, transformation of Mo (irrespective of origin) in MF and MoDC, PG-Y(-)Mo-T (of non-resident origin) in PG-Y(-)MF-T and PG-Y(-)MoDC-T, as well as Mo-R (of non-resident origin) in MF-R and MoDC-R takes place.

Formation of MF occurs constantly - with low intensity during phase 1, with average intensity during phases 2 and 3 and with high intensity during phases 4, 5 and 6 (Tab.2-1).

Active formation of MoDC begins during phase 3, under the influence of IFN-alpha and TNF-alpha secreted by PDC during LP4. Presence of PAMP (outside or inside) of Mo activates their transformation in MoDC. Activity of these formations in all subsequent phases 4, 5 and 6 depends on intensity of LP5 and LP8.

Mo (MF) endocytose and-or contact with LP2-content (including both Z-antigens and PAMP).

Having lost tolerance Mo-T (incl. Mo-R) and MF-T (incl. MF-R) also react to kPAMP endocytosed in blood flow. As a result, all of them are activated and secrete TNF-alpha, IL-1beta, IL-12 and IL-20 (Clark 2006b, Fuentes-Duculan 2010, Wang 2006).Mo (MF), which endocytose LP2-content, do not have special designation for their difference from inactive Mo (MF), because these differences are not principal for Y-model. During phase 6, the majority of active MF are MF-T and/or MF-R.

Under their characteristics PG-Y(-)MF-T and MF-R are indistinguishable, as after transformation in MF presence PG-Y in F-content ceases to play any role. Their secretory activity is defined by volume of kPAMP, containing in F-content.

### *Subprocess LP6.3. maDC-Z formation.*

Subprocess begins with average intensity during phase 3, high intensity during phases 4 and 5, weaken at the end of phase 5 and is absent during phase 6 (Tab.2-1).

DC and MoDC endocytose LP2-content (including PAMP and Z-antigens), secrete TNF-alpha, IL-12 and IL-23. Under the influence of IL-12 and IL-23 acceleration of their maturation (self-activation) takes place. Their maturation is also positively influenced by the combination IFN-gamma and IL-1beta. DC and MoDC process bacterial or virus fragments containing Z-antigen, form complexes of MHC class II with Z-antigen and transport them to the cell wall (Sabat 2007). After that, maDC-Z appear to be ready for the beginning of Z-antigen presentation. So, transformation of DC and MoDC (both of non-resident and resident origin) takes place in maDC-Z.

Loss of tolerance of PG-Y(-)DC-T and PG-Y(-)MoDC-T (LP6.1) gives the chance of their similar behavior, i.e. transformation in maDC-Z.

Z-antigen presentation to effector Tem-Z occurs during subprocess LP5.1. Final maturation of maDC-Z occurs during interaction with Tem-Z, i.e. LP6.3 and LP5.1 are interdependent.

### *Subprocess LP6.4. maDC-Y formation.*

The subprocess begins during phase 4 and proceeds with high intensity during phases 5 and 6 (Tab.2-1).

Having lost tolerance DC-R and MoDC-R secrete TNF-alpha, IL-12 and IL-23. Under the influence of IL-12 and IL-23 acceleration of their maturation (self-activation) takes place. Their maturation is also positively influenced by the combination of IFN-gamma and IL-1beta. DC-R and MoDC-R process F-content (containing kPAMP and PG-Y), form complexes of MHC class II with Y-antigen and transport them to the cell wall. After that, maDC-Y appear to be ready to begin Y-antigen presentation (Sabat 2007).

DC-R and MoDC-R having insufficient PG-Y content are incapable of transformation into maDC-Y. Such cells will show the activity similarly to MF-R. Only DC-R or Mo-R can be precursors of maDC-Y.

Y-antigen presentation to effector ThN-Y occurs during subprocess LP8.1. Final maturation of maDC-Y occurs during interaction with ThN-Y, i.e. LP6.4 and LP8.1 are interrelated.

Fig. 2-5 is continuation of Fig. 2-4. In this figure, basic fractions of maDC found in PLS-derma are illustrated (Angel 2006, Haniffa 2009, Zaba 2009a). This figure illustrates characteristics of maDC formed during subprocesses LP6.3 and LP6.4.

Mature dendritic cells maDC in PLS can be divided into two groups: BDCA-1+maDC and BDCA-1(-)maDC. CD208+maDC (CD208=DC-LAMP) are, basically, BDCA-1+maDC, and almost all TipDC are BDCA-1(-)maDC (Zaba 2009a).

Many cells from slanDC in PLS become maDC. At least those of them that actively secrete TNF-alpha, iNOS and IL-23 and fall under the definition of TipDC. Unfortunately, expression of CD80, CD83 (characterizing maturity receptors) in slanDC in PLS is not studied in these investigations (Hansel 2011, Schake 2006). However, fractions of mature slanDC (yellow oval) and TipDC (brown oval) are crossed only partly. Not crossed parts can be distinguished on expression CD16 (expression CD14 at maturation of DC vanishes). Mainly, slanDC are CD16+, and TipDC originate from blood CD14++CD16(-)Mo and CD14+CD16+Mo, and, hence, expression CD16 is not obligatory for them (Teunissen 2011).

Here, in contrast to (Zaba 2009a), based on the results of (Haniffa 2009), it is supposed that not all BDCA-1+maDC have resident origin, the part of them originate from non-resident BDCA-1+DC.

Besides, BDCA-1(-)maDC are CD14(low)CD163(low) and have the size 20% less than BDCA-1+maDC, which allows assumption of their different origin. Namely, BDCA-1+maDC apparently origin from BDCA-1+DC, and BDCA-1(-)maDC - from CD14+MoDC (larger part) and from BDCA-1(-)DC (smaller part).

Blood Mo, on the average, have the smaller size, than blood DC, and their transformation in MoDC at inflammation, as a rule, is fast (up to 18 hours) and accompanied by division (Eberl 2009). It is possible to assume that new cells formed in the course of fast division of Mo, their transformation in MoDC and subsequent maturation, do not have enough time for growth as in case with BDCA-1+DC at maturation without division. At maturation DC and MoDC lose receptors CD14 and CD163 and express receptors CD80, CD83, etc., inherent for maDC.

The larger part of TipDC (actively secreting TNF-alpha and iNOS) has non-resident origin, it is possible to assume that they originate mainly from blood CD14+CD16+Mo-T (incl. CD14+CD16+Mo-R). The fact that they undergo division before transformation in CD11c+maDC means decrease of F-content content (including PG-Y) per one cell. After all, at each division F-content is divided between daughter cells, and F-content degrades during division and transformation of Mo-T in MoDC-T (incl. Mo-R in MoDC-R). An essential part of CD208+maDC is also of non-resident origin, it is possible to suggest that they develop mainly from blood DC-R. At sufficient content of kPAMP (incl. PG-Y), transformation of DC-R into maDC-Y occurs quickly and without preliminary division.

On the average, F-content volume (including PG-Y) in Mo-R and in DC-R at the moment of their entering from blood flow into derma is approximately identical. Considering their further dermal transformations (with division and transformation and without division and transformation, respectively) it is possible to assert that F-content volume (including PG-Y) in one maDC-Y, derived from DC-R, will be several times higher, than in one maDC-Y, derived from Mo-R.

Possibly therefore mainly BDCA-1+maDC-Y provide presentation of Y-antigen to ThN-Y (LP8.1). The majority of BDCA-1(-)maDC are TipDC, which because of the smaller volume of F-content (including PG-Y), are more similar to active MF (Dominguez 2010). Or precursors of BDCA-1(-)maDC are PG-Y(-)Mo-T or PG-Y(-)DC-T and consequently they have simply nothing to present to ThN-Y. These assumptions are proved by the data on colocalization of CD208+maDC and ThN in PLS-derma and dermal lymphoid tissues (Guttman-Yassky 2011, Nograles 2010).

During phase 4, all maDC (irrespective of their origin) can present only Z-antigens (i.e. they are maDC-Z). During phase 5, maDC of resident origin can present only Z-antigens (i.e. can be only maDC-Z), and maDC of non-resident origin can present either Z-antigens, or Y-antigens (i.e. can be either maDC-Z, or maDC-Y). During phase 6, maDC of resident origin present nothing, and maDC of non-resident origin can present only Y-antigens (i.e. can be only maDC-Y).

DC are sensitive to the influence of ultraviolet type B (UV-B) and under its influence can be inactivated and even may die. As a result, LP6 stops and the vicious cycle B weakens. It determines the seasonal nature of psoriasis and efficiency of helium - and UV-therapy leading to transitory remission (Gudjonsson 2004, Male 2006).

As formation of maDC-Z (LP6.3) and maDC-Y (LP6.4) depends on the phase, it is also fair in respect of presentation (Fig. 2-10). These are conditional graphs of specific antigenic presentation by maDC (the amount of antigen presented by maDC in a unit volume of derma).

IYD - the graph for Y-antigen (presented by maDC-Y), and IZD – the graph for Z-antigen (presented by maDC-Z).

**Interdependence between LP6 subprocesses** (Fig. 2-7).

LP6.1 depends on LP6.2, LP6.3 and LP6.4. MoDC during its formation secrete IL-12, and both maDC-Z and maDC-Y release a lot of IL-12. It promotes secreting IFN-gamma by Th1, Tc1 and NK. Thereby, IL-12 (indirectly through IFN-gamma) influences loss of tolerance (Randow 1997).

LP6.2 depends on LP6.1. Intensity of formation of MF-T and MoDC-T (incl. MF-R and MoDC-R) depends on the rate of loss of tolerance of Mo-T, DC-T (incl. Mo-R, DC-R).

LP6.3 depends on LP6.2. The more MoDC is formed, the more mature maDC-Z can be formed.

LP6.4 depends on LP6.1. Intensity of formation of maDC-Y depends on the rate of loss of tolerance by Mo-R, DC-R and MoDC-R.

LP6.4 depends on LP6.2. The more MoDC-R is formed, the more mature maDC-Y can be formed.

**Other dependencies.**

LP6 depends on LP1.1. The more Mo-T and DC-T (incl. Mo-R and DC-R) arrive in derma, the more of them can be deprogrammed and lose tolerance (LP6.1). At inflammation, the most part of dermal MF and DC is formed from blood Mo and DC.

LP6.3 and LP6.4 depend from LP3.1. Self-RNA-LL37 complexes formed during LP3.1 cause activation of DC. It occurs through endosomal TLR8 and leads to secretion of TNF-alpha and IL-6 and maturation of DC in maDC.

Self-RNA-LL37 complexes are found in PLS-derma and colocalizated with maDC. Their quantity correlates with CD208+maDC quantity. PLS-biopsy samples were received from 11 psoriatics. The complexes were not found in three of them, in two subjects the quantity of these complexes did not differ from the norm, and only in 6 psoriatics the quantity of the complexes considerably exceeded the norm (Ganguly 2009, fig.8).

It is probable that excess over the norm is caused by PLS (in the biopsy site) being in phase 5, i.e. LP3.1 still continued. It can be, for example, LP3.1(HPV). The quantity of complexes within the norm (and also their absence) means that PLS (in the biopsy site) was in phase 6, i.e. LP2 (and also LP3, LP4 and LP5) terminated completely.

LP6 depends on LP4. Under the influence of IFN-alpha, Mo and DC increase expression of PAMP-receptors.

IFN-alpha promotes formation of MoDC (Farkas 2008, Farkas 2011). And cooperation of IFN-alpha with TNF-alpha accelerates this formation (Guzylack-Piriou 2006). IFN-alpha also accelerates maturation of DC, which, at that, secrete IL-12 more actively (Piccioli 2007).

After PDC have executed their role in LP4, they can be transformed into maDC to act further in LP6.3 (Zhang 2005).

LP6.1 depends on LP5. Dependence on LP5 is principal prior to phase 5 start, when LP8 has not begun yet. Without intensive LP5, LP6.1 start is impossible. During LP5, activated Th1-Z (Tc1-Z) secrete a lot of IFN-gamma. In particular, the Koebner effect is impossible at dermal only (without epidermal) or at epidermal only (without dermal) trauma at the absence of conditions for entering skin commensals into derma. Only simultaneous traumas of epidermis and derma create conditions, at which dermal adaptive response LP5 can begin (Weiss 2002). And at intensive LP5, IL-12, IFN-gamma and GM-CSF can be secreted in the amount big enough for LP6.1 start.

LP6.2, LP6.3 and LP6.4 depend from LP5. Intensity of secreted TNF-alpha, IL-1beta, IL-6 and IFN-gamma (at interaction maDC-Z with Tem-Z or mediated by FB and KC), determines the degree of transformation intensity of all Mo and DC. In particular, final maturation of maDC-Z occurs during interaction with Tem-Z (Guzylack-Piriou 2006).

LP6.1 depends on LP8. Activated Th1-Y secrete IFN-gamma and GM-CSF. This is the way, how process LP8 influences the intensity of process LP6.1. This dependence is principal for transition from phase 5 to phase 6 (Fig. 2-7, Fig. 2-8, letter B). Without this dependence, process LP8 will not become self-sufficient and, hence, phase 6 will not begin.

LP6.2, LP6.3 and LP6.4 depend from LP8. Intensity of secreted TNF-alpha, IL-1beta, IL-6 and IFN-gamma (at interaction of maDC-Y with ThN-Y or mediated by FB and KC), determines the degree of transformation intensity of all Mo and DC. In particular, maDC-Y finally mature during interaction with ThN-Y (Guzylack-Piriou 2006).

## *Process LP7. Lymph nodes. Clonal proliferation.*

This process is well-known and can occur during adaptive response. Subprocess LP7.1 takes place during LP5, and subprocess LP7.2 - during LP8. During phase 4 only LP7.1 can occur, during phase 5 – both LP7.1 and LP7.2, and during phase 6 – LP7.2 only (Fig. 2-13).

maDC (including maDC-Z and maDC-Y) by afferent lymph vessels come from area of inflammation in nearest regional lymph nodes LN.

Further (within the limits of LP7.1) the following events take place:

In lymph node, maDC-Z present Z-antigen to naive nTL (at primary response) and to central memory Tcm-Z (at secondary response), which provides their clonal proliferation into effector Tem-Z. Besides Tem-Z, Tcm-Z are also formed at primary and secondary responses.

Then Tem-Z enter blood flow through the efferent lymph vessel. The expression of homing-receptors on Tem-Z (in particular, CLA) provides their migration to postarterial venules, on which ligandic adhesive molecules (in particular E-selectin and ICAM-1) are expressed. In areas of inflammation this expression is promoted, in particular, by TNF-alpha. Migration of Tem-Z occurs mainly to the inflammation areas, from where maDC-Z have come to LN (Gudjonsson 2004, Male 2006).

Part of Tcm-Z enters blood flow similarly, but go to farther regional LN to proliferate there and differentiate in Tem-Z, which then enter other tissues (Fig. 2-12).

On the endothelium of postarterial venules, due to interaction of Tem receptors and ligandic adhesive molecules, infiltration of Tem from blood flow into derma, and then, their movement inside derma and epidermis takes place under the influence of chemokines.

Mostly (> 90%), in the absence of inflammation, Tem are constantly present in skin. The rest of Tem (<10%) circulate in blood flow. Skin Tem do not express CCR7 and L-selectin, which makes impossible their moving into LN. The total amount of Tem reaches 1 million per cm2 of skin (among them specific to all antigens, in respond to which SIS earlier have formed adaptive response). Such location of Tem provides the fastest and effective secondary adaptive response formed by SIS at repeated contact with antigens (Clark 2010).

### *Subprocess LP7.1. TL-Z formation.*

maDC-Z on afferent lymph vessels arrive from inflammation area into nearest regional LN. As a result of interaction of maDC-Z with naive nTL (at primary response) or with Tcm-Z (at secondary response) new TL-Z are formed, i.e. effector Tem-Z and central Tcm-Z.

At primary response, LP7.1 start precedes dermal interaction of maDC-Z with Tem-Z. It occurs at absence of Z-priming, i.e. if SIS has never earlier formed adaptive response against Z-antigen.

At secondary response, Tem-Z is already present in skin, and Tcm-Z - in LN. It occurs at Z-priming development, i.e. SIS has already previously formed adaptive response against Z-antigen. Such a variant is represented in Fig. 2-11.

In this case interaction of maDC-Z with Tem-Z (LP5.1) begins at once, and fast activation of LP7.1 occurs later at deficiency of Tem-Z in the inflammation area. If process LP2 terminates before development of such deficiency, LP7.1 may not occur at all (Clark 2010).

Irrespective of the order of events, we consider that with LP5 and-or LP7.1 start phase 4 begins.

***Subprocess LP7.2. TL-Y formation.***

maDC-Y arrive on afferent lymph vessels from an inflammation area into nearest regional LN. As a result of interaction of maDC-Y with naive nTL (at primary response) or with Tcm-Y (at secondary response) new TL-Y are formed, i.e. effector Tem-Y and central Tcm-Y.

At primary response, LP7.2 start precedes dermal interaction of maDC-Y with Tem-Y. It is possible only at primary PLS-plaque and only in case if SIS has never earlier formed adaptive response against PsB.

At secondary response, Tem-Y is already present in skin, and Tcm-Y in LN. In this case, interaction of maDC-Y with Tem-Y (LP8.1) begins at once, and fast activation of subprocess LP7.2 occurs later if in the inflammation area deficiency of Tem-Y develops.

It happens at any nonprime plaque, and, of course, on the external border of extending plaque (Vissers 2004).

It happens both at primary plaque if SIS has previously formed adaptive response, for example, against external PsB-infection.

Irrespective of the order of events, we consider that with LP8 and-or LP7.2 start phase 5 begins.

At LP2(IN), Tem-Z being formed consist mainly of ThN-Z, for Z – is a dominant antigen of commensal or pathogenic bacteria penetrated derma due to trauma.

At LP2(HPV), Tem-Z being formed consist mainly of ThN-Z2 and TcN-Z1.

KC-v process and present Z1-antigen through MHC class I (being thus transformed in KC-Z1). maDC-Z1 do the same by cross-presentation. maDC-Z1 and KC-Z1 cooperate with TcN-Z1, both resident and involved from blood flow (Black 2007, Male 2006, Nestle 2009a).

At LP2(HPV), CD4+Tem (mainly ThN-Z2) are actively formed and arrive to the area of inflammation (Stanley 2009). maDC-Z2 present Z2-antigen through MHC class II to ThN-Z2, both resident and involved from blood flow (see LP5 (HPV)).

Tem-Y mainly consists from ThN-Y as Y-antigen is a peptidoglycan fragment, endocytosed by Mo-R and DC-R in blood flow.

<u>LP7.1 depends on LP6.3.</u> The more dermal maDC-Z is formed, the greater possibility that the amount of dermal Tem-Z (cooperating with maDC-Z) will be insufficient, and the more uninvolved maDC-Z will enter LN.

<u>LP7.2 depends on LP6.4.</u> The more dermal maDC-Y is formed, the greater possibility that the amount of dermal Tem-Y (cooperating with maDC-Y) will be insufficient, and the more uninvolved maDC-Y will enter LN.

## *Process LP8. False adaptive response to imaginary PsB-infection.*

DC-R and MoDC-R, having lost tolerance, undergo maturation with the following transformation into mature dendritic cells maDC-Y (LP6.4). maDC-Y process PG-Y fragments contained in them, and after processing present Y-antigen to ThN-Y, both resident and involved from blood flow. The skin immune system SIS receives a false target – an imaginary PsB-infection, regarded by SIS by presented Y-antigens.

This process occurs at trauma (Fig. 2-9, Fig. 2-16) and at HPV-expansion (Fig. 2-18, Fig. 2-23), and is similar at other initiating and aggravating processes LP2.

The vicious cycle B = {LP1.1 > LP6 > LP8 > LP1} is initiated at expansion of LP2 (through LP3, LP4 and LP5) (Fig. 2-6). This cycle represents:

- Income of DC-R and Mo-R from blood flow to the area of developing inflammation (LP1.1).
- Loss of tolerance by DC-R and Mo-R and their transformation into maDC-Y (LP6.1, LP6.2 and LP6.4).
- Cooperation of maDC-Y with ThN-Y, causing process LP8, which, in its turn, supports LP1.

The vicious cycle B amplifies by the vicious cycle C = {LP6.4 > LP7.2 > LP1.2 > LP8 > LP6}. At excess of maDC-Y and shortage of ThN-Y movement of maDC-Y appears in the inflammation area into regional LN, where clonal proliferation of Tem-Y, mainly ThN-Y, takes place (LP7.2). Due to homing, ThN-Y arrive through blood flow to the inflammation area and adjacent areas (LP1.2), which causes intensification and expansion of plaque (LP8).

Process LP8 influences all subprocesses LP6 through secreted cytokines (Fig. 2-6).

For LP8 functioning, LP1, LP6 and LP7.2 actively supported by LP8 (Fig. 2-9; Fig. 2-18, Fig. 2-17), can be sufficient. It means that process LP2 (as well as LP3, LP4, LP5, LP6.3 and LP7.1) can come to the end, and LP8 will proceed (PLS will pass into phase 6).

If LP2 is preserved, or other secondary infections develop, PLS remains in phase 5.

The vicious cycle B can be interrupted in case of breach of SIS work, which can be realized for a while by application of biological drugs. However, the unique long-term decision eliminating the original cause of psoriasis is weakening or termination of SPP.

Within LP8 limits, KC hyperproliferation begins (physical protection of skin). At LP2(IN) it accelerates trauma healing. At LP2(HPV) it complicates possibility of the complete replicative cycle of virus. Thereby, processes LP3 and LP5 are being assisted.

If during LP5 it appears that the level of Mo-R and DC-R in skin is insufficient, or their loss of tolerance is insufficient, LP8 will not begin, and LP5 will constrain LP2 or lead to its complete elimination. Thus, skin in LP2-inflammation area from phase 4 will return to phase 1 (Fig. 2-25).

<u>LP8.1 depends on LP1.2.</u> LP8.1 intensity depends on income of ThN-Y from blood flow.

<u>LP8 depends on LP6.</u> Without preliminary and sufficient formation of maDC-Y, LP8.1 start is impossible, as well as its support without constant formation of new maDC-Y (LP6.4). Also, LP8 support is influenced by formation of MF-T and MoDC-T (incl. MF-R and MoDC-R) secreting proinflammatory cytokines (LP6.2).

It is shown that LP8 does not depend directly on LP4 (Bissonnette 2010). Process LP4 is transitory and necessarily preceding to development of new plaque, but not participating in its support. Due to LP4, subprocesses LP6 are initiated. And then subprocesses LP6 are supported by only LP5 and-or LP8.

### *Subprocess LP8.1. Y-antigen presentation by maDC-Y to effector ThN-Y.*

Due to loss of tolerance (LP6.1) and fast maturation (LP6.4) maDC-Y process PG-Y (as a part of F-content) and present Y (fragments of interpeptide bridges) to effector ThN-Y.

ThN-Y are activated, proliferate and secrete TNF-alpha, IFN-gamma, IL-17 and IL-22. Cytokine IFN-gamma is mainly secreted by Th1-Y. Cytokine IL-17 (IL-17A and IL-17F) is mainly secreted by Th17-Y. Cytokine IL-22 is mainly secreted by Th22-Y (Guttman-Yassky 2011, Nestle 2009a, Nestle 2009b, Tonel 2009).

IL-23 secreted by MF and maDC (LP6) plays important role in ThN stimulation for secretion of these cytokines. Genetic deviations in the condition of receptors for IL-12 and-or IL-23, can influence ThN-reaction intensity (Cargill 2007, Di Cesare 2009, Gudjonsson 2009, Meglio 2010, Tonel 2010).

Under the influence of TNF-alpha, expression of adhesive molecules on EC of postarterial venules increases, with the subsequent increase of permeability of endothelium (LP1).

Events occur variously, it depends on whether there are resident ThN-Y in skin or not, i.e. whether Y-priming have taken place (Clark 2010, Sabat 2007, Sabat 2011). If Y-priming has taken place, ThN-Y is present in derma. It means that the patient either already has psoriasis, or he\she has (or had) some skin, tonsillar (SP6) or generalized PsB-infections (Fig. 2-24).

In this case, maDC-Y, having executed PG-Y processing, presents Y to resident ThN-Y. LP8 begins quickly, at that, the threshold level of (PG-Y)-carriage by blood DC-R and Mo-R for its initialization can be lower.

ThN-Y amount in skin is enough to provide the active beginning and support of false adaptive response in case of interaction with maDC-Y. At that, the need in clonal proliferation of ThN-Y in LN (LP7.2) increases later, when maDC-Y amount starts to exceed ThN-Y amount essentially.

Y-priming (not because of psoriasis) increases probability of primary psoriasis and lowers the middle age of its occurrence. Such priming explains fast development (flash) in some patients of abundant primary plaques.

If Y-priming was absent, ThN-Y in derma is absent, i.e. the patient did not and does not have psoriasis, as well as skin, tonsillar or generalized PsB-infections. In this case, maDC-Y at the absence of ThN-Y, express CCR7 and begin chemotactic movement through afferent lymph vessels into nearest LN. In LN clonal proliferation of Tem-Y (mainly ThN-Y) takes place, then they get to blood flow through the efferent lymph vessels (LP7.2).

If such beginning of psoriasis is possible, it should be necessary conditioned by the combination of high level (PG-Y)-carriage of blood DC-R and Mo-R and intensive adaptive response LP5 during phase 4.

LP2, initiating primary plaques, should be more intensive, than LP2, initiating secondary plaques (when ThN-Y is already present in skin).

### *Subprocess LP8.2. KC hyperproliferation. Change of skin architecture. Growth of basal membrane area and vascularity increase.*

This subprocess is a consequence of subprocess LP8.1 and develops in time of its intensification. KC hyperproliferation begins at once in basal layers of epidermis at a skin site, where LP8.1 has begun.

However, visible pinpoint plaque appears on the surface of skin with a several-day delay. This natural delay is necessary for displacement of normal KC of the overbasal layers by incompletely differentiated KC, quickly rising from the basal layers. Thus, changes of architecture of skin begin, being definitively formed in process of expansion and aggravation of plaque. These changes cover epidermis and lower derma and are accompanied by the growth of basal membrane area and vascularity increase.

IFN-gamma, IL-17, IL-20, IL-22 and TNF-alpha influence KC and cause their hyperproliferation (Nestle 2009a, Nestle 2009b, Tonel 2009). Thus, IL-20 is secreted by KC, and then both IL-20 (in the autocrine way) and IL-22 promote much more intensive hyperproliferation (Guttman-Yassky 2011, Sabat 2011). TNF-alpha and IL-17 is synergic influence on KC (Chiricozzi 2010).

The hyperproliferation is realized by

- increase of the volume of the basal layer because of increase of the basal membrane area;
- increase of the fraction of growth (a part of proliferating KC in the basal layer) from 10-30% to 100%;
- increase of the amount of cycles of TA-keratinocytes division (TA - Transit Amplifying), which proceeds after their lifting to the overbasal layers.

In PLS, simultaneous increase of dermal papillas height and reduction of thickness of the layer above the papillas takes place (Fig. 2-17). At that, epidermis thickness as a whole is enlarged.

These changes are a consequence of accelerated proliferation (basal and overbasal layers), forwarded and incomplete differentiation (spinous and granular layers) and accelerated desquamation (cornual layer) of keratinocytes (Castelijns 2000, Grabe 2007, Iizuka 2004, Weinstein 1985).

Within the limits of Y-model, the accelerated desquamation of KC realizes physical protection of SIS against an imaginary external PsB-infection.

KC in PLS secrete cytokines TNF-alpha, IL-1beta, IL-6, TGF-beta, etc.

KC in PLS secrete chemokines (CCL2, CCL20, CX3CL1, CXCL1, CXCL3, CXCL5, CXCL8, CXCL9, CXCL10, CXCL11, etc.) and AMP (LL37, HBD1, HBD2, S100A7, S100A8, S100A9, etc.). It occurs under the influence of IFN-gamma, IL-17, IL-20, IL-22 and TNF-alpha.

In particular, IL-17 and IL-22 are capable of synergic influence on KC providing active secretion of HBD-2 (Ghannam 2011).

It is the way of VEGF secretion causing angiogenesis. At early stages of inflammation, CCL2, HBD-2 are more actively secreted, and, as a consequence, blood CCR2+DC and CCR2+Mo are mostly involved. At intensification of PLS-inflammation, KC secrete CCL20 more actively, and, as a consequence, blood CCR6+DC, CCR6+Tem are involved.

Secretion of chemokines and AMP, increase of vascularity and the basal membrane area, provides more intensive income of immunocytes in derma from blood flow (LP1) per unit of skin surface (Nestle 2009a, Nestle 2009b, Tonel 2009).

## Phases of psoriatic plaque development

Let's turn to the detailed description of each phase (Tab.2-1, Tab. 2-2, Fig. 2-8, Fig. 2-10, Fig. 2-19). Phases 1, 2, 3 and 4 correspond to conditions of a particular NLS site preceding plaque development, phases 5 and 6 correspond to initialization and support of plaque.

During all phases, SPP acts, without which development and existence of plaque are impossible.

In the phase name one or two marker local processes characterizing this phase are listed. Intensity of processes is illustrated by color: white (weak), beige (medium) and pink (high).

The processes taking place during LP2-inflammation, are bounded by the big lilac circle (LP1, LP2, LP3, LP4, LP5, LP6, LP7). The processes defining initialization and support of plaque, are framed by the big pink rectangle "PLS-inflammation" (LP1, LP6, LP7, LP8). The processes both in the circle and in the rectangle, are common for LP2-inflammation and PLS-inflammation (LP1, LP6, LP7). Subprocesses LP6.3 and LP7.1 take place only during LP2-inflammation, subprocess LP7.2 - only during PLS-inflammation.

The phases are also characterized by conditional graphs of unit antigenic presentation and logarithmic curves of unit quantity of effector ThN (Fig. 2-10 and Fig. 2-19).

Images and transformations of PG-Y(-)Mo-T and PG-Y(-)DC-T are specified only in the figures corresponding to phases 1, 2 and 6. In the figures corresponding to phases 3, 4 and 5, they are absent for the purpose of not raising complexity. Their possible transformations are identical to transformations of usual Mo and DC (Tab. 2-3).

***Phase 1. Prepsoriasis. Marker - homeostatic LP1.***

Phase 1 is the normal state of NLS. Phase 1 transfers into phase 2 at the moment of LP2 start (and LP3 begins once along with it). If LP4 begins simultaneously with LP3 (for example at LP2(IN)) phase 1 turns into phase 3.

Phase 1 precedes LP2 occurrence, it is common for all types of LP2. NLS can be Y-primed or not. In Fig. 2-11, Y-primed NLS is represented (resident ThN-Y are present). It means that future plaque is secondary, and if it is primary (i.e. the very first plaque in psoriatic's life ), hence, phase 0 occurred in the past (Fig. 2-24).

Y-primed NLS can become PLS with much greater probability, than Y-nonprimed NLS. LP2 (PsB) is an exception, it is the situation, when previous Y-priming does not influence further chain of events. At LP2 (PsB), during phase 0, immediate transition to phase 2 takes place.

As one of two possible variants, Tem-Z image also assumes previous Z-priming (Fig. 2-11).

During this phase, IZD and IYD are absent.

Renewal of pool of immunocytes (LP1) occurs homeostatically (the process intensity is weak). Mo-T and DC-T (incl. Mo-R and DC-R) degrade F-content faster, than they lose tolerance, and gradually transform into normal Mo and DC.

NLS differs from the norm by increased level of PDC. At LP2 (HPV), it is defined by latent HPV-carriage of KC.

***Phase 2. Prepsoriasis. Marker - LP2 and LP3.***

Phase 2 begins at the moment of LP2 start (along with LP3). Phase 2 turns to phase 3 at the moment of LP4 start. Phase 2 can return back to phase 1 in case of LP2 end (along with LP3) before LP4 start.

Phase 2 is at LP2(HPV), it includes medium-intensive LP1, LP2 and LP3 in the state of mutual balance (Fig. 2-20). During this phase, LP3 constrains diffusion of HPV among basal KC, and KC-v release special proteins constraining possible start of LP4 and LP5. LP6.2 takes place with average intensity, as MF necessary for this constrain are formed during this subprocess.

Tem-Z image also assumes (as one of two possible variants) previous Z-priming (Fig. 2-11). During this phase, IZD and IYD are absent.

On conditional graphs, phase 2 does not differ from phase 1, because of lack of adaptive response (Fig. 2-19).

After open trauma LP2(IN), phase 2 is absent, since LP4 (and phase 3) begins at once. At trauma that does not affect derma at all, or at trauma with damage of derma without epidermis injury, koebnerization is impossible, as Z-antigens do not get into derma (Weiss 2002).

***Phase 3. Prepsoriasis. Marker - LP4.***

Phase 3 begins at the moment of LP4 start. Phase 3 transfers to phase 4 at the moment of LP5 or LP7.1 start. This phase is transit and it includes high-intensive LP1, LP2, LP3 and LP4, and also medium-intensive LP6.2 and LP6.3.

***Phase 3(IN).***

The scheme of interaction of (sub)processes at LP2(IN) is illustrated in Fig. 2-9.
Phase 3 at LP2(IN) is illustrated in Fig. 2-14.
Phase 1 transfers to this phase right after open trauma of derma.

***Phase 3(HPV).***

The scheme of interaction of (sub)processes at LP2(HPV) is illustrated in Fig. 2-18.
Phase 3 at LP2(HPV) is illustrated in Fig. 2-21.
This phase is transit and begins in the result of active epidermal HPV-expansion (among basal layers) or attempts of dermal HPV-expansion (at HPV carriage of invasive types). During LP6.3 - maDC-Z1 and maDC-Z2 are formed.

***Phase 4. Prepsoriasis. Marker - LP5***

Phase 4 begins at the moment of LP5 or LP7.1 start. Phase 4 turns to phase 5 at the moment of LP8 start. Phase 4 can go to phase 1 if LP2 (along withLP3, LP4 and LP5) ends prior to LP8 start. Phase 4 can go to phase 2 (at LP2(HPV)) if LP4 and LP5 both come to end before LP8 begins.

Owing to LP4 providing abundant secretion of IFN-alpha, LP5 is quickly initiated - preventive adaptive response. Sharp growth of IZD - unit presentation of Z-antigen by maDC-Z takes place, and, as a consequence, exponential increase of Tem-Z quantity can be seen. Tem-Z proliferate both in skin (LP5) and in lymph nodes (LP7.1).

During phase 4, process LP4 gradually goes down (being suppressed), as the increased level of TNF-alpha suppresses PDC ability to secrete IFN-alpha. It is supposed that by phase 5 start, LP4 intensity becomes insignificant, that is why process LP4 is absent in the figures illustrating phases 5 and 6.

Phase 4 includes high-intensive LP1, LP2, LP3, LP5, LP6.2 and LP6.3 and medium-intensive LP4, LP6.1, LP6.4 and LP7.1. At that, LP7.1 intensity can be high if Z-priming was absent.

Owing to LP6.4 activity, IYD increases. If IYD reaches the critical level (depending on unit quantity of resident ThN-Y) than LP8.1 can begin (i.e. phase 5).

***Phase 4(IN).***

The scheme of interaction of (sub)processes at LP2(IN) is illustrated in Fig. 2-9.

Phase 4 at LP2(IN) is illustrated in Fig. 2-14.

During healing accompanied by LP3 and LP5 after trauma, IZD decreases gradually - unit presentation of Z-antigen.

During phase 5 at trauma healing, decrease of LP2 intensity takes place, and, as a consequence, LP3 and LP5 also weaken.

***Phase 4(HPV).***

The scheme of interaction of (sub)processes at LP2(HPV) is illustrated in Fig. 2-18.

Phase 4 at LP2(HPV) is illustrated in Fig. 2-21.

During LP6.3 - maDC-Z1 and maDC-Z2 are formed.

They present two dominant HPV-antigenes, accordingly: Z1 (through MHC class I) and Z2 (through MHC class II).

During LP5.1 - maDC-Z1 interacts with TcN-Z1, and maDC-Z2 with ThN-Z2. As a result of this interaction, TcN-Z1 and ThN-Z2 are activated, proliferate and secrete cytokines.

During LP5.2, interaction of KC-Z1 and TcN-Z1 takes place leading to destruction of KC-Z1.

## *Phase 5. Psoriasis. Marker - LP5 and LP8.*

Phase 5 begins simultaneously with LP8 start. Phase 5 turns to phase 6 at the moment of LP2 completion (along with LP3, LP4 and LP5).

Phase 5 will be able to go to phase 1 (remission) if it appears that after LP2 completion (along with LP3, LP4, LP5, LP6.3 and LP7.1) LP8 does not become self-sufficient and, therefore terminates quickly. Phase 5 corresponds to all processes in Fig. 2-9 (at LP2(IN)) or in Fig. 2-18 (at LP2(HPV)).

***Phase 5(IN).***

The scheme of interaction of (sub)processes at LP2(IN) is illustrated in Fig. 2-9.

Phase 5 at LP2(IN) is initiation and support of plaque after trauma (Koebner effect) is illustrated in Fig. 2-16. Simultaneously with trauma healing and decrease of unit presentation of Z-antigen, LP8 –false adaptive response against Y - subdominant antigen presented by maDC-Y can begin. It occurs, if IYD reaches critical level before the resolution of inflammation accompanying trauma healing.

***Phase 5(HPV).***

The scheme of interaction of (sub)processes at LP2(HPV) is illustrated in Fig. 2-18.

Phase 5 at LP2 (HPV) is initialization and support of plaque during LP5 against HPV.

It is illustrated in Fig. 2-23. During active adaptive response against Z1 and Z2 - dominant HPV-antigenes, initialization of LP8 - false adaptive response against Y - subdominant antigen presented by maDC-Y can begin. It can occur, if IYD reaches the critical level before LP5 – adaptive response against HPV-expansion stops.

If, despite LP3 and LP5 action, full release from HPV does not occur, HPV remains inside plaque aggravating it. In this case full transition to phase 6 does not occur.

## *Phase 6. Psoriasis. Marker - LP8.*

Phase 6 begins at the moment of LP2 end (along with LP3, LP4 and LP5, LP6.3 and LP7.1) at LP8 self-sufficiency.

Phase 6 can go to phase 1 (remission) that occurs at decrease of SPP severity.

Phase 6 can transfer to phase 5 at renewal of initial LP2 or joining of another LP2 (secondary infection). Plaque having arisen at trauma healing (Koebner effect) or HPV-expansion suppression remains (Fig. 2-17). The processes framed by the pink rectangle "PLS-inflammation" (Fig. 2-6, Fig. 2-9, Fig. 2-18) become self-sufficient. The processes that are out of this rectangle (as well as LP6.3 and LP7.1), terminate. It occurs if Y - subdominant antigen, presented by maDC-Y becomes dominant. For this purpose it is necessary provision by vicious cycles B and C sufficient quantity of maDC-Y and ThN-Y (per unit of PLS dermal volume).

Process LP8 will be self-sufficient until the decrease of unit income of Y-antigen (the quantity brought by DC-R and Mo-R from blood flow per unit of time per unit of dermal volume of PLS). LP8 severity in plaque, expansion or remission of this plaque depends only on unit income of Y-antigen. Unit income of Y-antigen is defined by SPP severity.

***Phases. Additions.***

At transition from phase 5 (or from phase 6) to phase 1, invisible Y-primed NLS-spot is preserved in the site of former plaque and around it. Inside this NLS-spot, there is high concentration of resident TL-Y (in the site of former plaque) gradually decreasing to the external border of the surrounding ring (Fig. 2-24, A, phase 1).

In plaque, there can be sites of phase 5 (LP2-inflammation and PLS-inflammation) and sites of phase 6 (only PLS-inflammation) without clear boundaries between them. LP2-inflammation intensification leads to PLS-inflammation intensification and expansion of phase 5 sites. Weakening of LP2-inflammation leads to weakening of PLS-inflammation and reduction of phase 5 sites. In Fig. 2-25, phase transitions possible at LP2-inflammation initialization, development and remission, as well as PLS-inflammation, are shown. In Fig. 2-26, variants of development of pinpoint plaque at LP2(HPV) are presented.

The type of dermal maDC depends on the phase (Fig. 2-5, Fig. 2-19). maDC appear in LP5 start (phase 4) and they are only maDC-Z. With LP8 start (phase 5), also maDC-Y appear, besides maDC-Z. If LP8 becomes self-sufficient (phase 6), all maDC are only maDC-Y. Total number of maDC is proportional to the sum of IZD and IYD (dark blue and red components on conditional graphs). maDC-Z precursors can be DC or Mo of any origin (resident or non-resident). Only blood DC-R or Mo-R can be precursors of maDC-Y.

# Discussion

## *Basic characteristics of Y-model*

Y-model of pathogenesis of psoriasis is based on three factors: one systemic and two local:

**Systemic necessary factor** (Part 1)

- SPP severity. It is estimated by unit Y-carriage of blood DC-R and Mo-R (quantity of Y-antigens in DC-R and Mo-R in 1 ml of blood). It is defined by a set of the interconnected subprocesses:
  - Hyperpermeability of intestinal walls for F-content (SP1)
  - Specific small intestine disbacteriosis (SP2)
  - Disorder of production and-or circulation of bile acids (SP3).
  - Overload and-or disorder of detoxicating systems (SP5)
  - Chronic PAMP-nemia (SP4)

**Local factors in specific skin site**

- Intensity and duration of LP2-inflammation.
- Y-priming level. It is estimated by concentration of TL-Y (Tem-Y - in derma, Tcm-Y - in lymph nodes);

If combination of these factors for specific NLS site appears critical, it is transformed to PLS-plaque. The role of any of these factors (processes and subprocesses) can be strengthened by genetic (Elder 2010, Gudjonsson 2009, Hollox 2008) and-or functional deviations.

Any PLS-plaque is caused by constant income in derma of blood DC-R and Mo-R containing endocytosed kPAMP (including PG-Y). DC-R and Mo-R arrive into skin together with other DC and Mo for renewal of dermal DC and MF pool. At LP2 and/or PLS-inflammation, this subprocess is intensified.

Tolerization of Mo-T and DC-T (incl. Mo-R and DC-R) is caused by longer stay under chronic kPAMP-load (SP8) (in comparison with DC and Mo).

PLS-plaque is a reaction of SIS caused by blood Mo-R and DC-R that, getting into derma, are transformed in maDC-Y and present Y-antigen to effector ThN-Y.

This reaction includes epidermal hyperproliferation, as one of mechanisms of false adaptive response of SIS on imaginary PsB-infection.

PLS-plaque initialization can occur only during LP2-inflammation including LP5 (adaptive response against LP2).

In particular, it is possible at LP2(IN) - open trauma of derma and at LP2(HPV) – HPV-carriage of KC.

Existence and severity of PLS-plaque is defined by intensity of income into derma of Y-antigen brought by Mo-R and DC-R.

Severity of PLS-plaque is aggravated by intensity of income into derma of kPAMP brought by Mo-T and DC-T.

These incomes are conditioned by SPP severity.

Severity of PLS-plaque is aggravated by intensity of LP2-inflammation if it continue to develop during PLS-inflammation.

SPP it is necessary for initialization and support of any PLS-plaque. SPP severity defines total severity of PLS-plaques as a whole. At decrease of SPP severity, remission of separate plaques takes place, up to total disappearance of all PLS-plaques.

Chemostatuses of blood Mo-T and DC-T (incl. Mo-R and DC-R) are similar to chemostatuses of nonactivated Mo and DC, which leads to analogousness of their migration behavior under the influence of chemokines.

Process LP2 can be initialized by any of triggers: trauma, HPV, S.aureus, Malassezia sp., Candida albicans, skin PsB-infection, etc. Processes LP2 can be different in different skin sites of a psoriatic.

Owing to processes LP3, LP5, LP6 and LP8, the level of cytokines IFN-gamma, GM-CSF, IL-1beta, IL-12, IL-17, IL-20, IL-22, IL-23, TNF-alpha secreted in PLS increases.

Under their influence, KC and FB actively secrete chemokines and AMP: CCL2, CCL20, CX3CL1, LL37, HBD-2, HBD-3, etc.

Chemokines and AMP get to blood flow, and even more Mo, DC, Mo-T, DC-T (incl. Mo-R, DC-R) and ThN-Y are attracted to the inflammation site.

Because of the increasing income from blood flow of Mo-R and DC-R, more maDC-Y, cooperating with effector ThN-Y are formed.

ThN-Y are activated, proliferate and secrete cytokines promoting hyperproliferation of KC, which is an attempt of skin to eliminate imaginary external PsB-infection through intensive renewal.

If SPP severity and-or intensity and duration of LP2-inflammation are insufficient, LP8 will not be able to begin. Or LP8 will begin, but at LP2 end it will terminate at once, as LP8 will not be able to become self-sufficient.

This alternative (beginning or impossibility of PLS-plaque beginning) also depends on the level of Y-priming in the site of LP2-inflammation.

If in some NLS-site LP2 it is not accompanied by LP5, or LP5 has weak intensity (for example at latent HPV-carriage) LP8 might not begin or begin with delay. The reasons of delayed initialization of LP8 can be: delayed activation of LP5 and-or increasing SPP severity and-or increasing level of Y-priming (for example because of tonsillar PsB-infection).

If an increase of SPP severity is temporary, plaques can be temporary, this exactly occurs at temporary guttate psoriasis.

During phase 5 (LP2-inflammation and PLS-inflammation), each LP2-activation aggravates and-or promotes plaque expansion. For this reason, process LP2 is called initiating and aggravating.

Similarly, secondary infections act (other than initiating LP2) that can join to already existing plaque.

If LP8 has already begun, the further existence of initiating LP2 in this plaque can appear unessential, i.e. phase 5 can go to phase 6. During phase 6, plaque severity and its expansion rate are defined by SPP severity (Fig. 2-6, vicious cycles B and C) and by Y-priming level in the site surrounding this plaque.

Y-priming level (resident placing ThN-Y - in skin, Tcm-Y - in lymph nodes), increases during any PsB-infection (tonsillar, skin or systemic) (Fig. 2-24).

The highest SPP severity is necessary for the primary plaque preceding Y-priming of the whole surface of skin. If Y-priming of skin has occurred, the primary plaque (especially in skin sites with increased level of Y-priming) can begin at lower SPP severity.

Psoriasis, once having begun, provides constant increased Y-priming of areas adjacent to plaques, as well as the whole skin surface. Hence, for expansion of primary and initializations of subsequent plaques, lower SPP severity is enough, than that for primary plaques initialization.

After plaque remission, an invisible NLS-spot with increased level of Y-priming remains on its place, which leads to increased probability of new PLS-plaque occurrence in the future in the same site (Clark 2011).

Primary guttate psoriasis (guttate psoriasis) approximately in 30% of cases completely resolves spontaneously, but in 70% it turns into chronic plaques immediately or after remission (Baker 2000, Baker 2006b). Guttate plaques is often preceded by intensive tonsillar PsB-infection when Y-priming of skin takes place and temporary, but intensive SPP occurs. And, if such patient, for example, has latent HPV-carriage in some sites, critical combination of all three factors is formed. Temporary plaques appear and then spontaneously disappearing at SPP end.

### *Comparison of Y-model with other models*

In App. 2-9, results of comparative analysis of Y-model and five other models of pathogenesis of psoriasis are presented in the tabular form: BF-model (Baker 2006b, Fry 2007b), N-model (Nestle 2009a, Nestle 2009b, Perera 2012), GK-model (Guttman-Yassky 2011, Nograles 2010), TC-model (Tonel 2009) and GL-model (Gilliet 2008, Lande 2010). The designation of each model by one or two Latin letters is introduced for the first time.

Other models (except BF) are based on the local concept, i.e. on the assumption that the reason of development and support of PLS-plaques is only of skin origin. Other models (except BF) do not consider any systemic processes.

The basic feature of Y-model is that, first, SPP is necessary for initialization and support of any plaque and, secondly, SPP severity defines psoriasis severity as a whole.

Let's list local (sub)processes noting their popularity or novelty (App. 2-9). Initial (sub)processes are a part of LP2-inflammation and do not enter vicious cycles.

**LP1. Attraction of immunocytes from blood flow.**

Known process. It is included in Y-model and other models as a vicious cycle link.

LP1.1 is known and included in Y-model (a link of the vicious cycle B). It is also included in other models also, but not as a vicious cycle link. The unique role of Mo-R and DC-R is considered only in Y-model.

Difference between Mo and DC of resident and non-resident origin (Zaba 2009a, Zaba 2009b) plays an important role in Y-model, but it is not considered in other models.

LP1.2 is known (attraction ThN, PDC, Neu, NK, etc.) and included in Y-model (ThN – in the vicious cycle C) and in other models.

**LP2. Initiating and aggravating process.**

Known initial process (Koebner effect).

LP2(IN) is included in Y-model and other models.

In GK-model at the initialization stage derma-epidermal antigen plays the leading role (we will designate it - X) (Guttman-Yassky 2011). X is antigen of commensal or pathogenic bacteria which have got to derma as a result of trauma (Nograles 2010) (Fig. 2-28).

LP2(HPV) has been for the first time included in Y-model.

**LP3. Innate response against LP2.**

Known initial process.

LP3.1(IN) is included in Y-model and other models.

In GK-model it is obviously not mentioned. At the initialization stage such reaction to X-antigen which immediately includes adaptive response is supposed. It is provided by LC and dermal BDCA-1+DC of resident origin carrying X-antigen into regional LN (Fig. 2-28).

LP3.1(HPV) has been for the first time included in Y-model.

LP3, as a whole, as a separate process is included for the first time in Y-model. In other models, the role of LP3, as a whole, is not reflected.

**LP4. Trigger of adaptive response against LP2.**

The known transit initial process is included in Y-model and other models (except BF).

In N-, GK and GL-model, this process is considered not transit, but as constant and is included in vicious cycles. However, this assumption does not correspond to the recently received results (Albanesi 2010, Bissonnette 2010).

**LP5. Adaptive response against LP2.**

Known initial process.

LP5 as a separate process has been for the first time included in Y-model.

In GK-model adaptive response against unknown X-antigen is partly represented (the initialization stage passing into the acute stage). Here the author shows the reaction of LC and dermal BDCA-1+DC which endocytose X-antigen and then carry it into regional LN for processing and presentation to nTL or Tcm. As a result effector Th1, Th17 and Th22 are formed in LN, which arrive subsequently in derma (Fig. 2-28).

Its role for initialization of LP8 has been for the first time formulated in Y-model.

LP5.2(HPV). One of the initial events preceding initialization of pinpoint plaque (and then promoting its aggravation) is attraction from derma into epidermis of CD8+Tem (N-model and TC-model). In Y-model, at LP2(HPV), during phase 4, such attraction of TcN-Z1 (Z1-specific CD8+TcN) occurs because of HPV-expansion. Such attraction really precedes initialization of LP8 (phase 5) and occurs during phases 4 and 5.

**LP6. Mo and DC transformations.**

LP6.1. Loss of tolerance of Mo-R, DC-R and MoDC-R to kPAMP.

Known (but not at psoriasis) subprocess.

For the first time it has been included (as a necessary link of the vicious cycle B) in Y-model.

LP6.2. MF and MoDC formation.

Known subprocess.

The subprocess is a link of the vicious cycle B and for the first time (for MoDC) it has been included in Y-model. In other models, the role of MoDC is not considered.

LP6.3. maDC-Z formation..

Known subprocess (but not at psoriasis).

As a separate process, for the first time it has been included in Y-model.

LP6.4. maDC-Y formation.

Known subprocess is included in other models, but without antigen definition. Subprocess with Y-antigen is included in Y-model and it is a necessary link of the vicious cycle B.

Y-model novelty is that Y-antigen is defined and the mechanism of its appearance in dermal DC-R and Mo-R is suggested. Y-model novelty is that the mechanism of loss of tolerance DC-R and Mo-R and their transformations in maDC-Y is suggested.

**LP7. Lymph nodes. Clonal proliferation.**

LP7.1. TL-Z formation. This is known initial subprocess that has been included in Y-model (and in GK-model also).

LP7.2. TL-Y formation. This is known subprocess that has been included (as a link of the vicious cycle C) in Y-model and other models (except GL).

In GK-model at the chronic stage effector TL are formed at interaction with CD208+DC in lymphoid structures (Fig. 2-28).

The characteristic feature of Y-model is that Y-priming is considered a significant local factor (as well as in (Sabat 2011).

**LP8. False adaptive response to imaginary PsB-infection.**

LP8.1. Presentation of Y-antigen by maDC-Y to effector ThN-Y.

This known subprocess has been included in other models, but without the definition of antigen (except BF-model) and without the concept of imaginary PsB-infection. The subprocess with Y-antigen is included in Y-model and is a necessary link of the vicious cycles B and C.

LP8.2. KC hyperproliferation. Change of skin architecture. Growth of basal membrane area and vascularity increase.

This known subprocess is included in Y-model and in other models.

**Other differences**

In GK-model, exit of self-DNA and self-RNA from the apoptotic psoriatic KC is considered as a vicious cycle link. It is supposed that complexes self-RNA-LL37 being formed do stimulate secretion of TNF-alpha, IL-6, and IL-23 by dermal DC, and their maturation in CD208+maDC lately interacting with TL. However, in GK-model, no assumptions concerning antigen, presented by CD208+maDC are made.

In Y-model, similar dependence (LP3.1 –> LP6.3, LP6.4) is regarded when the source of self-RNA are injured KC (LP3.1(IN)) or destroyed KC-v (LP3.1(HPV)).

In GL-model the exit of self-DNA from apoptotic psoriatic KC is considered as a vicious cycle link. It is supposed that complexes self-DNA-LL37 formed later stimulate secretion of IFN-alpha by PDC.

In the last version of N-model (Perera 2012), the exit of self-DNA and self-RNA from incompletely differentiated psoriatic KC is regarded as a vicious cycle link. It is supposed that self-DNA-LL37 and self-RNA-LL37 complexes formed later stimulate dermal PDC to secretion of IFN-alpha.

In Y-model, dependence (LP3.1 –> LP4) is considered as transit (phase 3). The basic source of self-DNA during this phase is considered not apoptotic, but injured KC or destroyed KC-v. Dependence (LP3.1 –> LP6) can become evident while LP2 (in particular, LP2(HPV)) and LP6 operate (i.e. during phases 4 and 5). However, in Y-model, these dependencies are not considered as vicious cycle links.

In other models (except BF), Neu and KC secrete LL37 and it is regarded as a vicious cycle link. In Y-model, LL37 secretion is a part of process LP3 that certainly influences initialization and support of LP4, but it is not considered as a vicious cycle link.

**Y-model novelty**

- systemic psoriatic process SPP - as the main necessary factor
- Y-antigen is defined and the mechanism of its appearance in dermal DC-R and Mo-R is suggested.
- mechanism of loss of tolerance DC-R and Mo-R and their transformation in maDC-Y is suggested.
- role of Mo-R and DC-R in LP1.1 as a necessary link of the vicious cycle B
- LP2(HPV) as an initiating process
- LP5 as an process necessary for LP8 initialization
- LP6.1 as a necessary link of the vicious cycle B
- role of MoDC in LP6.2
- mechanism of initialization of PLS-inflammation during LP2-inflammation (Koebner effect) is suggested.
- Y-priming role
- PLS-inflammation as a reaction of SIS to an imaginary external PsB-infection

## *Conclusions*

In Y-model (as well as in other models) it is admitted that genetic deviations can define the severity and the form of psoriasis manifestation, but they are not the reason of initialization or support of PLS-plaques.

Systemic psoriatic process (SPP) has been included in Y-model as a necessary factor for initialization and support of plaques.

Comparison of Y-model with other models shows that the majority of local processes included in Y-model have been earlier formulated in one or several models. However, as a whole, Y-model is principally new.

Concepts not borrowed from other models and formulated for the first time are collected in the block «Y-model novelty».

The systemic approach allows closer coming to construction of the model of pathogenesis of psoriatic arthritis as joint manifestation of SPP, and then to the model of pathogenesis of psoriatic disease as a whole.

The new Y-model requires check up. Although numerous facts from practice of researches and treatment confirm and-or do not contradict Y-model, yet does not mean its validity. The results of future researches will allow investigators to confirm and-or specify Y-model, define more precisely the role of intestine PsB, quantitative parameters of dependencies between processes (both systemic and local).

I hope that this work will stimulate joint researches of the dermatologists, rheumatologists, gastroenterologists and microbiologists and approximate complete solution of the riddle of psoriatic disease.

In the future, treatment of psoriatic disease will be directed not to cosmetic and-or anti-inflammatory correction of local manifestations, rather elimination and-or decrease of action of original causes, i.e. in most cases, to the treatment of intestine dysfunctions (SP1, SP2). The efficiency of such treatment will depend on the patient, his/her desire and possibility to control the way of life and diet. As consequence, remission will be prolonged (or life-long!), probably supported by regular or periodic intake of medicines (bacteriophages, pre - and probiotics).

**The author declares no conflict of interest.**

# Appendices

***Appendix 2-1. Abbreviations (except TL) (added sample from Appendix 1, Part 1)***

| | | Abbr. | Term and comments |
|---|---|---|---|
| | | AMP | Antimicrobial proteins (peptides) |
| * | | BS | Beta-hemolytic streptococci |
| * | | BSPG | Peptidoglycan of beta-hemolitic streptococci |
| * | | CPs | Chronic psoriasis |
| | #2 | DC | Dendritic cells |
| * | #2 | DC-T | Tolerized DC (App.2-2) |
| * | #2 | DC-R | Reprogrammed (tolerized) and repleted by PG-Y dendritic cells are subfraction of tolerized fraction DC-T. (App.2-2) |
| | | DDC | Dermal dendritic cells |
| | | EC | Endothelial cells |
| * | | F | Fragments of bacterial products with PAMP (including kPAMP) |
| | #1 | FB | Fibroblasts |
| | #1 | HPV | Human Papilloma Virus |
| * | | IB-Y | Interpeptide bridges from peptidoglycan of Str.pyogenes: L-Ala(2-3) or L-Ala-L-Ser |
| | | iNOS | Inducible nitric oxide synthase |
| * | | IYD | Unit Y-presentation by dermal maDC-Y (quantity of Y-antigens, presented in unit volume of derma) |
| * | | IZD | Unit Z-presentation by dermal maDC-Z (quantity of Z-antigens, presented in unit volume of derma). At LP2(HPV): Z is Z1 or Z2. |
| | #1 | KC | Keratinocytes |
| * | #1 | KC-v | HPV-carring keratinocytes |
| * | #1 | KC-Z1 | HPV-carring keratinocytes presenting Z1-antigen |
| | | LC | Langerhans cells |
| | | LP | Local process |
| | #1 | LPS | Lipopolysaccharide (endotoxin) |
| | | maDC | Mature DC |
| * | #2 | maDC-Y | Mature DC derived from DC-R or from MoDC-R, presenting Y-antigen |
| * | #2 | maDC-Z | Mature DC, presenting Z-antigen |
| | | MDP | Muramyl dipeptide - component Gram+ and Gram(-) PG, ligand NOD2 |
| | | MC | Mast cells |
| | | MHC | Major histocompatibility complex |
| | #2 | Mo | Monocytes |
| * | #2 | Mo-T | Tolerized Mo (App.2-2) |
| * | #2 | Mo-R | Reprogrammed (tolerized) and repleted by PG-Y monocytes are subfraction of tolerized fraction Mo-T. (App.2-2) |
| | | MoDC | DC, derived from Mo |
| * | #2 | MoDC-T | DC-T, derived from Mo-T |
| * | #2 | MoDC-R | DC-R, derived from Mo-R |
| | #2 | MF | Macrophages |
| * | #2 | MF-T | Macrophages, derived from Mo-T |
| * | #2 | MF-R | Macrophages derived from Mo-R |

**App. 2-1. (continuation)**

| | | Abbr. | Term and comments |
|---|---|---|---|
| | #2 | Neu | Neutrophils |
| * | #2 | Neu-T | Tolerized Neu (App.2-2) |
| * | #2 | Neu-R | Reprogrammed (tolerized) and repleted by PG-Y neutrophiles are subfraction of tolerized fraction Neu-T. (App.2-2) |
| | #2 | NK | Natural killers |
| | | NKT | Natural Killer T cell |
| | | NLS | Non-lesional (prepsoriatic, uninvolved) skin – psoriatic skin without symptoms |
| | | NOD2 | Intracellular receptor - ligand to MDP |
| | | PAMP | Pathogen-associated molecular patterns |
| | #2 | PDC | Plasmacytoid dendritic cells |
| | #1 | PG | Peptidoglycan (in particular PG-Y) |
| * | #1 | PG-Y | Peptidoglycan A3alpha with interpeptide bridges IB-Y (but can contain and others also) |
| | | PLS | Psoriatic lesional skin |
| * | #1 | PsB | Psoriagenic bacteria – Gram+ bacteria with peptidoglycan PG-Y (Part 1, SP2) |
| * | | SIS | Skin Immune system |
| | | slanDC | 6-sulfo LacNAc-expressing dendritic cells (Hansel 2011, Schakel 2006, Teunissen 2011). A subfraction of blood dendritic cells BDCA-1(-)DC fraction. It is very close to CD14(-)CD16+Mo in terms of characteristics. It can be found in blood flow and derma in psoriatics. A part of slanDC becomes TipDC in PLS. |
| * | | SP | Subprocess |
| * | | SPP | Systemic psoriatic process |
| | | TipDC | maDC actively secreting TNF-alpha and iNOS (in particular maDC-Y or maDC-Z) |
| | | TLR4 | Membranous receptor - ligand to LPS |
| | | TLR9 | Intracellular receptor – ligand to CpG |
| * | #1 | Y | Y-antigen = part(s) of interpeptide bridge IB-Y |
| * | | Y-model | Model of pathogenesis of psoriasis, offered in this monograph |
| * | | Z1 | Dominant antigen, presented through MHC class I for TcN-Z1. At LP2(HPV): Z1 is HPV-antigen. |
| * | | Z2 | Dominant antigen, presented through MHC class II for ThN-Z2. At LP2(IN): Z2 is antigen of commensal or pathogenic bacteria transferred into derma as a result of trauma. At LP2 (HPV): Z2 is HPV-antigen. |
| * | #1 | Z | At LP2(IN): Z is Z2, at LP2(HPV): Z is Z1 or Z2. |

* - new abbreviation, images: #1 – Fig. 2-1 and #2 – Fig. 2-2,
Links are given for good articles in Wikipedia.

### *Appendix 2-1a. Abbreviations for TL*

New designations ThN (for Th1/Th17/Th22) and TcN (for Tc1/Tc17/Tc22) are introduced. It has been made for the next reason: besides TL with clearly allocated secretion of IFN-gamma or IL-17 or IL-22, transitive TL, secreting more than one of these cytokines also exist (Kagami 2010, Nograles 2010, Zenewicz 2011). At that, Th1 can be transformed in Th17 and vice versa (Annunziato 2010, Kurschus 2010).

| | | Abbr. | Term and comments |
|---|---|---|---|
| | #3 | TL | T-lymphocytes (any) |
| | #3 | nTL | Naive TL. Non-effector and nonspecific to any of antigens, they can become effector and specific after interaction with maDC in a lymph node. |
| | | Tem | Effector memory TL (here called effector TL). They are formed from nTL (at primary response) and from Tcm (at secondary response) in a lymph node. Then they move to blood flow and into tissues to the site of inflammation to participate in adaptive response and, subsequently, stay there constantly. |
| | | Tcm | Central memory TL (here called memory TL). They are formed from nTL (at primary response) in regional LN. Then they move to blood flow and then to farther LN of the human organism to stay there constantly. At secondary response Tem are quickly formed in lymph nodes from Tcm. |
| | | Th1 | CD4+Tem, characterized by secretion: IFN-gamma(+)IL-17(-)IL-22(-) |
| | | Th17 | CD4+Tem characterized by secretion: IFN-gamma(-)IL-17(+)IL-22(-) |
| | | Th22 | CD4+Tem, characterized by secretion: IFN-gamma(-)IL-17(-)IL-22(+) |
| * | | ThN | CD4+Tem secreting one or more of three cytokines IFN-gamma, IL-17 and IL-22. In particular, these are Th1, Th17 and Th22, but also intermediate TL (sometimes designated as Th1/Th17, etc.) |
| | | Tc1 | CD8+Tem characterized by secretion: IFN-gamma(+)IL-17(-)IL-22(-) |
| | | Tc17 | CD8+Tem, characterized by secretion: IFN-gamma(-)IL-17(+)IL-22(-) |
| | | Tc22 | CD8+Tem, characterized by secretion: IFN-gamma(-)IL-17(-)IL-22(+) |
| * | | TcN | In particular, these are Tc1, Tc17 and Tc22, and intermediate TL as well (sometimes designated as Tc1/Tc17, etc.) |
| | | | |
| * | #3 | ThN-Y | Y-specific ThN |
| * | #3 | Tem-Y | Y-specific Tem (basically ThN-Y) |
| * | #3 | Tcm-Y | Y-specific Tcm |
| * | #3 | TL-Y | Tem-Y and Tcm-Y |
| * | #3 | TcN-Z1 | Z1-specific TcN |
| * | #3 | ThN-Z2 | Z2-specific ThN |
| * | #3 | ThN-Z | Z-specific ThN |
| * | #3 | Tem-Z | Z-specific Tem (basically TcN-Z1 and ThN-Z2) |
| * | #3 | Tcm-Z | Z1-specific and Z2-specific Tcm |
| * | #3 | TL-Z | Tem-Z and Tcm-Z |

* - new abbreviation, images: #3 – Fig. 2-3.

***Appendix 2-2. New terms (from Part 1, App. 2)***

| Term | Description | Bibliography |
|---|---|---|
| Chemostatus | Range of chemokine receptors expressed by cell (CCR1, CCR2, CCR3 etc.) including the amount of each. Chemostatus of cell defines it's behavior in reply to homeostatic and/or inflammatory chemokines. | Bachmann 2006, Male 2006, Sozzani 2005 |
| PAMP-load | Phagocyte-dependent consumption (linkage, endocytosis) of PAMP and contact with PAMP. | Part 1, App.9 |
| Tolerized phagocytes | The activated phagocytes under chronically increased PAMP-load are tolerized (reprogrammed).<br>Tolerized phagocytes are designated as Neu-T, Mo-T and DC-T. It is supposed that tolerized phagocytes (except known properties) have:<br><br>**<u>Property 1.</u> Despite contact, linkage with PAMP and endocytosis of PAMP (as a part of endocytosed F-content) their chemostatuses are similar to nonactivated.**<br><br>**<u>Property 2.</u> As tolerization occurs under chronic kPAMP-load then there is kPAMP in endocytosed F-content. Its degradation occurs slowly and not completely, i.e. kPAMP-carriage takes place.**<br><br>Properties 1 and 2 seem to depend on the blocking intracellular protein IRAK-M. Level of this protein in tolerized phagocytes becomes more than critical as a result of chronically increased PAMP-load.<br>Property 2 depends on parity between rate of endocytosis F and intracellular proteolytic activity.<br><br>In process of aging chemostatus of tolerized Neu changes similarly to chemostatus of nonactivated Neu.<br>Those and others start to express actively CXCR4 that provides their homing into bone marrow (Part 1, SP9). | Biswas 2009, Buckley 2006, Hedl 2007, Medvedev 2006, Nakatani 2002, Savina 2007, van't Veer 2007,<br><br>Part 1, App. 4, App.10 |
| R-phagocytes = Reprogrammed (tolerized) and repleted by PG-Y phagocytes | R-phagocytes are tolerized phagocytes which also are (PG-Y)-carriers. They are designated as Neu-R, Mo-R and DC-R and have additional:<br><br>**<u>Property 3.</u> There is PG-Y in endocytosed F-content. Its degradation occurs slowly and not completely, i.e. takes place (PG-Y)-carriage.**<br><br>Property 3 (as well as Property 2) depends on parity between rate of endocytosis F and intracellular proteolytic activity.<br>As endocytosis and conservation of PG-Y (as a part of F-content) by phagocytes occurs in a random way, also subfraction<br>R-phagocytes in tolerized fraction is formed in a random way. | |

**App. 2-2 (continuation)**

| Term | Description | Bibliography |
|---|---|---|
| kPAMP =<br>key PAMP | Not all PAMP but only key PAMP (kPAMP) provide reprogramming of tolerized phagocytes. Tolerized phagocytes were likely to contact with kPAMP than with other PAMP and/or endocytosed F-content contained more kPAMP than other PAMP. | |
| kPAMP-load | Phagocyte-dependent consumption (linkage, endocytosis) of kPAMP and contact with kPAMP | Part 1, App.9 |
| PAMP-nemia | Chronic increasing of kPAMP-load on blood phagocytes resulting in<br>a) formation of essential share of tolerized phagocytes;<br>b) kPAMP-level increasing in blood.<br>c) increased kPAMP-carriage of tolerized phagocytes | |
| Cytokines-deprogrammers | Normal phagocytes aren't tolerized (reprogrammed) and tolerized phagocytes (including R-phagocytes) are quickly lose tolerance (deprogrammed) in the presence of such cytokines.<br>IFN-gamma, GM-CSF and IL-12 are known. IL-12 stimulates non-monocytic cells to produce IFN-gamma. | Adib-Conquy 2002, Cavaillon 2008, Kylanpaa 2005, Randow 1997 |

*Appendix 2-3. Causative dependencies between processes and subprocesses*

<table>
<tr><th colspan="2" rowspan="2"></th><th colspan="15">Cause</th></tr>
<tr><th>SPP</th><th>LP1.1</th><th>LP1.2</th><th>LP2</th><th>LP3</th><th>LP4</th><th>LP5</th><th>LP6.1</th><th>LP6.2</th><th>LP6.3</th><th>LP6.4</th><th>LP7.1</th><th>LP7.2</th><th>LP8.1</th><th>LP8.2</th></tr>
<tr><td rowspan="15">Effect</td><td>SPP</td><td></td><td></td><td></td><td></td><td></td><td></td><td></td><td></td><td></td><td></td><td></td><td></td><td></td><td></td><td></td></tr>
<tr><td>LP1.1</td><td>+</td><td></td><td></td><td></td><td rowspan="2">+</td><td></td><td rowspan="2">+</td><td></td><td></td><td></td><td></td><td></td><td></td><td rowspan="2" colspan="2">+B</td></tr>
<tr><td>LP1.2</td><td></td><td></td><td></td><td></td><td></td><td></td><td></td><td></td><td></td><td>+</td><td>+C</td></tr>
<tr><td>LP2</td><td></td><td></td><td></td><td></td><td>#</td><td></td><td>#</td><td></td><td></td><td></td><td></td><td></td><td></td><td></td><td>#</td></tr>
<tr><td>LP3</td><td></td><td></td><td>+</td><td>+</td><td></td><td></td><td>+</td><td></td><td>+</td><td></td><td></td><td></td><td></td><td></td><td></td></tr>
<tr><td>LP4</td><td></td><td></td><td>+</td><td>+</td><td>+</td><td></td><td>#</td><td></td><td></td><td></td><td></td><td></td><td></td><td></td><td></td></tr>
<tr><td>LP5</td><td></td><td></td><td>+</td><td>+</td><td>+</td><td>+</td><td></td><td></td><td>+</td><td></td><td></td><td></td><td></td><td></td><td></td></tr>
<tr><td>LP6.1</td><td></td><td>+B</td><td></td><td></td><td></td><td></td><td rowspan="4">+</td><td></td><td colspan="3">+</td><td></td><td></td><td rowspan="4" colspan="2">+C</td></tr>
<tr><td>LP6.2</td><td></td><td rowspan="3">+</td><td></td><td></td><td></td><td rowspan="3">+</td><td>+B</td><td></td><td></td><td></td><td></td><td></td></tr>
<tr><td>LP6.3</td><td></td><td></td><td></td><td>+</td><td></td><td>+</td><td></td><td></td><td></td><td></td></tr>
<tr><td>LP6.4</td><td></td><td></td><td></td><td>+</td><td>+B</td><td>+</td><td></td><td></td><td></td><td></td></tr>
<tr><td>LP7.1</td><td></td><td></td><td></td><td></td><td></td><td></td><td>#</td><td></td><td></td><td>+</td><td></td><td></td><td></td><td></td><td></td></tr>
<tr><td>LP7.2</td><td></td><td></td><td></td><td></td><td></td><td></td><td></td><td></td><td></td><td></td><td>+C</td><td></td><td></td><td>#</td><td></td></tr>
<tr><td>LP8.1</td><td></td><td></td><td>+C</td><td></td><td></td><td></td><td></td><td></td><td>+B</td><td></td><td>+B</td><td></td><td></td><td></td><td>+</td></tr>
<tr><td>LP8.2</td><td></td><td></td><td></td><td></td><td></td><td></td><td></td><td></td><td></td><td></td><td></td><td></td><td></td><td>+</td><td></td></tr>
</table>

**Notes:**

The table is made according to Y-model of pathogenesis (Fig. 2-6).
Colors in the first column are used further for highlighting a specific (sub)process and-or causative dependence designated by an arrow beginning from this (sub)process.
+ - positive influence;
+B or +C - vicious cycles;
# - negative influence (suppression);
Empty table cells mean that connections are either absent or insignificant.

*Appendix 2-4. Local processes, references and hypotheses*

| Process or subprocess | | References with facts | | Hypotheses |
|---|---|---|---|---|
| | | **With psoriasis** | **Without psoriasis** | |
| **LP1** | Attraction of immunocytes from blood flow. | | | |
| **LP1.1.** | Attraction Mo and DC, Mo-T and DC-T (incl. Mo-R and DC-R) from blood flow. | *, but without mentioning Mo-T and Dc-T (incl. Mo-R and DC-R) | * | LP1-H1. Attraction of Mo-R and DC-R as a necessary vicious cycle link |
| **LP1.2.** | Attraction of other immunocytes from blood flow. | * | * | |
| **LP2** | Initiating and aggravating process. | | | |
| **LP2(IN)** | Open trauma of derma | Kalayciyan 2007, Weiss 2002 and other models | Gillitzer 2001, Ishida 2008 | |
| **LP2(HPV)** | HPV–carriage KC | Fomina 2009, Cronin 2008, De Koning 2011, Favre 1998, Fry 2007b, Mahe 2003, Majewski 2003, Prignano 2005, Viviano 2005, Weissenborn 1999 | De Koning 2005, Hazard 2007, Hebner 2006, Stanley 2009, Woodworth 2002 | LP2-H1. Not only epidermal, but dermal HPV-expansion of invasive types can become a trigger of PLS-plaque. |
| **LP3** | Innate response against LP2. | * | * | |
| **LP4** | Trigger of adaptive response against LP2. | | | |
| **LP4(IN)** | LP4 after trauma. | Other models except BF-model | Gregorio 2010 | |
| **LP4(HPV)** | LP4 against HPV. | no | Tang 2010 | see LP2-H1 |
| **LP5** | Adaptive response against LP2. | | | LP5-H1. LP5 is a necessary process for LP8 initialization. |
| **LP5(IN)** | Adaptive response after trauma. | see LP2(IN) | see LP2(IN) | |
| **LP5(HPV)** | Adaptive response against HPV. | see LP2(HPV) | see LP2(HPV) | LP5-H2. Attraction of TcN-Z1 from derma into epidermis as an event preceding LP8 initialization. |

**App. 2-4 (continuation)**

| Process or subprocess | | References with facts | | Hypotheses |
|---|---|---|---|---|
| | | **With psoriasis** | **Without psoriasis** | |
| **LP6** | Mo and DC transformations. | | | |
| **LP6.1** | Loss of tolerance to kPAMP. | no | Buckley 2006, Cavaillon 2008, Kylanpaa 2005, Medvedev 2006 | LP6-H1. LP6.1 - is a necessary vicious cycle link. |
| **LP6.2** | MF and MoDC formation. | Fuentes-Duculan 2010, Zaba 2009a (except MoDC) | Angel 2006, Haniffa 2009 | LP6-H2. MoDC-R formation is a probable vicious cycle link. |
| **LP6.3** | maDC-Z formation | Zaba 2009a | Angel 2006, Haniffa 2009 | |
| **LP6.4** | maDC-Y formation | Zaba 2009a | Angel 2006, Haniffa 2009 | LP6-H4. LP6.4 - is a necessary vicious cycle link. |
| **LP7** | Lymph nodes. Clonal proliferation. | Nestle 2009a, Vissers 2004 | Clark 2010, Stanley 2009 | |
| **LP8** | False adaptive response to imaginary PsB-infection. | Immune response in other models without specific antigen. Against Y-antigen in BF-model. | no | |

* - (sub)process is known, references are available;
LPn-Hm = hypothesis Hm, proposed in process LPn.

## *Appendix 2-5. Chemokine receptors, chemokines and AMP, certainly or possibly involved in LP1.*

| Chemokin receptors / Chemokins and AMP (names) | | CCR1 | CCR2 | CCR3 | CCR4 | CCR5 | CCR6 | CCR7 | CCR10 | CXCR1 | CXCR2 | CXCR3 | CXCR4 | CXCR5 | CXCR6 | CX3CR1 | FPRL1 | ChemR23 |
|---|---|---|---|---|---|---|---|---|---|---|---|---|---|---|---|---|---|---|
| New | Old | | | | | | | | | | | | | | | | | |
| CCL2 | MCP-1 | 9,4 | 10,2 A | 7,8 | | 7,5 | | | | | | | | | | | | |
| CCL3 | MIP-1alpha | 10,2 | | | | 9,0 | | | | | | | | | | | | |
| CCL4 | MIP-1beta | 7,7 | | | | 9,6 | | | | | | | | | | | | |
| CCL5 | RANTES | 8,2 | | 9,3 | | 9,7 B | | | | | | 6,4 | | | | | | |
| CCL17 | TARC | | | | 8,7 L | | | | | | | | | | | | | |
| CCL19 | MIP-3beta | | | | | | | 9,0 | | | | 5,9 | | | | | | |
| CCL20 | MIP-3alpha | | | | | | 8,5 C | | | | | 6,8 | | | | | | |
| CCL21 | 6Ckine/SLC | | | | | | | 9,3 | | | | | | | | | | |
| CCL27 | CTACK | | | | | | | | * O | | | | | | | | | |
| CX3CL1 | Fractalkine | | | | | | | | | | | | | | | * D | | |
| CXCL1 | Gro-alpha | | | | | | | | | 6,4 P | 9,7 P | | | | | | | |
| CXCL8 | IL-8 | | | | | | | | | 9,5 Q | 9,5 Q | | | | | | | |
| CXCL9 | MIG | | | | | | | | | | | 8,3 E | | | | | | |
| CXCL10 | IP-10 | | | | | | | | | | | 9,8 F | | | | | | |
| CXCL11 | ITAC | | | | | | | | | | | 10,5 G | | | | | | |
| CXCL12 | SDF-1alpha | | | | | | | | | | | 6,4 | 8,2 H | | | | | |
| CXCL16 | | | | | | | | | | | | | | | * I | | | |
| HBD-2 | | | * J | | | | * K | | | | | | | | | | | |
| HBD-3 | | | * J | | | | | | | | | | | | | | | |
| Chemerin | | | | | | | | | | | | | | | | | | * M |
| LL37 | | | | | | | | | | | | | | | | | * N | |

**Notes to App. 2-5.**

Basic chemokines and AMP associated with local psoriatic processes in skin are included in the table (Nickoloff 2007b, Nestle 2009a). In case if a chemokine receptor is a ligand of chemokine or AMP, the value of maximal affinity between them is specified in the table cell according to IUPHAR (Murphy 2010). In case the value in IUPHAR is absent, mark "*" is given in the table and information on their interaction is received from another source. Green color – the receptor/chemokine work at homeostasis, pink - during attraction of immunocytes in PLS (Mantovani 2004). Ovals including certain list of chemokines and AMP are present only on schemes of Fig. 2-9 and Fig. 2-18. References in the form of capital Latin letter are made to the following notes:

**A)** CCL2 (MCP-1) and CCR2.

CCL2 is secreted in the norm (Ginhoux 2007). CCL2 is secreted by KC in PLS significantly above the norm (Vestergaard 2004). KC also secrete CCL2 in NLS near to plaque. In PLS and NLS secretion is concentrated in the basal layer (Gillitzer 1993).

**B)** CCL5 (RANTES) and CCR5.

CCL5 is secreted both in the norm and at inflammation. The level of CCL5 in PLS is significantly higher than in NLS, CCR5+ prevail among CD68+Mo found out in PLS compared with similar Mo in NLS. However, CCR5 receptor does not play an essential role in psoriasis pathogenesis (De Groot 2007). In PLS only KC secrete CCL5, the level of which is 3-4 times more than in the norm (Raychaudhuri 1999); TNF-alpha in PLS actively promotes secretion of CCL5 by basal KC (Giustizieri 2001). Attraction of NK in PLS occurs basically due to CXCL10 and CCL5. At that, 80% of involved NK are CD56(bright)CD16(-) (Ottaviani 2006).

**C)** CCL20 (MIP-3alpha) and CCR6.

CCL20 is secreted in the norm for attraction of blood Mo - precursors of LC (Ginhoux 2007); CCL20 promotes chemotaxis of CCR6+DC in derma to epidermis (Homey 2000, Hedrick 2010).

**D)** CX3CL1 (fractalkine) and CX3CR1.

CX3CL1 is secreted by EC for attraction of CX3CR1+CD16+Mo from blood flow (Ancuta 2003). CX3CL1+DDC are found out in PLS-derma in the number of 187 against 6 (per sq. mm) in normal derma. Proinflammatory DC and MoDC actively secrete CX3CL1 promoting attraction of CX3CR1+Mo and CX3CR1+DC from blood flow – their own potential precursors (Raychaudhuri 2001). CX3CL1 is actively secreted by KC in PLS, and much less - in NLS (Sugaya 2003).

**E)** CXCL9 (MIG) and CXCR3.

Secretion of CXCL9 in PLS is higher than in NLS, where it exceeds the norm (Chen 2010). CXCL9 is highly expressed in capillaries and in papillaes. It is secreted by MF and by endothelial cells EC. In norm and in NLS, it is practically absent (Goebeler 1998). It is secreted in PLS mainly by EC (in dermal vessels) and by mononuclear cells. In PLS-epidermis (secreted by KC) and in near NLS much lesser amount of it is present, and in distant NLS it is completely absent (Rottman 2001).

**F)** CXCL10 (IP-10) and CXCR3.

CXCL10 is secreted in PLS in greater amount than in NLS, where it exceeds the norm (Chen 2010). It is secreted in PLS mainly by EC (in dermal vessels) and by mononuclear cells. In PLS-epidermis and near NLS only insignificant amount can be found, and in distant NLS it is absent completely (Rottman 2001). Attraction of NK in PLS occurs basically due to CXCL10 and CCL5. The basic source of CXCL10 in inflamed skin are activated KC, at that psoriatic KC are capable of secreting CXCL10 more than the normal ones (Ottaviani 2006).

**G)** CXCL11 (ITAC) and CXCR3.

Secretion of CXCL11 in PLS is higher than in NLS, where it exceeds the norm (Chen 2010).

**H)** CXCL12 (SDF-1alpha) and CXCR4 are a homeostatic pair.

CXCL12 is constantly secreted in normal skin (Bogunovic 2006, Urosevic 2005).

**I)** CXCL16 and CXCR6.

Expression of CXCL16 is very active in basal layers of KC in PLS, and is increased in all capillary EC in PLS-derma (Oh 2009).

**J)** HBD-2, HBD-3 and CCR2.

Earlier it has been assumed that HBD-2 is a ligand of CCR6 - receptor expressed by blood DC, but practically not expressed by Mo (Yang 1999). However, later, it has been shown that CCR6 receptor is not involved in chemotaxis of Mo supported by HBD-2 and HBD-3 (Soruri 2007). Recently it has been established that HBD-2 and HBD-3 support chemotaxis through CCR2 receptor and, hence, they can support attraction of blood CCR2+Mo and CCR2+DC. Preliminary incubation of Mo with CCL2 has disturbed such migration (Rohrl 2010). It is necessary to note that secretion of all AMP both quantitatively and spatially depends of LP2 and the depth of skin damage as well (Diamond 2009, Poindexter 2005). Probably, this is the reason of essentially various results observed at mRNA HBD-2 psoriasis: 2-time excess of the norm (Li 2004), 5000 times (Gambichler 2008), and 20000 times (De Jongh 2005). In PLS-epidermis, the most part of HBD-2 is secreted by KC in the top spinous and granular layers, it reaches maximal concentration in the intercellular space in the cornual layer, where it envelops all KC completely (Huh 2002). In normal skin the maximal concentration of HBD-2 is observed in the basal layer (Poindexter 2005).

**K)** HBD-2 and CCR6.

HBD-2 is strongly expressed in PLS (De Jongh 2005). HBD-2 is a ligand of CCR6 (Yang 1999). HBD-2 is a chemoattractant for Mo and MF, but not through CCR6, rather through CCR2 (Soruri 2007, Rohrl 2010).

**L)** CCL17 (TARC) and CCR4.

At psoriasis modeling in mice, CCL27 blocking did not affect attraction of TL into skin, only additional blocking of interaction between CCL17 and CCR4 stopped this attraction (Sabat 2007).

**M)** Chemerin and ChemR23 (CMKLR1).

ChemR23 is present in PLS in large amounts (Skrzeczynska-Moncznik 2009a, Skrzeczynska-Moncznik 2009b). ChemR23 promotes attraction of PDC and NK in PLS and promotes colocalization of NK and DC.

**N)** LL37 and FPRL1.

LL37 is a ligand of FPRL1 - receptor expressed by blood Mo and DC. Interaction of LL37 and FPRL1 supports chemotaxis of Mo and DC to the inflammation site (Sozzani 2005).

**O)** CCL27 (CTACK) and CCR10.

CCL27 is constantly secreted by KC in normal skin (Baker 2006c, Bogunovic 2006). Chemokine CCL27 is important for attraction of TL in PLS (Sabat 2007).

**P)** CXCL1 and CXCR1, CXCR2.

CXCR1 is secreted by KC, it is important for attraction of Neu in PLS (Baker 2006c).

**Q)** CXCL8 (IL-8) and CXCR1, CXCR2.

CXCR8 is secreted by KC, it is important for attraction of Neu and CD8+TL in PLS (Baker 2006c).

*Appendix 2-6. Chemokine receptors and blood immunocytes.*

| Chemokin receptors / Immunocytes | CCR1 | CCR2 | CCR3 | CCR4 | CCR5 | CCR6 | CCR7 | CCR10 | CXCR1 | CXCR2 | CXCR3 | CXCR4 | CXCR5 | CXCR6 | CX3CR1 | FPRL1 | ChemR23 | References |
|---|---|---|---|---|---|---|---|---|---|---|---|---|---|---|---|---|---|---|
| **CD14++Mo** | 95 | 98 | 10 | | 10 | | | | 50 | 95 | | 100 | | | 15 | | 85<br>/100<br>**D** | **A** |
| **CD14+CD16+Mo** | 90 | 87<br>**B** | 20 | | 35<br>**B** | | | | 45 | 65 | | 100<br>**B** | | | 50<br>**B** | | 60<br>/400<br>**D** | |
| **CD14(low)CD16+Mo** | 10 | 10 | 10 | | 10 | | | | 18 | 15 | | 95 | | | 86 | | 62/<br>1200<br>**D** | |
| **Mo** | 21<br>/13 | 72<br>/58 | | | 11<br>/5 | 15<br>/8 | 0 | | 61<br>/46 | 42<br>/20 | 32<br>/10 | 32<br>/10 | 0 | | | *<br>**E** | | **C** |
| **DC** | 1<br>/9 | 69<br>/25 | | | 5<br>/5 | 16<br>/24 | | | 40<br>/17 | 41<br>/13 | 21<br>/6 | 36<br>/33 | 3<br>/4 | | | *<br>**E** | | |
| **PDC** | 4<br>/9 | 45<br>/22 | | | 5<br>/9 | 16<br>/5 | | | 54<br>/43 | 36<br>/10 | 30<br>/4<br>**F** | 40<br>/13<br>**F** | 0 | | | | 93<br>**D**<br>**F** | |
| **BDCA-1+DC** | | | | | | | | | | | | | | | | | 42 | **D** |
| **BDCA-3+DC** | | | | | | | | | | | | | | | | | 25 | |
| **DC** | 0 | ++ | | - | ++ | 0 | 0 | | 0 | - | -/+ | ++ | 0 | | *<br>**G** | | | **H** |
| **PDC** | -- | ++ | | - | ++ | 0 | ++ | | 0 | - | + | ++ | 0 | | | | | |
| **CD16+NK** | | | | | | | | | * | | | | | | * | | *<br>**J** | **I** |
| **CD16(-)NK** | | | | | ++ | | | | | | ++ | | | | + | | | **K** |
| **Neu** | * | *<br>**M** | | | | * | | | ++<br>**O** | *<br>**N** | | * | | | | *<br>**E** | | **L** |
| **Neu naive** | | | | | | | | | ++ | ++ | | - | | | | | | **O** |
| **Neu senescent** | | | | | | | | | ++ | + | | + | | | | | | |
| **Neu apoptotic** | | | | | + | | | | ++ | - | | ++ | | | | | | |
| **CD4+TL** | | | | 90<br>**Q** | | 72<br>**R** | | *<br>**T** | | | 16<br>**Q** | | | | | *<br>**E** | | **P** |
| **CD8+TL** | | | | 50<br>**Q** | | 51<br>**R** | | | | | 30<br>**Q** | | | | | *<br>**E** | | |
| **Tem** | | | | **S** | | **R** | | **S** | | | | | | | | | | |
| **Th1** | | | | - | | 26 | | - | | | * | | | | | | | **T** |
| **Th17** | | | | * | | 81 | | - | | | | | | | | | | |
| **Th22** | | | | * | | - | | * | | | | | | | | | | |

**Notes to App. 2-6.**

Green color - receptor works at homeostasis, pink - mainly during attraction of immunocytes in PLS. If in a reference there is no information either on a share of expressing cells, or on expression intensity, designation stands «*» in the table. If comments are necessary - a capital Latin letter directing to one of notes is added in the table. A Latin letter in the last column refers to all lines opposite to it.

**A)** Mo - three fractions according to the nomenclature of blood monocytes (Ziegler-Heitbrock 2010).

For each receptor, the percentage of cells expressing it in a certain fraction is specified. An empty table cell means that the data are absent (Ancuta 2003).

**B)** CD14+CD16+Mo.

This fraction expresses CCR2, CCR5, CXCR4 and CX3CR1 (Serbina 2008). In the absence of inflammation their homeostatic traffic in derma provides chemokine CXCL12 (SDF-1alpha), which is a ligand of CXCR4 receptor (Urosevic 2005). Chemokine CXCL12 is constantly secreted by dermal endothelial cells, and CXCR4 receptor is expressed by all fractions of blood monocytes (Strauss-Ayali 2007). Besides CXCL12, dermal endothelial cells constantly secrete chemokine CX3CL1 (fractalkine) (Ziegler-Heitbrock 2007), which also promotes homeostatic attraction of blood CD16+Mo with increased expression of CX3CR1 (Gautier 2009).

**C)** CD33+CD14+Mo; CD33+CD14(-)CD11c+DC and CD33(dim)CD14(-)pDC.

For each receptor, the percentage of cells expressing it and /MFI - average intensity of fluorescence are specified. Mo and PDC, but not myeloid DC, are attracted by high concentrations of CXCL10 (Cravens 2007). At the present, full data characterizing chemokine receptors expression by fractions of blood BDCA-1+ and BDCA-1(-)DC are unknown. Recently a new chemokine XCR1 receptor has been found in blood BDCA-1(-)BDCA-3+DC, however there is no information about secretion of its ligand - XCL1 - at psoriasis (Crozat 2010).

**D)** Three fractions of monocytes and three fractions of dendritic cells (BDCA-1+DC, BDCA-1(-)BDCA-3+DC and PDC) according to the nomenclature (Ziegler-Heitbrock 2010).

For ChemR23, the percent of cells expressing it and /MFI - average intensity of fluorescence are specified. At comparison with migration, in reply to CXCL12, chemerin induced the same migration of DC, but only 50% of migration of PDC (Vermi 2005).

**E)** Mo, DC, Neu and TL.

LL37 is a ligand of FPRL1 - receptor expressed by blood Mo and DC. Interaction of LL37 and FPRL1 supports chemotaxis of Mo, DC, Neu and TL to the inflammation site (Sozzani 2005).

**F)** PDC.

Attraction of PDC in NLS-derma takes place due to ChemR23 receptor – a ligand of chemerin (Albanesi 2010). Traffic of PDC at homeostasis and inflammation in skin, besides ChemR23 receptor, is also provided by receptors CXCR3 (ligand of CXCL10) and CXCR4 (ligand of CXCL12) (Sozzani 2010).

**G)** DC.

As to their attraction by chemokine CX3CL1 (fractalkine), experimental data concerning its ligand expression (CX3CR1 receptor) are ambiguous. This receptor expresses by maDC, but there is diverse information on its expression by DC. It is shown that mRNA CX3CR1 expression in immature and mature MoDC is almost identical, and chemotaxis of all dendritic cells (initiated CX3CL1) is essential and almost identical (Dichmann 2001). On the other hand, CX3CR1 expression is found only in maDC (Liu 2011). The level of CX3CR1 expression in maDC defines their movement inside derma after their maturation. All chemokine receptors responsible for DC attraction in inflamed tissues are listed in (Lukacs-Kornek 2008).

**H)** DC and PDC.

In an early work, chemokine receptors of two fractions of blood dendritic cells were investigated. In the same work it is noticed that expression of chemokine receptor not always defines chemotaxis to the influence of a ligand of this receptor. In particular, in chemotactic tests PDC reacted only to CXCL12 (SDF-1alpha) (Penna 2001).

**I)** CD56(low)CD16+NK.

Attraction of CD16+NK in inflamed tissues occurs through their chemokine receptors ChemR23 (ligand of chemerin), CX3CR1 (ligand of CX3CL1 - fractalkine) and CXCR1 (ligand of CXCL1, CXCL6 and CXCL8). Getting into inflamed tissues, NK, besides destruction of infected cells, also cooperate with DC accelerating their maturation, and own cytotoxicity as well. Besides attraction of PDC, NK promotes colocalization of NK and DC in inflamed tissues (Moretta 2008).

**J)** NK

Attraction of NK is regulated by chemerin (it is shown in vitro), which can define their excess in PLS (and also excess of PDC) (Skrzeczynska-Moncznik 2009a).

**K)** CD56++CD16(-)NK.

These cells accumulate in PLS under the influence of CCL5 and CX3CL1 as CCR5 and CX3CR1 are well expressed by them. Besides, CXCR3 are well expressed by them (Ottaviani 2006).

**L)** Neu. All chemokine receptors expressed by Neu are listed in (Kobayashi 2008).

**M)** Neu express CCR2 (Rohrl 2010).

**N)** Neu well express CXCR2 receptor (it is a ligand of cytokine-chemokine CXCL8 and a ligand of CXCL1) defining their attraction to inflammation sites (Reddy 2010).

**O)** Neu changes expression of its chemokine receptors along with aging (more details are in Part 1, SP9). Consecutive change of expression of receptors involved in release of new Neu from bone marrow, their aging and partial return to bone marrow is analysed in (Rankin 2010).

**P)** CD4+TL and CD8+TL.

In the table cell, % of epidermal TL in PLS with the corresponding receptor is specified. In PLS-epidermis, the majority of CD4+TL and about a half of CD8+TL express CCR4, whereas only 30% of CD8+TL and 16% of CD4+TL express CXCR3. In PLS-epidermis, 72% of CD4+TL and 51% of CD8+TL express CCR6, which is significantly higher than in blood TL (Teraki 2004). The same information is given in (Sabat 2007).

**Q)** CCR4 and CXCR3 on TL.

In PLS, CCR4+ cells were CD3+TL or CD68+Mo or CD68+MF. In PLS, CXCR3+ cells are mainly CD3+TL (Rottman 2001). The same information is given in (Sabat 2007).

**R)** CCR6 on CD4+TL and CD8+TL

Among all blood immunocytes, CCR6 expression is found mainly on CD4+TL and CD8+TL, and it is more intensive on CD4+TL. At more detailed study, it was found out that CCR6 expression is mainly observed on Tem (Liao 1999).

**S)** CCR4 and CCR10 on Tem.

After interaction of CLA+Tem with E-selectin, expressed on EC, Tem get under the influence of CCL27 as their part expresses CCR10. Besides, Tem also express CCR4 and this is the reason for its attraction by CCL17 (Sabat 2007, Clark 2010).

**T)** There is no any unique marker to separate Th17 from Th1, therefore several markers were used.

CCR6 is expressed on 81% of IL-17A+ cells and only on 26% of IFN-gamma+ cells, being the best marker distinguishing Th17 from Th1. Recently, CCR4 and CCR10 have been identified on Th22 (Kagami 2010). The same data for Th1 and Th17 are presented in (Harper 2009).

## *Appendix 2-7. Cytokines and cells. Secretion and influence.*

| Cells | | Mo | MF | PDC | DC | ma DC | Neu | NK | ThN | | | TcN | | | EC | MC | FB | KC |
|---|---|---|---|---|---|---|---|---|---|---|---|---|---|---|---|---|---|---|
| | | | | | | | | | Th1 | Th17 | Th22 | Tc1 | Tc17 | Tc22 | | | | |
| **EGF** | secr. | | **A** | | | | | | | | | | | | | | **A** | |
| | infl. | | | | | | | | | | | | | | | | | **A** |
| **GM-CSF** | secr. | | | | | | | | | | | | | | | | **B** | **B** |
| | infl. | **B** | | | **B** | | | | | **B** | | | **B** | | | | | |
| **IFN-alpha** | secr. | | | **C** | | | | | | | | | | | | | | |
| | infl. | **D** | | | **D** | **D** | | **E** | | **E** | | | **E** | | | | | |
| **IFN-gamma** | secr. | | | | | | | **H** | **E,H** | **–** | **–** | **E,H** | **–** | **–** | | | | |
| | infl. | **B** | | | **B,F** | **F** | | **E** | | | | | | | | | | **F** |
| **IL-1beta** | secr. | **F** | **F** | | **F** | **F** | | | | | | | | | | | | |
| | infl. | | | | **F,P** | **F,P** | | | | | | | | | | | | **F** |
| **IL-6** | secr. | | | **C** | **P,G** | **P,G** | | | | **G** | | | | | **G** | | **G** | **G** |
| | infl. | | | | ***G*** | | | | | **G** | | | **G** | | | | | |
| **IL-12** | secr. | **H** | **H** | | **H** | **H** | | | | | | | | | | | | |
| | infl. | | | | **H** | **H** | | **H** | **H** | | | **H** | | | | | | |
| **IL-17** | secr. | | | | | | **I** | | **–** | **I** | **–** | **–** | **I** | **–** | | **I** | | |
| | infl. | | | | | | | | | | | | | | | | **I** | **I** |
| **IL-20** | secr. | | **J** | | | **J** | | | | | | | | | | | | **J** |
| | infl. | | | | | | | | | | | | | | | | | **J** |
| **IL-22** | secr. | | **K** | | **K** | | | **K** | **–** | **–** | **K** | **–** | **–** | **K** | | | | |
| | infl. | | | | | | | | | | | | | | | | | **K** |
| **IL-23** | secr. | | | | **L** | **L** | | | | | | | | | **L** | | | |
| | infl. | | | | | | | | **L** | **L** | **L** | | | | | | | **L** |
| **iNOS** | secr. | | **M** | | | **M** | | | | | | | | | | | | |
| | infl. | | | | | | | | | | | | | | | | | |
| **KGF** | secr. | | | | | | | | | | | | | | | | **N** | |
| | infl. | | | | | | | | | | | | | | | | | **N** |
| **TGF-beta** | secr. | | | | | | | | | | | | | | | | **O** | **O** |
| | infl. | | | | | | | | | | | | | | | | **O** | **O** |
| **TNF-alpha** | secr. | **P** | **P** | **C** | **P** | **P** | | **P** | **P** | | | **P** | | | | **P** | | **P** |
| | infl. | **D** | | **D** | **D,P** | **D,P** | | | | | | | | | **P** | | | **F** |

**Notes to App. 2-7.**

Basic cytokines and cells associated with processes in PLS are included in the table. Each table cell is divided into the upper and lower parts. The upper part of a table cell corresponds to ability of biological cell to secrete a specific cytokine, and lower part – to the ability of cytokine to influence the biological cell actively. If a certain type of cells is also defined by its disability to secrete a specific cytokine (for example, ThN and TcN), designation «-» can be seen in the table. An empty table cell means that secretion (influence) is absent or insignificant. If comments are necessary - one or several Latin letters is given in the table cell meaning references to the notes:

**A)** EGF (epidermal growth factor)

It is secreted by MF and FB, influences KC and FB, in particular, stimulating GM-CSF secretion (Mascia 2010, Nestle 2009b). It can be secreted by MF.

**B)** GM-CSF, IFN-gamma.

Tolerized Mo-T, DC-T and MoDC-T (incl. Mo-R, DC-R and MoDC-R) lose tolerance to PAMP-content under the influence of cytokines-deprogrammers GM-CSF, IFN-gamma (LP6.1). KC and FB secrete GM-CSF (Koga 2008, Mascia 2010). Activated ThN and TcN also secrete GM-CSF (Male 2006, Shi 2006).

**C)** IFN-alpha, TNF-alpha, IL-6.

PDC secrete IFN-alpha in great amount, and slightly less amounts of TNF-alpha and IL-6 (LP4).

**D)** IFN-alpha, TNF-alpha.

Under the influence of IFN-alpha, Mo and DC increase expression of PAMP-receptors (in particular, TLR4). Under cooperative influence of IFN-alpha and TNF-alpha transformation of Mo in MoDC (LP6.2), as well as maDC formation (LP6.3 and LP6.4) is accelerated. Increased level of TNF-alpha suppresses ability PDC to secrete IFN-alpha.

**E)** IFN-alpha, IFN-gamma.

LP5 depends from LP4. IFN-alpha influences activation of ThN and TcN and formation of Tcm, especially, if LP2 is a virus (Seo 2010, Zhang 2005). IFN-alpha also promotes activation of TL (in particular, of Th1) and increases their secretion of IFN-gamma (Eriksen 2005).

IFN-alpha promotes formation of Th22 (Nestle 2009a).

Th1 and Tc1 secrete IFN-gamma under the influence of IFN-alpha, during activation due to interaction with maDC as well.

Under the influence of IFN-alpha and IFN-gamma, NK increase their cytotoxic (cytolytic) activity.

**F)** IFN-gamma, IL-1beta, TNF-alpha.

Mo, MF, DC and maDC secrete IL-1beta at activation and during interaction with antigens.

IFN-gamma and IL-1beta together influence maturation of maDC (LP6.3 and LP6.4).

IFN-gamma and TNF-alpha influence expression of adhesive molecules on KC, which promotes interaction of KC and activated ThN (LP8.2).

Influence of TNF-alpha and IL-1beta on KC promotes active secretion of CCL2, CCL20 and HBD-2.

**G)** IL-6.

All ThN and TcN involved and present in PLS get under the influence of a high level of IL-6 secreted by EC, DC, and Th17. It allows ThN and TcN to avoid suppression from Treg and gives Th17 the chance to participate actively in inflammation (Goodman 2009). KC and FB also secrete IL-6 activating DDC (Nestle 2009a).

**H)** IL-12.

It is secreted by DC and maDC at activation and maturation. IL-12 accelerates transformation of DC in maDC (autocrine influence). At maturation, maDC secrete even more IL-12. Mo (to a lesser degree), and CD163+MF both actively secrete IL-12 (Fuentes-Duculan 2010). IL-12 influences Th1, Tc1 and NK and they secrete IFN-gamma (Randow 1997).

**I)** IL-17.

Activated ThN (basically Th17) and TcN (basically Tc17) secrete IL-17A and IL-17F (Nestle 2009a, Nestle 2009b). MC and Neu in PLS secrete IL-17 (Lin 2011, Res 2010). IL-17 influences KC and FB secreting GM-CSF (Koga 2008, Mascia 2010). IL-17 promotes hyperproliferation of KC also secreting chemokines and AMP. Increased secretion of IL-17 is regarded as one of vicious cycle links (App.2-9).

**J)** IL-20.

It is secreted by maDC and MF (Wang 2006), and by KC as well (Baker 2006c). It causes hyperproliferation and acanthosis influencing (including autocrine way) KC (Guttman-Yassky 2011, Sabat 2011).

**K)** IL-22.

It is secreted by activated ThN (basically Th22) and TcN (basically Tc22), but also by NK (Kagami 2010, Sabat 2011, Zenewicz 2011). IL-22 in PLS is also secreted by MF and DC (Res 2010).

IL-22 stimulates KC to secretion of chemokines and AMP; IL-22 stimulates KC to secretion of IL-20;

IL-22 (together with IL-20) inhibits differentiation of KC, which causes hyperproliferation and acanthosis; IL-22 causes production of MMP1 and MMP3, which weakens extracellular matrix (Guttman-Yassky 2011, Sabat 2011, Wolk 2009). Increased secretion of IL-22 is considered as one of vicious cycle links (App.2-9).

**L)** IL-23.

IL-23 is secreted basically by dermal DC and maDC (at activation and maturation) and activates ThN. EC also secrete IL-23 (Meglio 2010, Tonel 2010). IL-23 influences Th1 and stimulates Th17 to secretion of IL-17, and Th22 - to secretion of IL-22 (Guttman-Yassky 2011, Zenewicz 2011). Besides, IL-23 influences KC promoting formation of STAT3 (Nestle 2009a). Increased secretion of IL-23 is considered as one of vicious cycle links (App.2-9).

**M)** iNOS (inducible nitric oxide synthase).

It has antimicrobial and antitumoral properties. It is secreted by TipDC, in which some researchers find more signs of MF (Dominguez 2010).

**N)** KGF (keratinocyte growth factor).

It is secreted by FB, and can influence proliferation of KC (Lowes 2007, Nestle 2009b).

**O)** TGF-beta.

It is secreted and interacts with KC and FB (Nestle 2009b).

**P)** TNF-alpha.

Activated MF secrete the basic part of TNF-alpha (Clark 2006b).

Also, TNF-alpha is secreted by Mo and DC during activation, and by DC during maturation.

Self-RNA-LL37 complexes formed during LP3.1 cause activation and maturation of DC. It takes place involving endosomal TLR8 and stimulates DC and maDC to secretion of TNF-alpha and IL-6 (Guttman-Yassky 2011).

MC (mast cells) also secrete TNF-alpha (Sabat 2011).

Activated Th1, Tc1, NK and KC secrete TNF-alpha (Nestle 2009b).

TNF-alpha together with IL-1beta activates DC and promotes maturation.

TNF-alpha influences EC increasing expression of E-selectin (a ligand of CLA) and ICAM-1, which provides attraction of Tem from blood flow (Tonel 2009, Sabat 2011).

Increased secretion of TNF-alpha is one of vicious cycle links (App.2-9).

***Appendix 2-8. Psoriasis on NLS-transplants placed onto AGR129-mice.***

Investigations of the psoriasiform plaques arising on mouse's skin immediately or on NLS-transplants after their transplantation have been conducted for a long time (Gudjonsson 2007).

However, only in 2004 it become possible to create the line of knockout mice AGR129, in which spontaneous development of typical PLS-plaques on NLS-transplants was seen for the first time. Not only good engraftment of transplant appeared to be characteristic for this line of mice, but also conservation of donor resident immune cells, first of all TL and DC, the most part of which, as a rule, is destroyed by NK of a recipient mouse. These studies and their subsequent analysis also showed that under certain circumstances the initialization and development of PLS-plaque can occur at participation of only resident immune cells (Boyman 2004, Boyman 2007, Nestle 2005a, Tonel 2009, Tonel 2010).

AGR129-mice are RAG-2(-)/(-), which provides inferiority of mouse B-lymphocytes and TL. Deficiency on receptors to IFN-alpha and IFN-gamma provides significantly lowered cytotoxic (cytolytic) activity of mouse NK cells both in vitro and in vivo. AGR129-mice appeared to be excellent recipients of human NLS-transplants. For initialization and development of PLS-plaque on NLS-transplant no exogenous cells or factors are required, except for those contained in NLS-transplant itself. One of defining factors for PLS-plaque development is activation and proliferation of donor Tem, being already in NLS at the moment of transplantation.

In psoriatics with the confirmed diagnosis (12 patients) fragments of NLS (6 x 2 x 0,04 cm) from the lower part of the back or breech (31 samples) were taken. These fragments were replaced onto AGR129-mice. In 28 cases from 31 (90%) since the 4$^{th}$ week after transplantation, plaques spontaneously appeared on NLS-transplants, having completely developed by the 6-8$^{th}$ week. Histological condition of NLS as of the date of transplantation was similar to normal human skin. The histology of mature plaque developed on NLS-transplant after its transplantation, was comparable to biopsy of PLS-plaque of a donor with psoriasis.

Various combinations have been checked up at transplantation of donor human skin and a recipient line of mice. It has appeared that development of plaque of psoriatic phenotype takes place only at NLS-transplantation onto AGR129-mice. Papillomatosis and acanthosis revealed at different moments of PLS-plaque development on NLS-transplants onto AGR129-mice, were absent on the control transplants ($p < 0{,}0002$) (Boyman 2004).

Inflammation in NLS-transplant, immediately after its engraftments, is similar to LP2(IN) - dermal trauma of NLS, which create conditions for LP4 start (Gregorio 2010, Tang 2010). Absence of any donor cells in infiltrate (including NK) limits possibilities of innate response LP3(IN) on restriction and elimination of LP2-inflammation. It makes inevitable LP4 start and fast transition to adaptive response LP5(IN) to trauma. It was exactly the phenomenon observed at the detailed time-dependent analysis of events in NLS-transplants (Boyman 2004, Nestle 2005a).

In these works, the role of PDC (actively secreting IFN-alpha) in PLS-plaque initialization was proved for the first time. It was shown that the quantity of PDC in NLS exceeds the norm, and in PLS it is greater than in NLS.

Besides, it was revealed that a share of PDC in blood of psoriatics is decreased to 0,18% in comparison with 0,32% in the norm.

It became possible to show at research of NLS-transplants that in dermal department PDC begins active secretion of IFN-alpha in 7 days after transplantation, this secretion reaches the maximum by the 14$^{th}$ day, but then quickly decreases. Visible psoriatic epidermal activation begins only on the 21$^{st}$ day after transplantation, and developed PLS-plaque is formed definitively on the 35$^{th}$ day. Active PDC-secretion of IFN-alpha in derma is an early and transient event preceding plaque occurrence. It is proved that PDC do secrete IFN-alpha, and the most important thing is that this secretion is an indispensable condition of the subsequent appearance of plaque (Nestle 2005a).

It is known that PDC actively secrete IFN-alpha at interaction of the endosomal receptor TLR9 with CpG – endocytosed fragments of virus DNA (Cao 2007). However, the role of antimicrobial protein

LL37 in intensification of influence of CpG on TLR9 as well as influence of fragments of self-DNA (own DNA) on TLR9 were elucidated in the late years (Lande 2007, Gilliet 2008, Hurtado 2010).

At NLS-transplantation, a trauma of donor derma takes place and, as a consequence, in extracellular dermal space donor DNA and RNA appear. Due to the trauma in NLS-transplant, donor KC, Neu and FB actively secrete LL37 (Dorschner 2001). Self-DNA-LL37 and self-RNA-LL37 complexes are formed. Donor PDC endocytose these complexes, deliver them to endosomal TLR7 and TLR9. Then complexes cooperate with receptors (self-DNA-LL37 with TLR9, self-RNA-LL37 with TLR7). As a consequence, PDC actively secrete IFN-alpha (Ganguly 2009, Nograles 2010, Tang 2010).

On the other hand, donor KC can have latent HPV-carriage (Favre 1998). Restraint of HPV-expansion in NLS is provided by NK, however, after NLS-transplantation, replenishment of NK pool from blood flow occurs only at the expense of mouse NK, cytolytic activity of which in AGR129-mice is essentially decreased. As a consequence, HPV-expansion begins, the quantity of extracellular virus CpG is increased. PDC endocytose them. CpG also as well as self-DNA-LL37 complexes influence on TLR9. And LL37 strengthens this influence. Thereby, active secretion of IFN-alpha can be caused not only by trauma, but also latent HPV-carriage. In the absence of HPV-carriage and at normal engraftment of NLS-transplant, consequence of trauma disappear gradually (LP2, LP3, LP5 processes are completed) and plaque on NLS-transplant appears to be in phase 6. If HPV-carriage on NLS-transplant takes place, such plaque remains in phase 5 (Fig. 2-7, Fig. 2-8).

Within the limits of Y-model, NLS-transplant contains Mo-T and DC-T (incl. Mo-R and DC-R), arrived at homeostatic renewal of dermal Mo and DC of non-resident origin before transplantation (Fig. 2-4, Fig. 2-11). Mo-R and DC-R arrive from blood flow together with F-content (including PG-Y) and, being reprogrammed, keep tolerance. They gradually degrade F-content, therefore those Mo-R and DC-R preserve the possibility to be transformed into maDC-Y that arrived from blood flow immediately before transplantation.

Donor derma contains an overwhelming amount (> 90%) of Tem in comparison with their quantity present in blood flow (< 10%) (Clark 2006a, Clark 2010). As activity of LP8 is caused also by LP7.2, an essential part of Tem in NLS-transplant should be ThN-Y. ThN-Y after interaction with maDC-Y can be activated and proliferate, enlarging the total amount of CD3+TL in NLS-transplant (Tonel 2010).

AGR129-mice are knockout on the receptors for IFN-alpha and IFN-gamma. It means that secreted in NLS-transplant IFN-alpha (initialization) and IFN-gamma (support) will mainly influence donor cells.

Under the influence of IFN-alpha, secreted by PDC, transformation of Mo-R into MoDC-R (LP6.2) takes place (Farkas 2011).

Under the influence of IFN-gamma and GM-CSF, loss of tolerance of part of Mo-R and DC-R to kPAMP-content (LP6.1) occurs and their transformation into maDC-Y takes place (LP6.4). The same phenomenon is promoted by self-RNA-LL37 complexes influencing DC-R and MoDC-R via endosomal TLR8 (Ganguly 2009).

Thereby, preconditions to LP8 start appear. In this experiment, only the vicious cycle B takes place, as LP7.2 (for donor TL-Y) is impossible (Fig. 2-6).

PLS-plaques started on the 21st day and prevailed until the termination of observation of mice-recipients (56 days). Long (more than 30 days) action of LP8 without attraction of new DC-R and Mo-R (LP1.1) from blood flow can have the following reasons:

- ability of maDC to present antigens continuously;
- kPAMP-preactivation of Mo-R and DC-R providing increased resistance of maDC-Y to influence of cytotoxic cells (Mueller 2006);
- insufficient quantity of the cytotoxic cells capable of elimination of maDC-Y in the transplant;

As a result, it appears that there are enough of maDC-Y (formed from dermal Mo-R and DC-R present in NLS at the moment of transplant capture) and effector ThN-Y (proliferating from ThN-Y present in NLS at the moment of transplant capture) for initialization and long maintenance of LP8 (more than 30 days).

It is possible to conclude that IYD level - unit Y-presentation, necessary for initialization and support of LP8 in NLS-transplant on AGR129-mice, is lower than the level necessary for initialization and support of LP8 in PLS (Fig. 2-10).

It is possible to assume that at further observation of NLS-transplant (after the $56^{th}$ day) PLS-plaque would have gradually regressed and disappeared. Or it could have been transformed into LP2(HPV)-plaque in case of HPV-carriage on the transplant. It should occur because of reduction of maDC-Y quantity that are not replenished from blood flow and gradually die. LP8.1 subprocess (similarly to LP8, as a whole) will be completed gradually because of IYD decrease.

***Appendix 2-9. The comparative analysis of models of psoriasis pathogenesis.***

After SPP local (sub)processes of Y-model are listed in the table.

First three columns contain the information about (sub)processes and dependencies of Y-model. In the first column, color designates the group of lines associated with the concrete process. In the second column, causative dependencies are shown. In the third column, comments are given.

The last five columns contain the information about other models of pathogenesis, grouped according to (sub)processes and dependencies of Y-model.

An empty table cell means that the model does not contain any information about (sub)process and-or dependence. Figure under the model name, contains original illustration.

Abbreviations. ChemA = chemokines and AMP, Cyt = Cytokines.

| | | **Other models** | | | | |
|---|---|---|---|---|---|---|
| **Y-model** | | **BF-model** | **N-model** | **GK-model** | **TC-model** | **GL-model** |
| Cause > Effect | (Sub)process. Comments. | Baker 2006b, Fry 2007b | Nestle 2009a, Nestle 2009b, Perera 2012 | Guttman-Yassky 2011 | Tonel 2009 | Gilliet 2008 |
| Fig. 2-6 | | Part 1, fig.2 | Fig. 2-27 | Fig. 2-28 | Fig. 2-29 | Fig. 2-30 |
| **Systemic processes** | | | | | | |
| | **SP1** | | Systemic processes are not considered. | | | |
| | **SP2** | Intestine. Beta-streptococcal carriage. | | | | |
| | **SP3** | | | | | |
| | **SP4** | (PG-Y)-load on blood Mo | | | | |
| | **SP5** | | | | | |
| | **SP6** | Tonsils. Beta-streptococcal infection. | | | | |
| | **SP7** | | | | | |
| | **SP8** | (PG-Y)-carriage of blood Mo | | | | |
| | Similar process in Y-model is Y-priming (phase 0), when Tem-Y and Tcm-Y are formed. | Most activated TL become anergic or perish, whereas TL-Y remain because of contact with PG-Y(+)Mo. It occurs in tonsillar (or intestine) regional LN. | | | | |
| SPP > LP1.1 | Systemic psoriatic process (SPP) is necessary and it defines the severity of psoriasis as a whole. | Systemic process defined by the presence of PG-Y(+)Mo at blood flow and in LN. | | | | |

Continued on next page.

| Y-model | | Other models | | | | |
|---|---|---|---|---|---|---|
| Cause > Effect | (Sub)process. Comments. | BF-model | N-model | GK-model | TC-model | GL-model |
| **Local processes** | | | | | | |
| **LP1**. Attraction of immunocytes from blood flow. | | | | | | |
| ChemA > LP1.1 | LP1.1 – link of the vicious cycle B. | PG-Y(+)Mo migrate into skin to activate TL-Y. | It is mentioned, but without participation of Mo-T and DC-T (incl. Mo-R and DC-R). It is not considered as a vicious cycle link. | | | |
| ChemA > LP1.2 | LP1.2 – link of the vicious cycle C. | Attraction of TL-Y is one of the reasons of psoriatic inflammation. | Th1 (expressing CLA, CXCR3 and CCR4) and Th17 (expressing CLA and CCR4 and CCR6) migrate from blood flow into derma due to chemokines CCL20, CXCL9, CXCL10 and CXCL11, etc. | Attraction of ThN and PDC is a vicious cycle links. | | |
| ChemA > LP1.2 | LP1.2 | | Neu are involved by CXCL8, CXCL1, etc. into epidermis through derma from blood flow (2009b). | | | CXCL8 involves Neu from blood flow. |
| | LP1.2 | | NKT also participate (2009b). | Are not considered. | | |
| **LP2**. Initiating and aggravating process. | | The external trigger (Fry 2007b) or Ker-antigen (one of keratins) | Stress or infection or medicinal preparations or trauma influencing skin. | Trauma or skin infection. Derma-epidermal X-antigen. | Trauma or skin infection. | Mechanical trauma. |
| | LP2(IN). Self-RNA and self-DNA get to the extracellular space from damaged KC. | | Self-RNA and self-DNA get to the extracellular space from damaged KC. | | Self-DNA get to the extracellular space from damaged KC. | |
| | LP2(HPV). Self-DNA, HPV-DNA and self-RNA get to the extracellular space from destroyed KC-v. | | | | | |

Continued on next page.

| Y-model | | Other models | | | | |
|---|---|---|---|---|---|---|
| Cause > Effect | (Sub)process. Comments. | BF-model | N-model | GK-model | TC-model | GL-model |
| **LP3.** Innate response against LP2. | | It is not regarded separately (except GK-model). | | | | |
| LP3 > Cyt > LP3 | - | | KC secrete IL-1alpha/beta, IL-18, TNF-alpha activating other KC. | | | |
| Cyt > LP6 | - | | These cytokines activate both LC and DDC. | | | |
| LP3 > ChemA | Neu and KC secrete LL37. | | Neu and KC secrete LL37. | | | |
| LP3.1(IN) > LP4 | | | In the extracellular space self-RNA-LL37 and self-DNA-LL37 complexes are formed. Then, PDC endocytose these complexes, which cooperate with TLR7 and TLR9 accordingly. | | In the extracellular space self-DNA-LL37 complexes are formed. Then, PDC endocytose these complexes, which cooperate with TLR9. | |
| LP3.1 > (LP6.3, LP6.4) | Self-RNA-LL37 complexes stimulate DC to secretion of TNF-alpha, IL-6 and IL-23 and to maturation DC in maDC. | | Self-RNA-LL37 complexes activate DC and promote their maturation in maDC. | Self-RNA-LL37 complexes stimulate DC to secretion of TNF-alpha, IL-6 and IL-23 and to maturation DC in maDC. | | |
| LP3.1(HPV) > LP4 | PDC endocytose CpG-LL37 complexes, which cooperate with endosomal TLR9. | | | | | |
| **LP4.** Trigger of adaptive response against LP2. | | | | | | |
| LP4 > Cyt | PDC actively secretes IFN-alpha and a little of TNF-alpha. | | PDC actively secretes IFN-alpha. | | | |
| LP4 > LP6 | | | IFN-alpha activates maturation and differentiation of DC. | | | |
| **LP5.** Adaptive response against LP2. | | The immune response after trigger influence or against Ker-antigen. | It is not regarded separately | Not completely | It is not regarded separately | |
| **LP6.** Mo and DC transformations. | | PG-Y(+)MF formation. | Mo-T and DC-T (incl. Mo-R and DC-R) - are absent. Mo and DC of resident and non-resident origin are not separated. LP6.1 - is absent. LP6.2 - only MF formation, MoDC formation - is not considered. LP6.3 and LP6.4 - are not parted (except GK-model). | | | |
| LP6 > Cyt > LP8.1 | | | DC secrete IL-23. TipDC secrete IL-23, iNOS and TNF-alpha. | TipDC secrete TNF-alpha, iNOS, IL-20, IL-12 and IL-23 activating Th1 and Th17. | TNF-alpha is secreted by activated MF, DDC and, to a lesser degree, by KC and TL. | |
| Cyt > LP6 | | | IL-1beta, *IL-6 and* TNF-alpha activate DDC. | | High levels of TNF-alpha promote maturation of DC. | |

Continued on next page.

| Y-model | | Other models | | | | |
|---|---|---|---|---|---|---|
| Cause > Effect | (Sub)process. Comments. | BF-model | N-model | GK-model | TC-model | GL-model |
| **LP7.** Lymph nodes. Clonal proliferation. | | | | | | |
| | LP7.1. maDC-Z. Migration into LN. | | Activated DDC migrate to draining LN for presentation of an unknown antigen (of own or microbial origin) to nTL and promote their differentiation in Th17 and-or in Th1 (2009a). Also, in Tc17 and-or in Tc1 (2009b). | LC and BDCA-1+DDC endocytose X-antigen and carry it into LN. | | |
| | LP7.2. maDC-Y. Migration into LN. | | | | | |
| | LP7.1. Interaction maDC-Z and TL-Z in LN. | | | As a result of interaction with maDC in LN effector Th1, Th17 and Th22 are formed. | | |
| | LP7.2. Interaction maDC-Y and TL-Y in LN. | | DC and TL form perivascular groups and lymphoid structures around blood vessels due to chemokines, such as CCL19 secreted by MF (2009b). | CD208+maDC are colocalized with TL in the lymphoid structures including CCR7+ cells and cells secreting CCL19. | | |
| Cyt > LP7.2 > LP1.2 | | | Effector Th17, Tc17, Th1 and Tc1 leave from LN into blood flow, circulate there and are slowed down in skin capillaries because of interactions ligand-receptor controlled by selectin and integrin (2009b). | | TNF-alpha promotes EC to express E-selectin (CLA-ligand) and ICAM-1, which provides attraction of Tem from blood flow. | |

Continued on next page.

| Y-model | | Other models | | | | |
|---|---|---|---|---|---|---|
| Cause > Effect | (Sub)process. Comments. | BF-model | N-model | GK-model | TC-model | GL-model |
| **LP8.** False adaptive response to imaginary PsB-infection. | | The immune response to imaginary beta-streptococcal infection | | | | |
| | LP8.1. Y-antigen presentation by maDC-Y to effector ThN-Y. | Y-antigen presentation by maDC-Y to TL-Y. | Dermal DC present unknown autoantigens to TL. | | | Dermal DC present unknown autoantigens to autoreactive TL. |
| Cyt > LP8.1 > Cyt | | Cytokines (including IFN-gamma) secreted by TL-Y cause proliferation KC and development of psoriatic plaque. | Both Th1 and Tc1 secrete IFN-gamma and TNF-alpha (2009b). Th17 (and Tc17 - 2009b) secrete IL-17A, IL-17F and IL-22. | Th1 secrete IFN-gamma.<br><br>IL-23 stimulates Th17 and Th22 to secretion of IL-17 and IL-22. | Th1 and Th17 secrete IFN-gamma, IL-17 and IL-22. These cytokines influence KC and lead to psoriatic changes. | Activated Th17 secrete IL-22 and IL-17. |
| | In Y-model, it is considered as a part of LP5.2(HPV). | | CD8+Tem express VLA-1 binding with collagen IV (via alpha1beta1-integrin). It allows them to get from derma to epidermis through the basal membrane. | | alpha1beta1-integrin is expressed only on epidermal TL. They define psoriatic changes as increase of their number (and not TL as a whole) correlates with plaque beginning. | |
| Cyt > LP8.2 | | | IL-17A, IL-17F, IL-22, IFN-gamma, TNF-alpha influence KC and stimulate their hyperproliferation. | IL-22 and other cytokines of IL-20 family cause hyperproliferation and acanthosis. | IFN-gamma, IL-17 and IL-22 induce in KC expression of ICAM-1, CD40 and MHC-II. Under the influence of these cytokines, KC hyperproliferate. Which leads to acanthosis. IL-22 influence KC through receptor IL-22R > STAT3 > hyperprolife-ration. | IL-22 and IL-17 induce KC hyperprolife-ration.<br><br>*They activate Neu.* |

Continued on next page.

| Y-model | | Other models | | | | |
|---|---|---|---|---|---|---|
| Cause > Effect | (Sub)process. Comments. | BF-model | N-model | GK-model | TC-model | GL-model |
| Cyt > LP8.2 > Cyt | | | KC secrete IL-1beta, IL-6, TNF-alpha and TGF-beta. FB secrete KGF, EGF and TGF-beta. KC and FB cooperate through secreted cytokines, and that leads to tissue reorganization and deposition of extracellular matrix (for example, collagen and proteoglycans). | | KC secrete IL-1beta, IL-6 and TNF-alpha. | |
| Cyt > LP8.2 > ChemA | KC secrete LL37, but it is not accepted by vicious cycle link. | | IL-17A, IL-17F, IL-22, IFN-gamma, TNF-alpha influence KC and stimulate secretion of LL37, HBD1, HBD2, S100A7, S100A8, S100A9, and chemokines CCL20, CXCL1, CXCL3, CXCL5, CXCL8, CXCL9, CXCL11. | IFN-gamma influences KC-secretion of chemokines CXCL1, CXCL2, CXCL3, CXCL5, CXCL8, CXCL9, CXCL10, CXCL11 and VEGF causing angiogenesis IFN-gamma and IL-17 influence KC-secretion of chemokines, AMP and lipokalin-2. In particular KC secrete LL37 that is accepted by a vicious cycle link. | | IL-22 and IL-17 influence KC-secretion of chemokines (for example CXCL8) and AMP.<br><br>In particular, KC secrete LL37, which is accepted by a vicious cycle link. |
| | | | | | | |
| | | | **In addition** | | | |
| During LP3.1 from damaged KC or destroyed KC-v self-DNA and self-RNA get to the extracellular space, and LL37-complexes are formed.<br><br>These events are considered as transit and-or aggravating, but not as vicious cycle links. | | | Self-DNA and self-RNA come from incompletely differentiated KC in the extracellular space. Self-DNA-LL37 and self-RNA-LL37 complexes are endocytosed by PDC and influence TLR9 (Perera 2012) | Self-DNA and self-RNA get from apoptotic KC to the extracellular space. Self-RNA-LL37 complexes (self-RNA from apoptotic KC) stimulate DC to secretion of TNF-alpha, IL-6, and IL-23 and their maturation in CD208+maDC. It is accepted by vicious cycle links. | | Self-DNA get to the extracellular space from apoptotic KC. Self-DNA-LL37 complexes (self-DNA from apoptotic KC) stimulate PDC to secretion of IFN-alpha. It is accepted by vicious cycle links. |
| Genetic deviations can define severity and the type of psoriasis, but they are not the reason of its initialization or support. However, they can affect the intensity of any of processes. | | Genetic predis-position. | Environment factors can cause psoriasis in genetically predisposed people carrying alleles of susceptibility. | | Genetic predisposition. | Possible dependence on the genetic variations of IL-23R influencing Th17-LL37 axis. |

**Notes.** Light blue color designates main pathogenetic links according to N-model. Italics designates the information which is not supported by other references.

## *Appendix 2-10. List of essential changes and additions*

This appendix is intended for readers who are familiar with the previous edition of this book (e1.2) and allows to get briefly acquainted only with essential changes and additions (Tab. 2-4). In the text the significant new or revised fragments are marked with a vertical line on the right. The new works included into the bibliography are also marked the same way.

**Tab. 2-4. Changes and additions**

| What | Where |
|---|---|
| Tolerized phagocytes are designated Neu-T, Mo-T and DC-T.<br>Also derived from Mo-T macrophages and dendritic cells are designated MF-T and MoDC-T.<br>For all tolerized phagocytes special images are offered (Fig. 2-2) .<br>In figures these images as a rule are used for PG-Y(-) tolerized phagocytes.<br>Concerning chemostatus of senescent Neu-T specification is made. | App. 2-1 и 2-2. |
| Definition of tolerized phagocytes is specified:<br>kPAMP-carriage is recognized by obligatory (Property 2).<br>Definition of R-phagocytes is specified:<br>(PG-Y)-carriage is recognized by obligatory (Property 3).<br>As a result there was actually renaming entities:<br>Before: a) tolerized b) R-phagocytes and c) PG-Y(+) R-phagocytes;<br>Now: a) and b) tolerized and c) R-phagocytes.<br>It has entailed the most part from changes listed further. | App. 2-1 и 2-2. |
| In all fragments where earlier there were materials about R-phagocytes (Mo-R and DC-R) corrections were made:<br>Before: "R-phagocytes",<br>Now: «Tolerized phagocytes (incl. R-phagocytes)».<br>Before: «Mo-R and DC-R»,<br>now: «Mo-T and DC-T (incl. Mo-R and DC-R)», etc. | In several places. |
| Illustrations. Fig. 2-1 and Fig. 2-2 are corrected and added so that to become identical with fig. 4 and fig. 5 of Part 1.<br>The Fig. 2-6, Fig. 2-7, Fig. 2-8, Fig. 2-9, Fig. 2-11, Fig. 2-17, Fig. 2-18, Fig. 2-20 are corrected | |
| In figures Fig. 2-11 (phase 1), Fig. 2-17 (phase 6), Fig. 2-20 (phase 2) images of Mo-T, DC-T and MoDC-T are added. Thus it is meant that all of them are PG-Y(-). | |

**Tab. 2-4 (continuation)**

| What | Where |
|---|---|
| Name SPP is changed:<br>Before:<br>«Increased kPAMP-carriage of R-phagocytes.»<br>Now:<br>«Increased kPAMP-carriage of tolerized phagocytes.<br>Increased (PG-Y)-carriage of R-phagocytes».<br>It is made in connection with change of the formulation of subprocess SP8. | Part 1, SPP<br>Fig. 2-8,<br>Fig. 2-9,<br>Fig. 2-18 |
| Name of LP1.1 is changed.<br>Before:<br>«Attraction of Mo, Mo-R, DC, DC-R from blood.»<br>Now:<br>«Attraction of Mo and DC, Mo-T and DC-T (incl. Mo-R and DC-R) from blood.»<br>It is made in connection with specification of concept of R-phagocytes.<br>The description of role Mo-T and DC-T in psoriatic inflammation is added. | LP1.1,<br>Fig. 2-8,<br>Fig. 2-9,<br>Fig. 2-18 |
| Name of LP6.1 is changed.<br>Before:<br>"Loss of tolerance of Mo-R, DC-R and MoDC-R to kPAMP."<br>Now:<br>"Loss of tolerance to kPAMP."<br>It is made in connection with addition of Mo-T, DC-T and MoDC-T in process LP6.<br>Tab. 2-3 and Fig. 2-7 are corrected and added. | LP6 |

# Figures

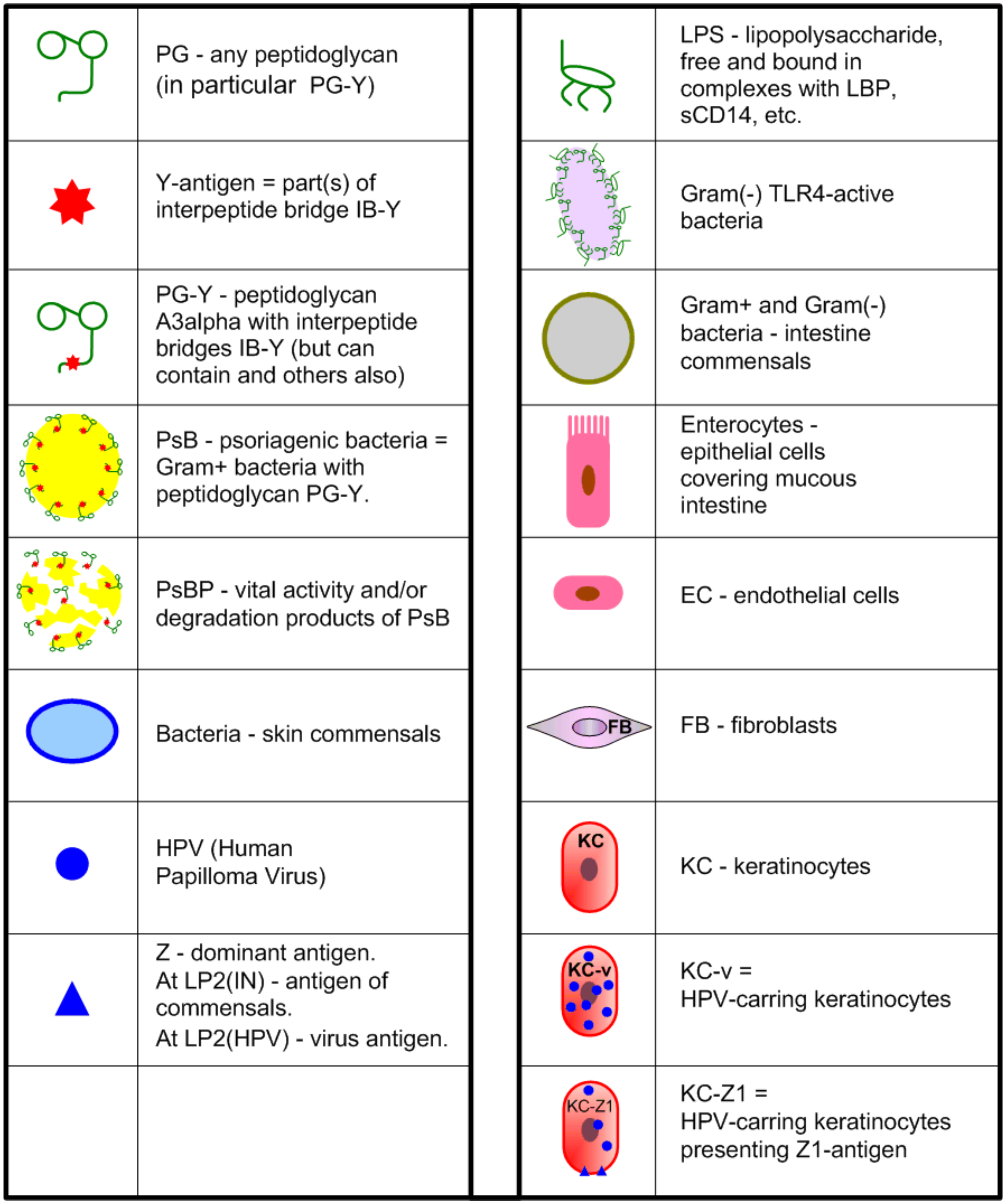

**Fig. 2-1. Bacteria, bacterial products, viruses and tissue cells (symbols).**
(Part1, fig.4). Appendix 2-1.

| | | | |
|---|---|---|---|
| 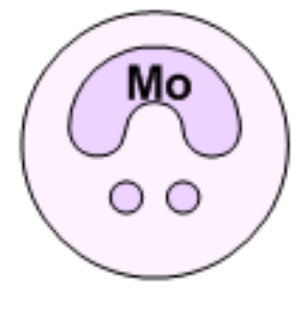 | Mo - monocytes | 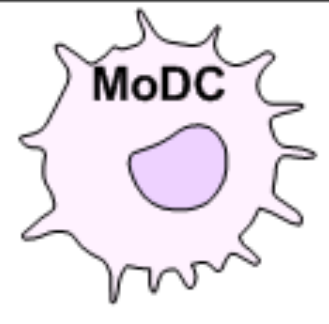 | MoDC - dendritic cells, derived from Mo |
| 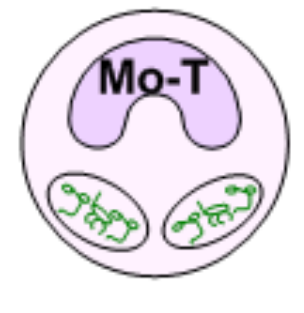 | Mo-T -tolerized monocytes | 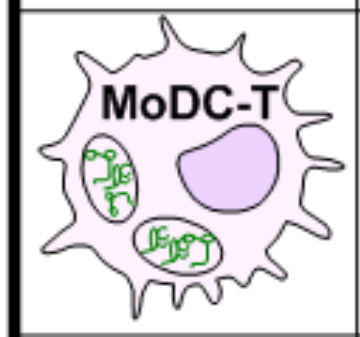 | MoDC-T - dendritic cells, derived from Mo-T |
| 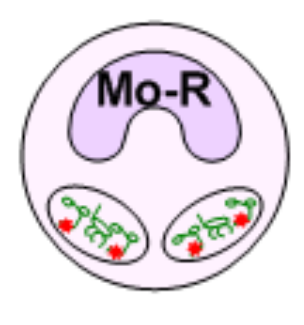 | Mo-R – reprogrammed (tolerized) and repleted by PG-Y monocytes | 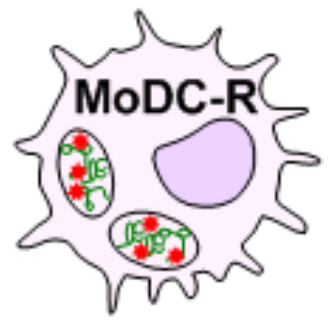 | MoDC-R - dendritic cells, derived from Mo-R |
| 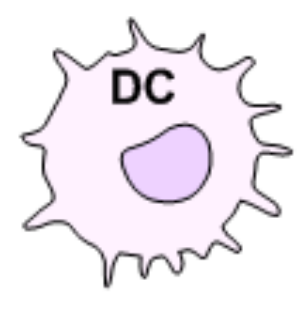 | DC – dendritic cells | 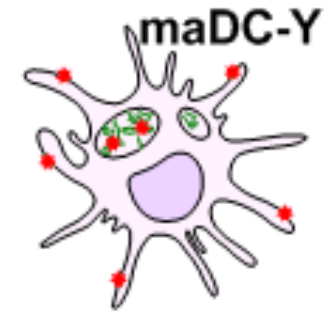 | maDC-Y = mature dendritic cells, presenting Y-antigen |
| 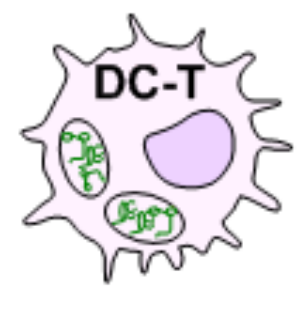 | DC-T – tolerized dendritic cells | 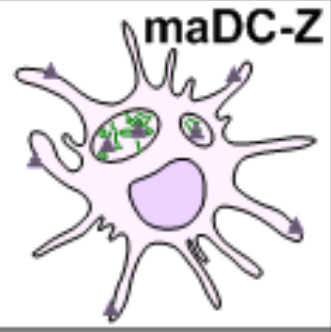 | maDC-Z = mature dendritic cells, presenting Z-antigen |
| 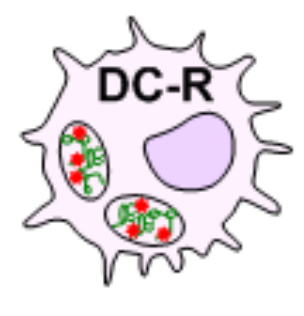 | DC-R – reprogrammed (tolerized) and repleted by PG-Y dendritic cells | 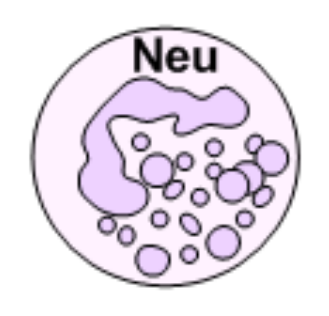 | Neu - neutrophils |
| 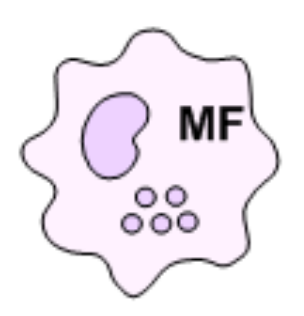 | MF - macrophages, derived from Mo | 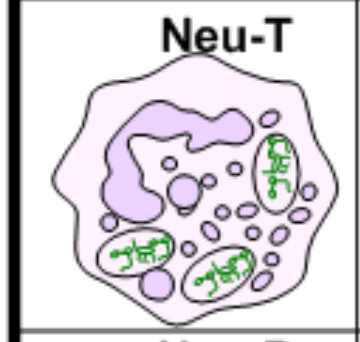 | Neu-T – tolerized Neu |
| 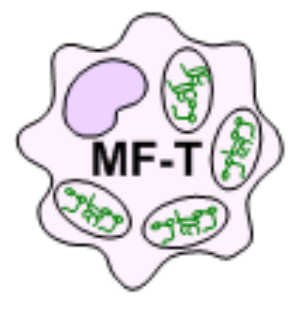 | MF-T - macrophages, derived from Mo-T | 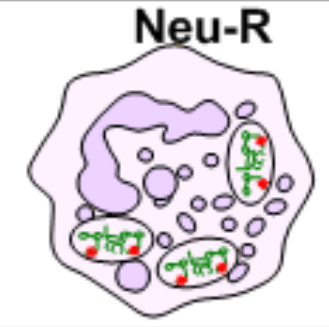 | Neu-R – reprogrammed (tolerized) and repleted by PG-Y neutrophils |
| 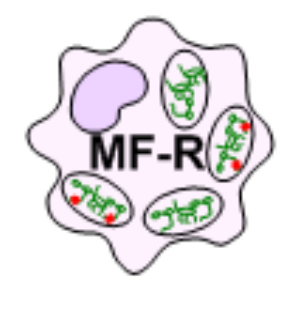 | MF-R - macrophages, derived from Mo-R | 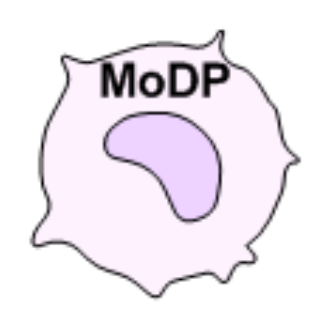 | MoDP - CD34+ cells - precursors of monocytes and immature dendritic cells of bone marrow |
| 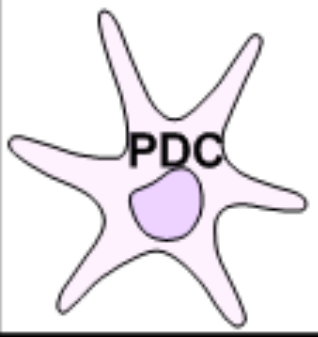 | PDC – plasmacytoid dendritic cells | 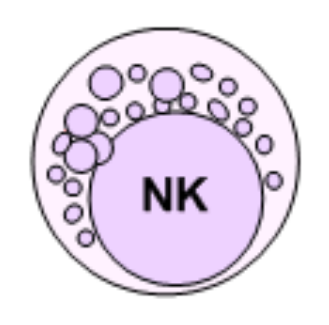 | NK – natural killers |

**Fig. 2-2. Immune cells (symbols).**
(Part1, fig.5). Appendix 2-1.

| | | | |
|---|---|---|---|
| 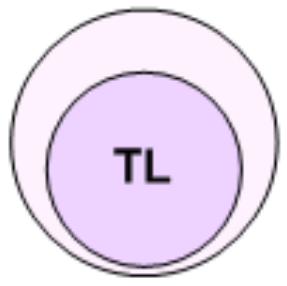 | TL = any T-lymphocytes | 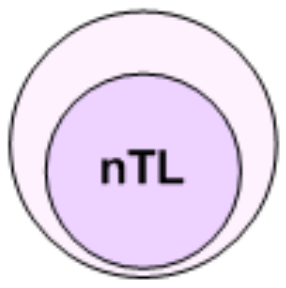 | nTL = naive T-lymphocytes |
| 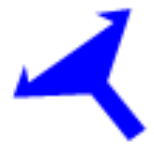 | TCR-receptor of TL-Z | 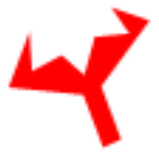 | TCR-receptor of TL-Y |
| 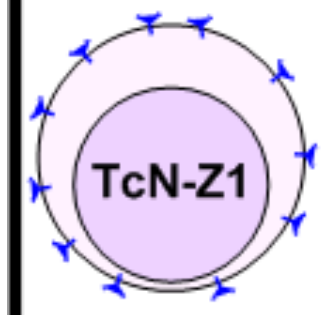 | TcN-Z1 = Z1-specific TcN | | |
| 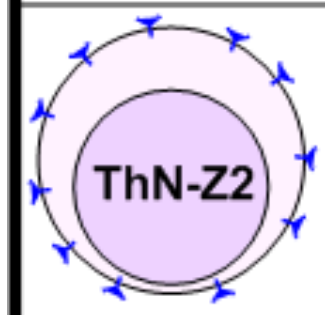 | ThN-Z2 = Z2-specific ThN | | |
| 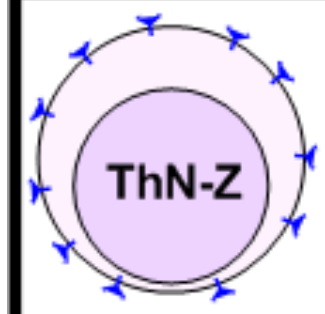 | ThN-Z =Z-specific ThN | 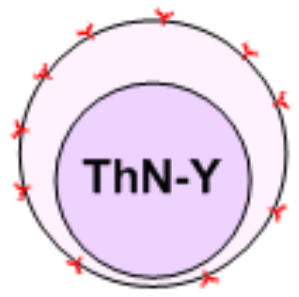 | ThN-Y = Y-specific ThN |
| 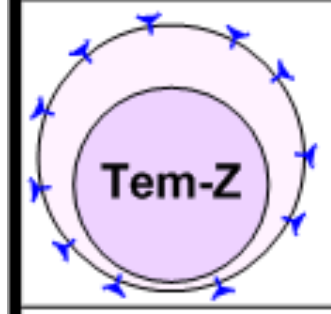 | Tem-Z = Z-specific Tem (mainly TcN-Z1 and ThN-Z2) | 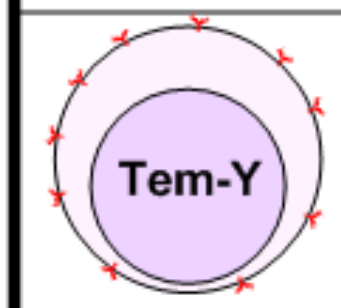 | Tem-Y = Y-specific Tem |
| 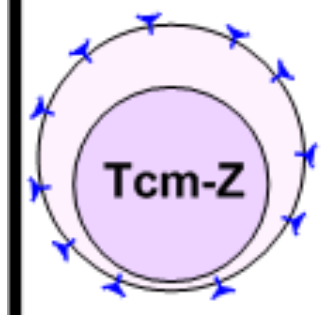 | Tcm-Z = Z1-specific and Z2-specific Tcm | 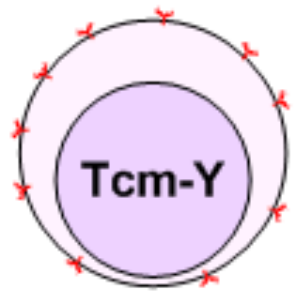 | Tcm-Y = Y-specific Tcm |
| 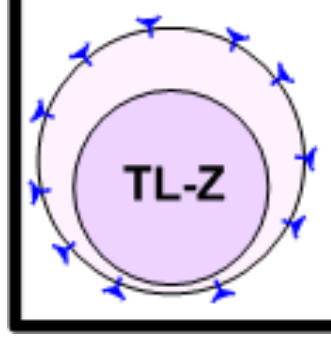 | TL-Z = Tem-Z and Tcm-Z | 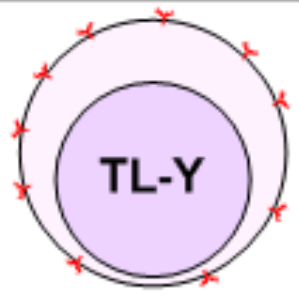 | TL-Y = Tem-Y and Tcm-Y |

**Fig. 2-3. T-lymphocytes (symbols).**

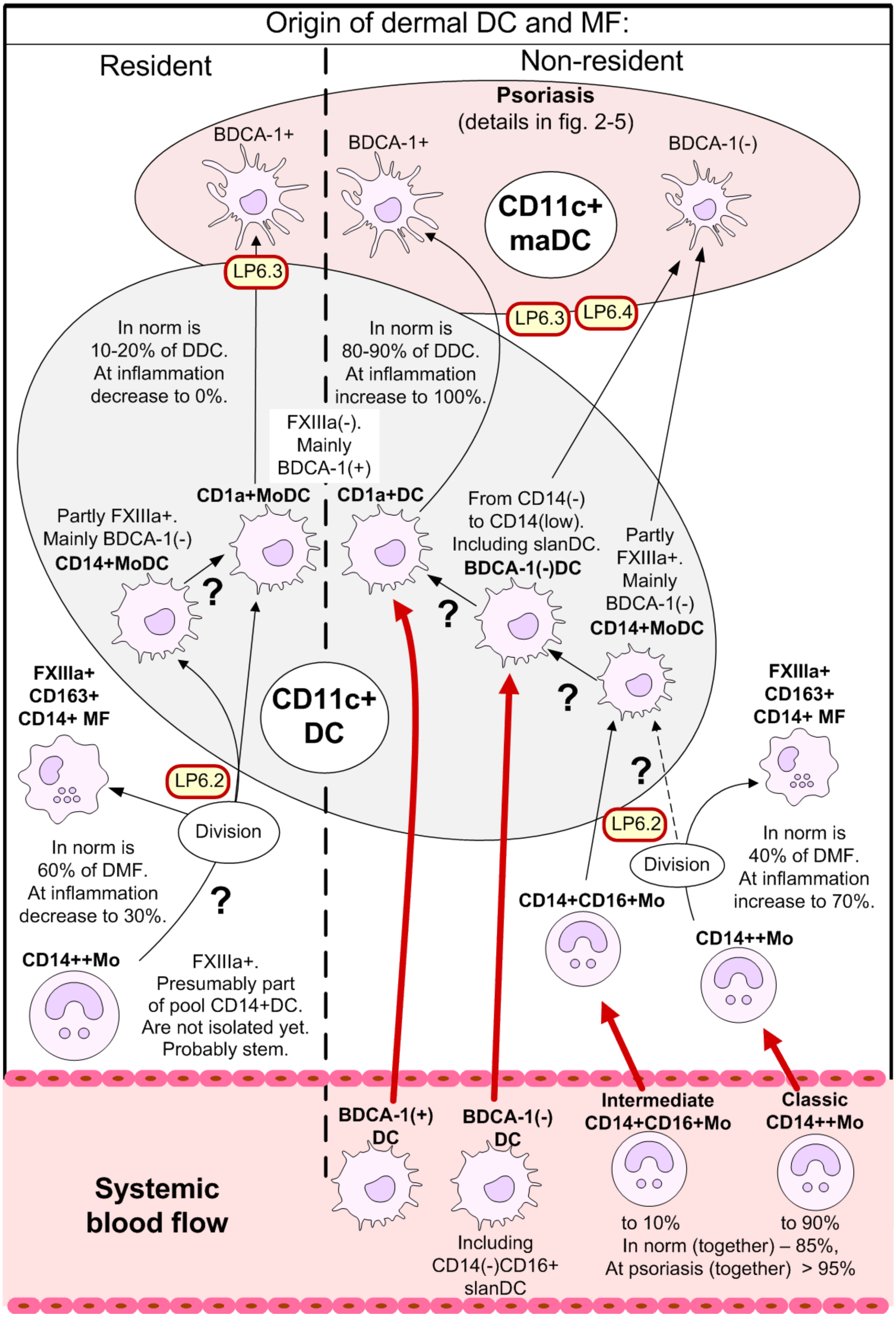

**Fig. 2-4. Origin of dermal DC and MF.**

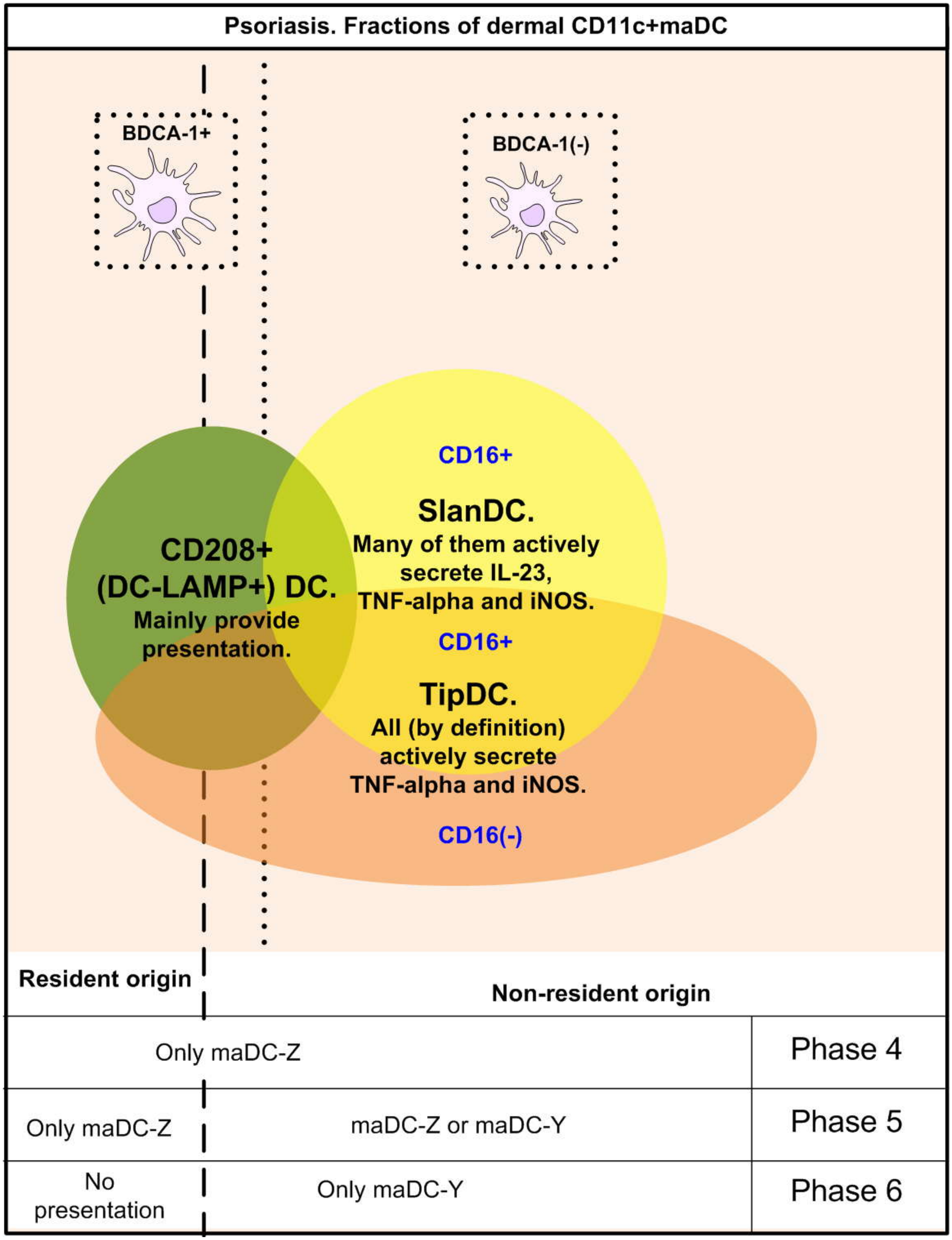

**Fig. 2-5. Fractions CD11c+maDC in PLS-derma.**

CD208+(DC-LAMP+)DC - green oval, SlanDC - yellow oval, TipDC - brown oval.
A part of maDC belongs to two or even three fractions at once.

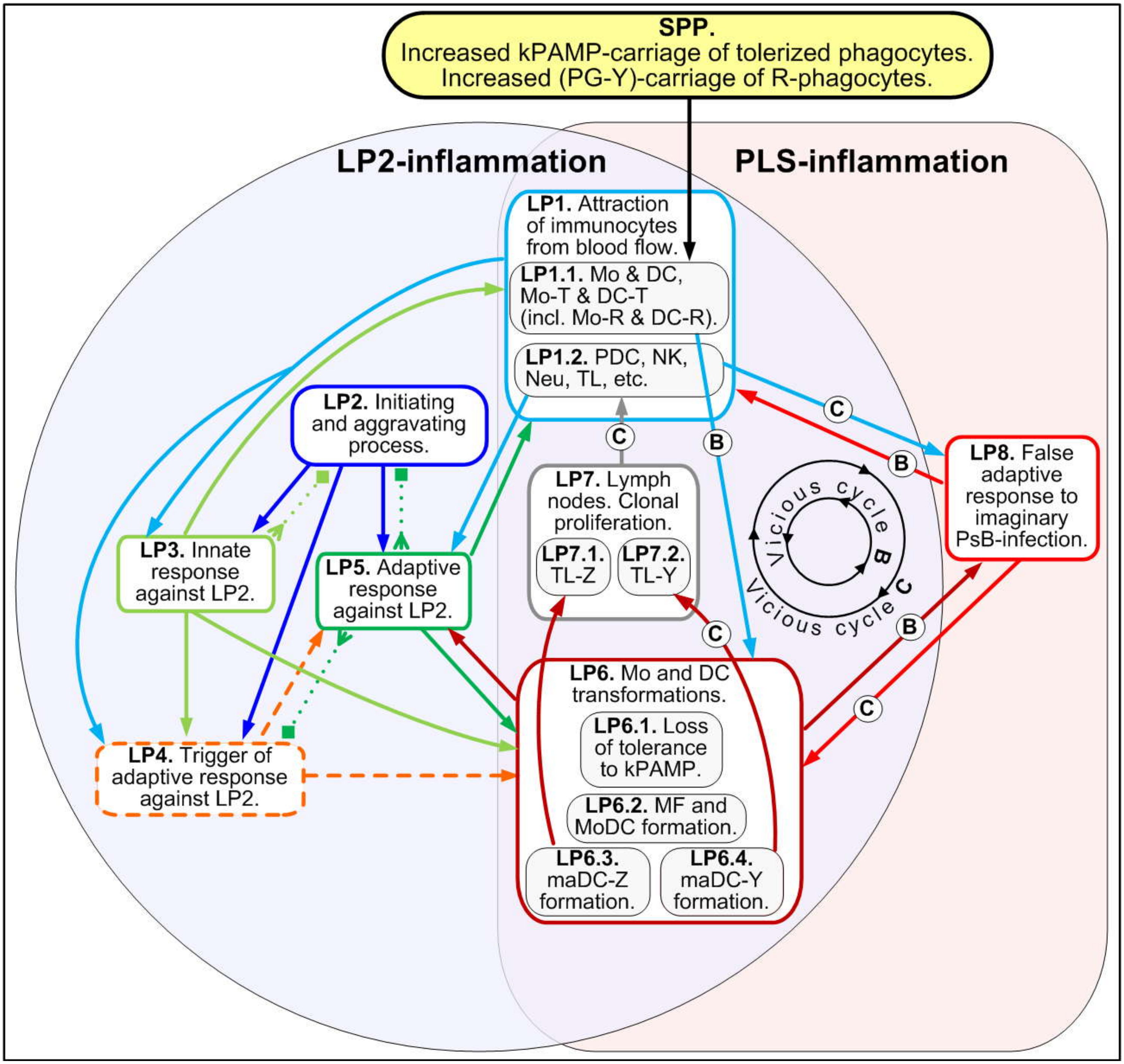

**Fig. 2-6. Y-model of pathogenesis of psoriasis (LP2 is not specified). Interaction of local processes.** Dashed lines – transit process LP4 and influences connected to it. Dotted arrows with small squares - suppression. Letters B and C - vicious cycles.

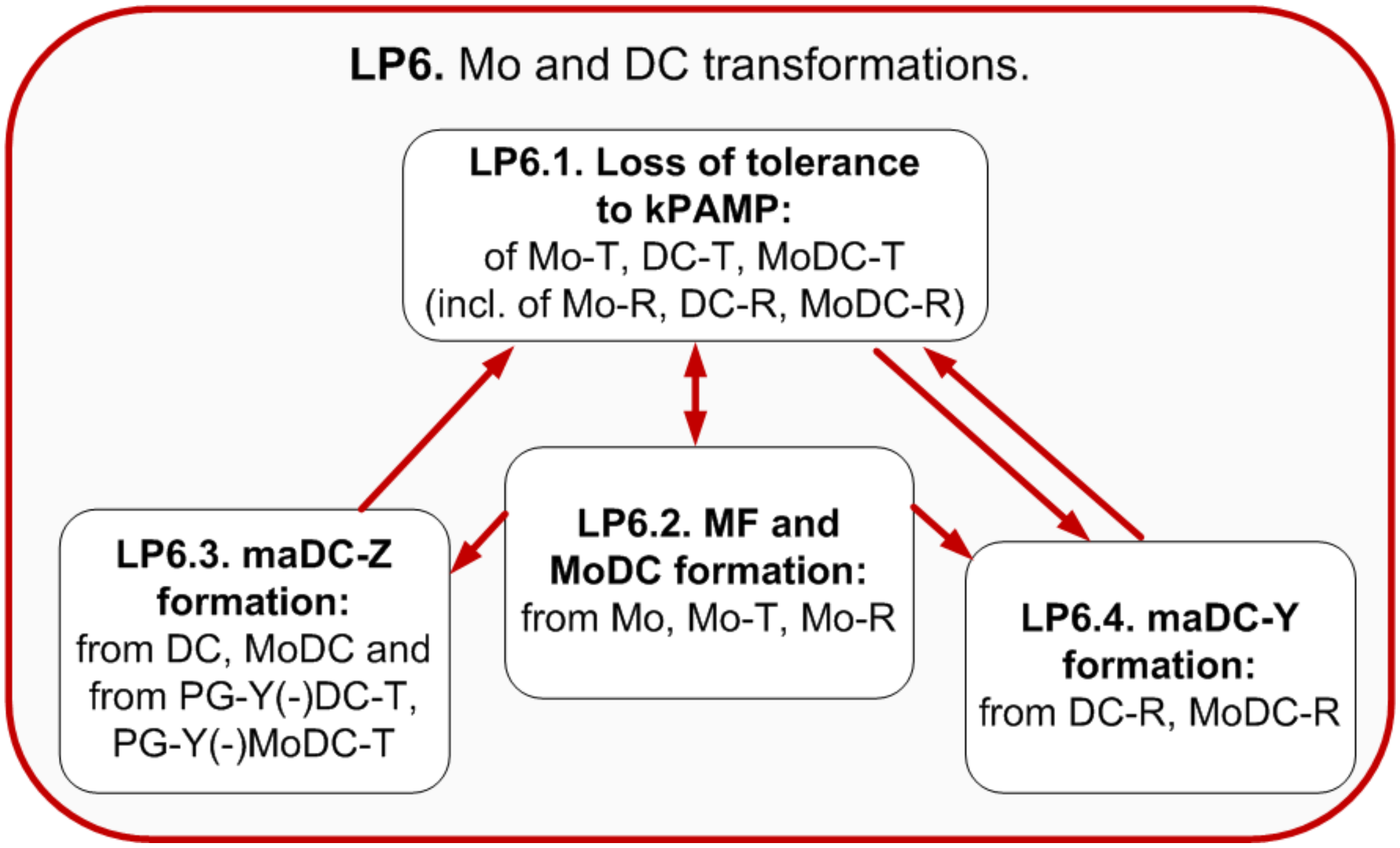

**Fig. 2-7. Interdependence of subprocesses of Mo and DC transformations.**

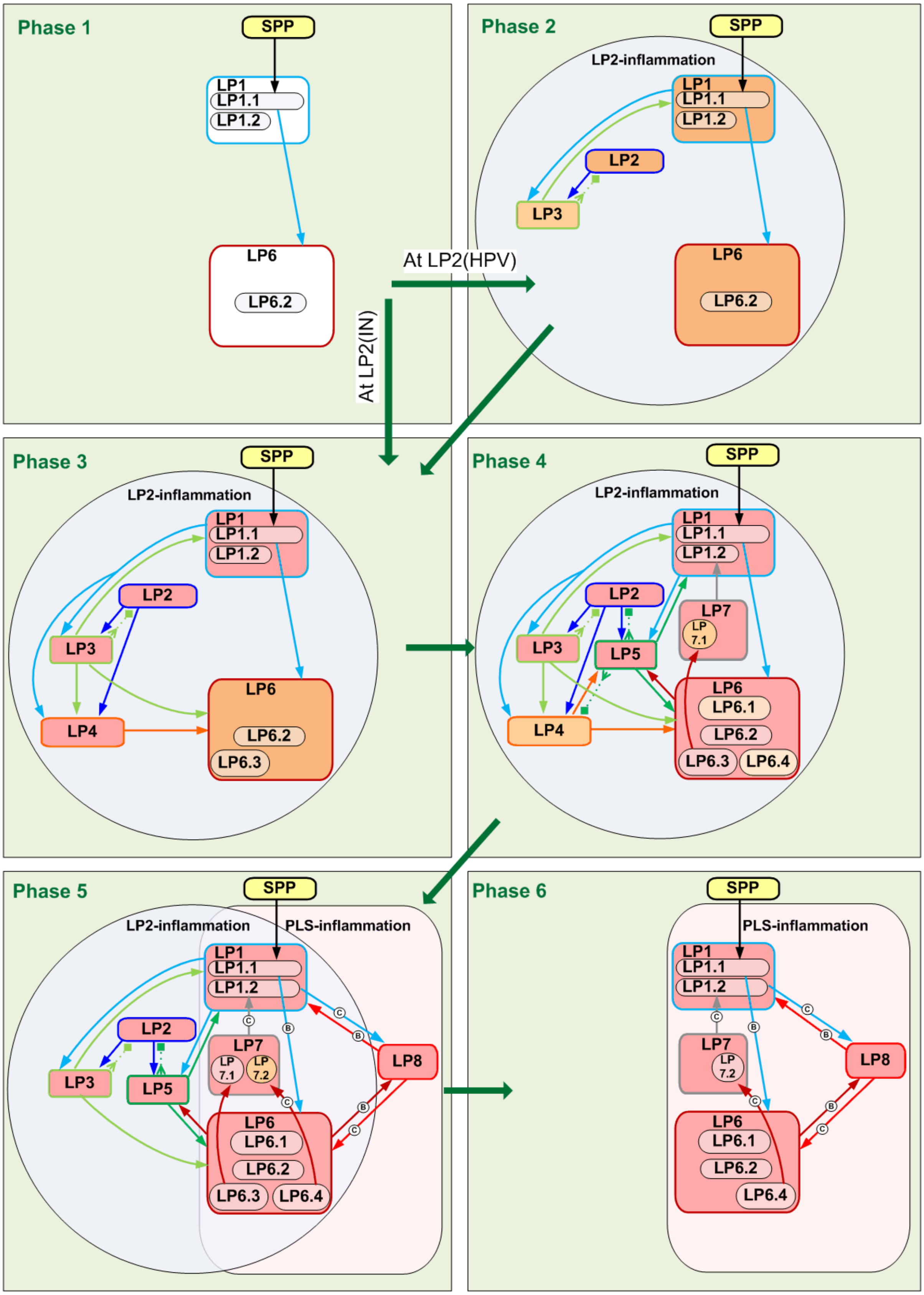

**Fig. 2-8. Y-model of pathogenesis of psoriasis (LP2 is not specified). Phases of psoriatic plaque development.**
Dotted arrows – suppression. Letters B and C - vicious cycles. White color - process occurs with weak intensity; beige - inflammatory, average intensity; pink - inflammatory, high intensity.

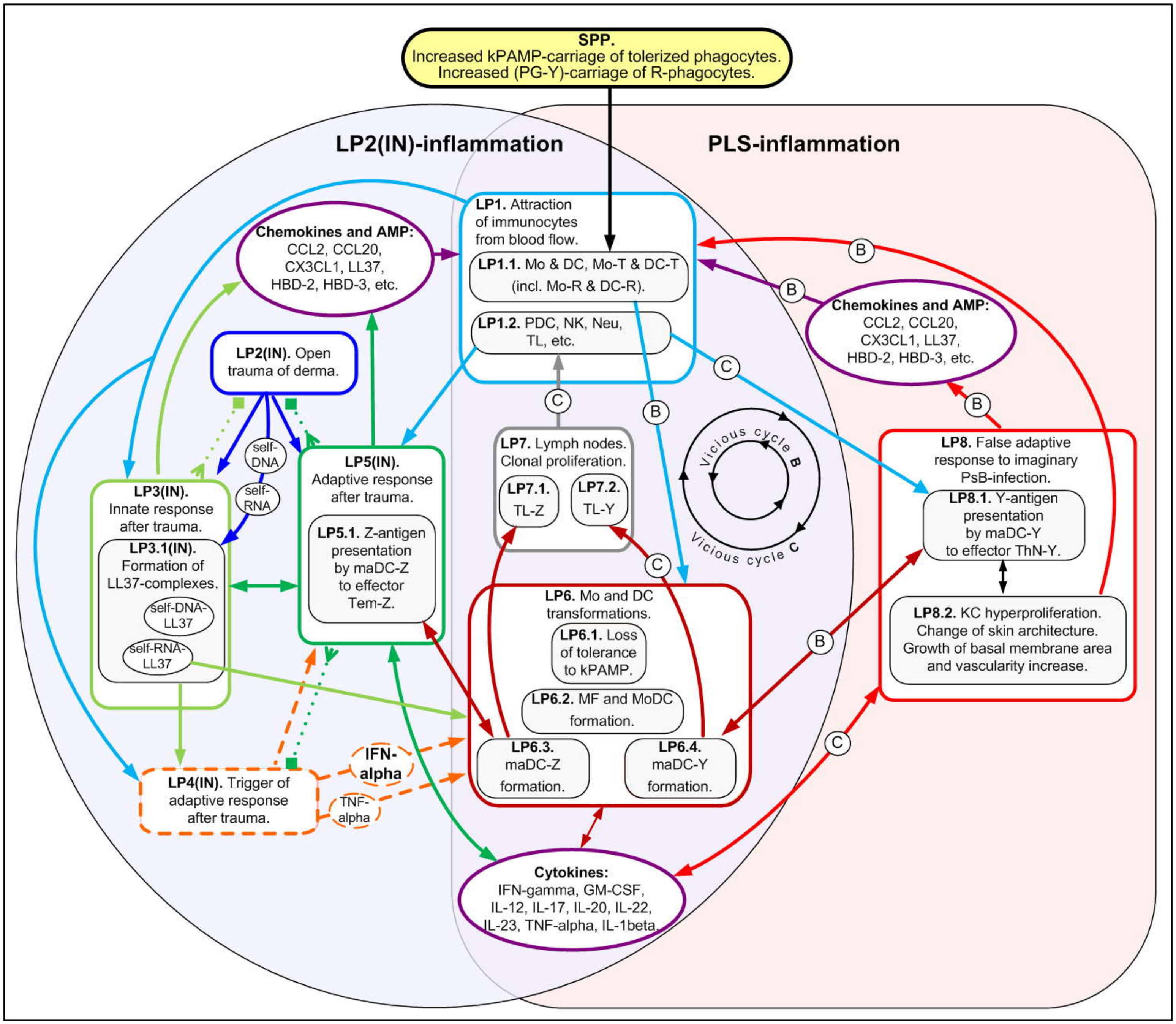

**Fig. 2-9. Y-model of pathogenesis of psoriasis at LP2(IN) - open trauma of derma.**
Dashed lines - suppression. On LP2 end (phase 6) only the processes framed by the pink rectangle "PLS-inflammation" remain.

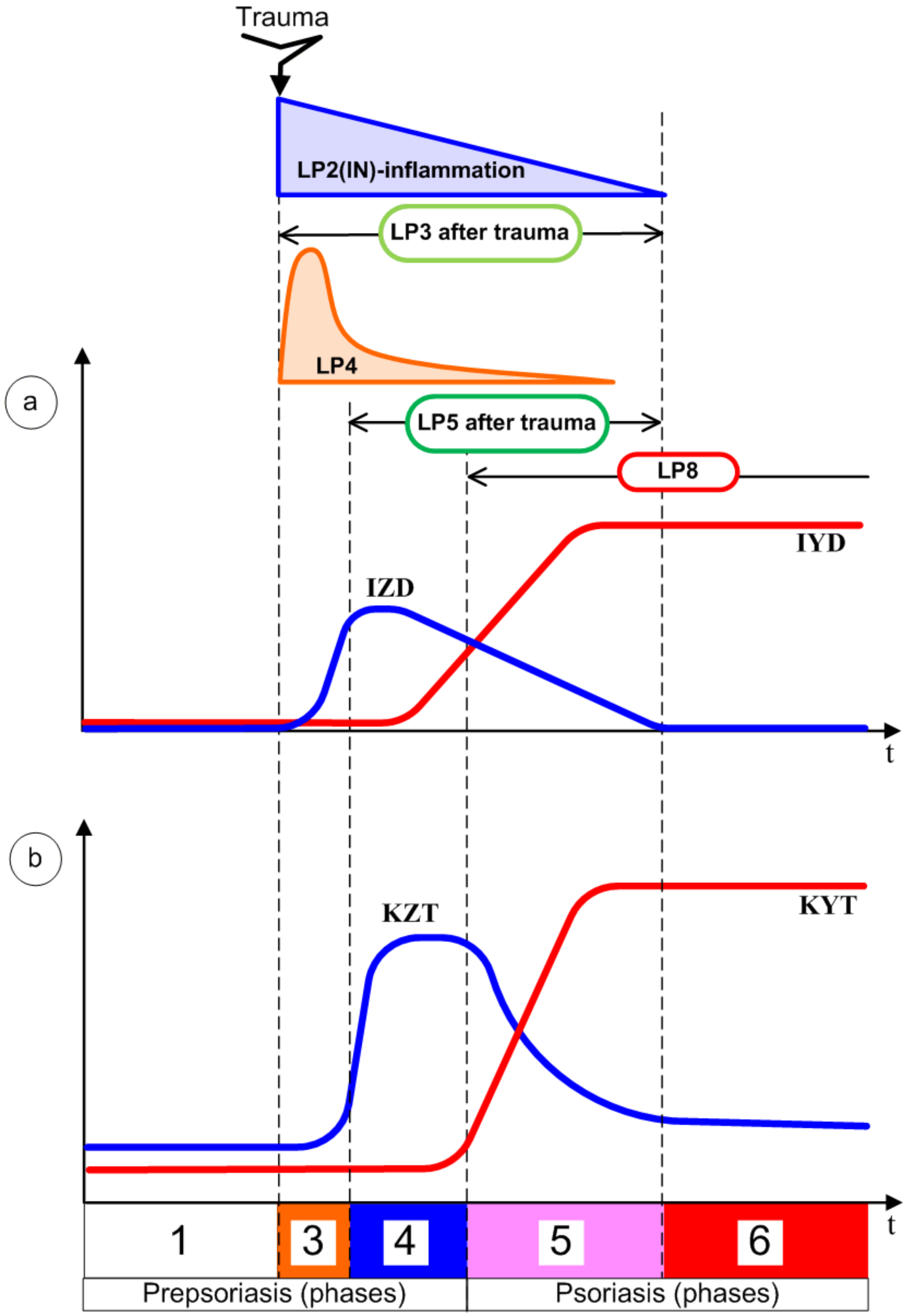

**Fig. 2-10. Phases of psoriatic plaque development at LP2(IN) and previous Y-priming. Conditional graphs.**

LP4 - adaptive response trigger; LP3 - innate response, LP5 - adaptive response against dominant Z-antigen of commensals, LP8 - false adaptive response against Y-antigen.

a) Unit antigenic presentation of maDC (the quantity of antigen presented by maDC in the unit of volume of derma): IYD - defined by Y-antigen (presented by maDC-Y), IZD - defined by Z-antigen (presented by maDC-Z).

b) Logarithmic curves of unit quantity of effector ThN in the unit of volume of derma:
KYT = ln (unit of quantity of ThN-Y), KZT = ln (unit of quantity of ThN-Z).

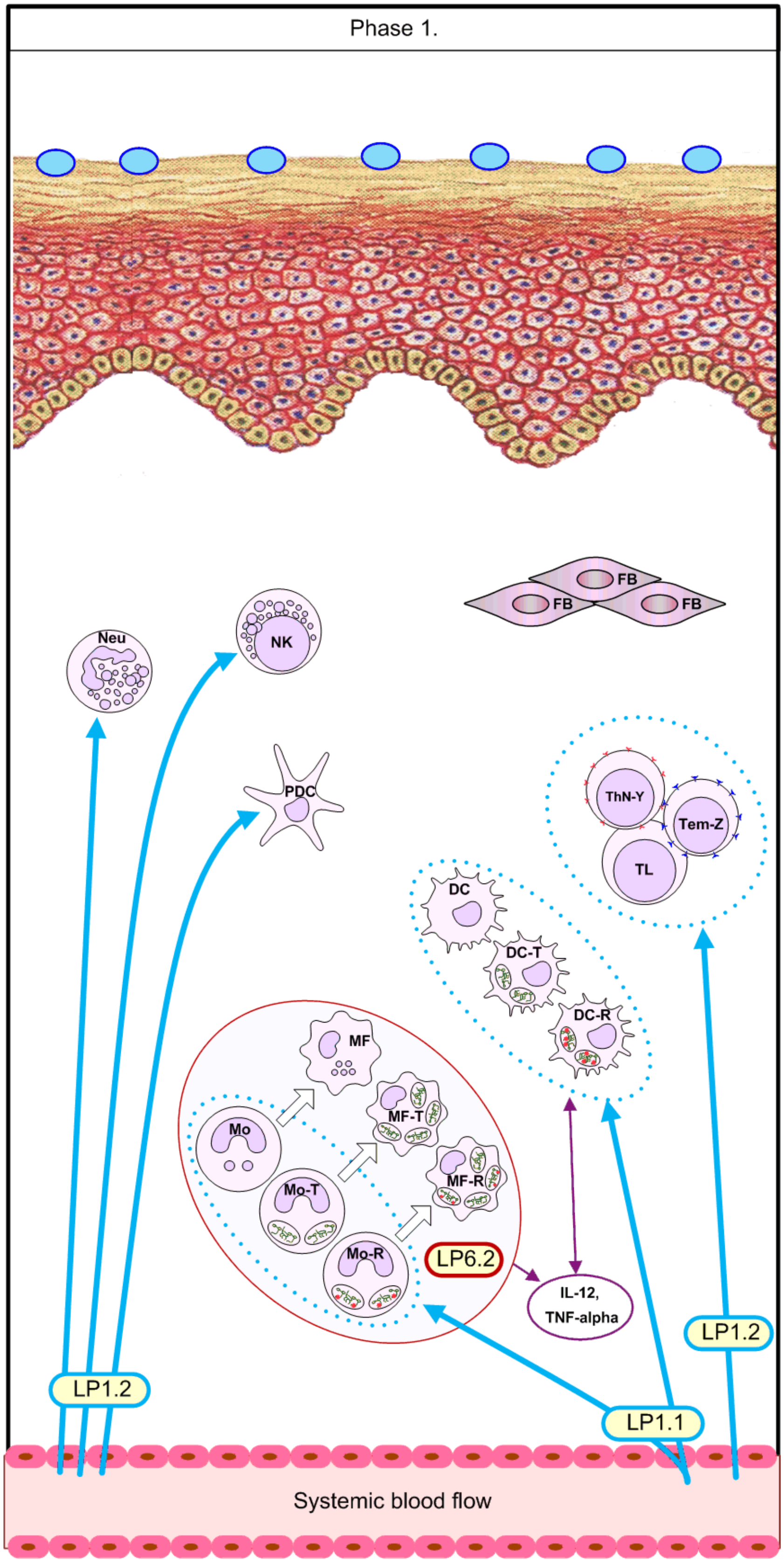

**Fig. 2-11. Prepsoriasis. Phase 1, common for all LP2 (as LP2 has not started yet).**
Here (as anywere): Mo-R = PG-Y(+)Mo-T; MF-R = PG-Y(+)MF-T ; DC-R = PG-Y(+)DC-T;

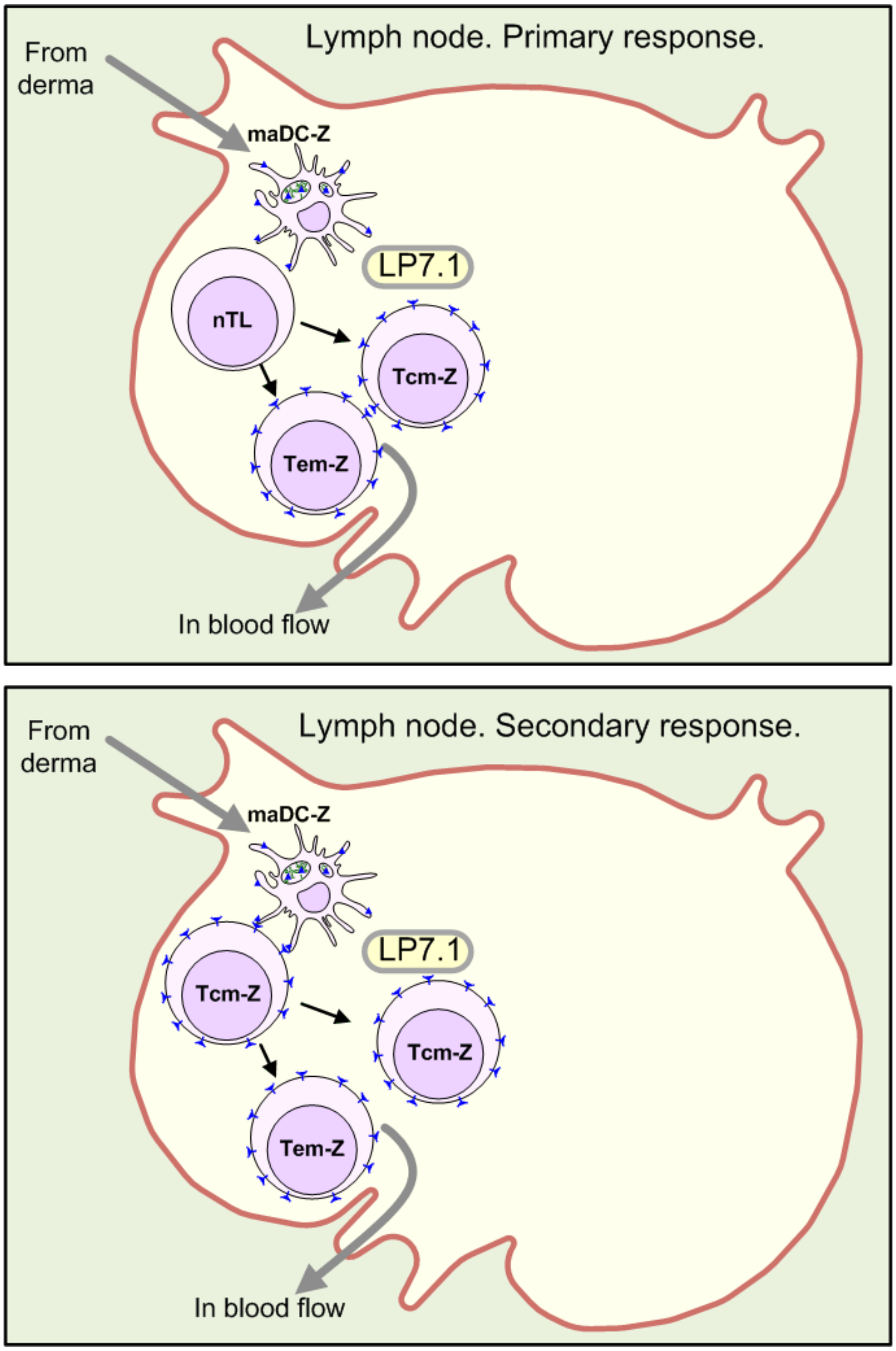

**Fig. 2-12. Process LP7.1. Lymph node. Primary and secondary responses. Formation of TL-Z, i.e. Tem-Z and Tcm-Z.**

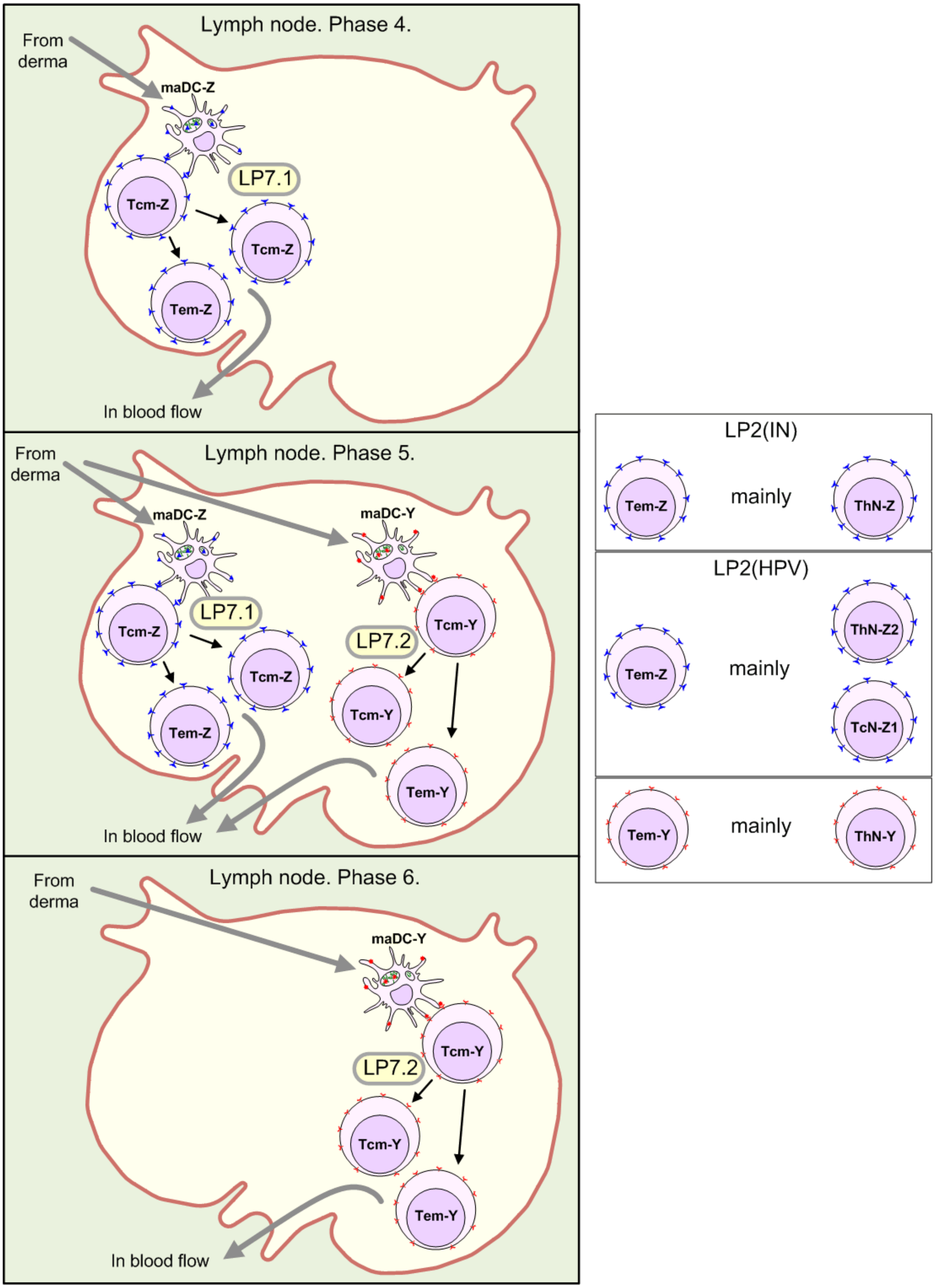

**Fig. 2-13. Process LP7. Lymph node. Clonal proliferation. Phases 4, 5 and 6.**

It is supposed that priming by Z-antigens and Y-antigens took place, i.e. secondary response developed both to Z-antigen and Y-antigen. Tem-Z are quickly formed from Tcm-Z, and Tem-Y are quickly formed from Tcm-Y. At LP2(IN) formed Tem-Z consist basically from ThN-Z. At LP2(HPV), formed Tem-Z consist basically from ThN-Z2 and TcN-Z1. Tem-Y consist basically from ThN-Y.

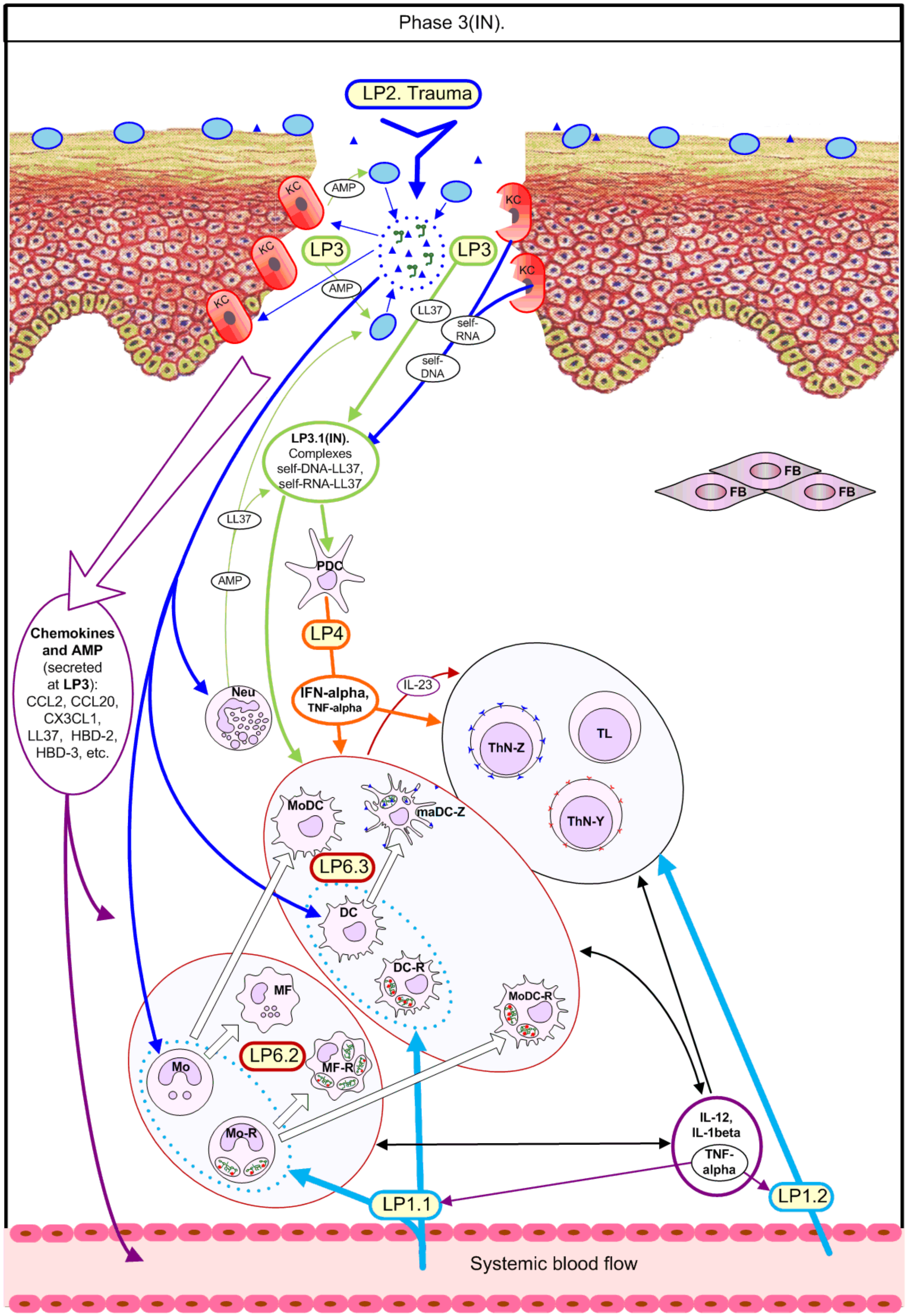

**Fig. 2-14. Prepsoriasis. Phase 3 at LP2(IN) - open trauma of derma.**
Possible transformations of PG-Y(-)Mo-T and PG-Y(-)DC-T are identical to transformations of Mo and DC, therefore their images are absent.

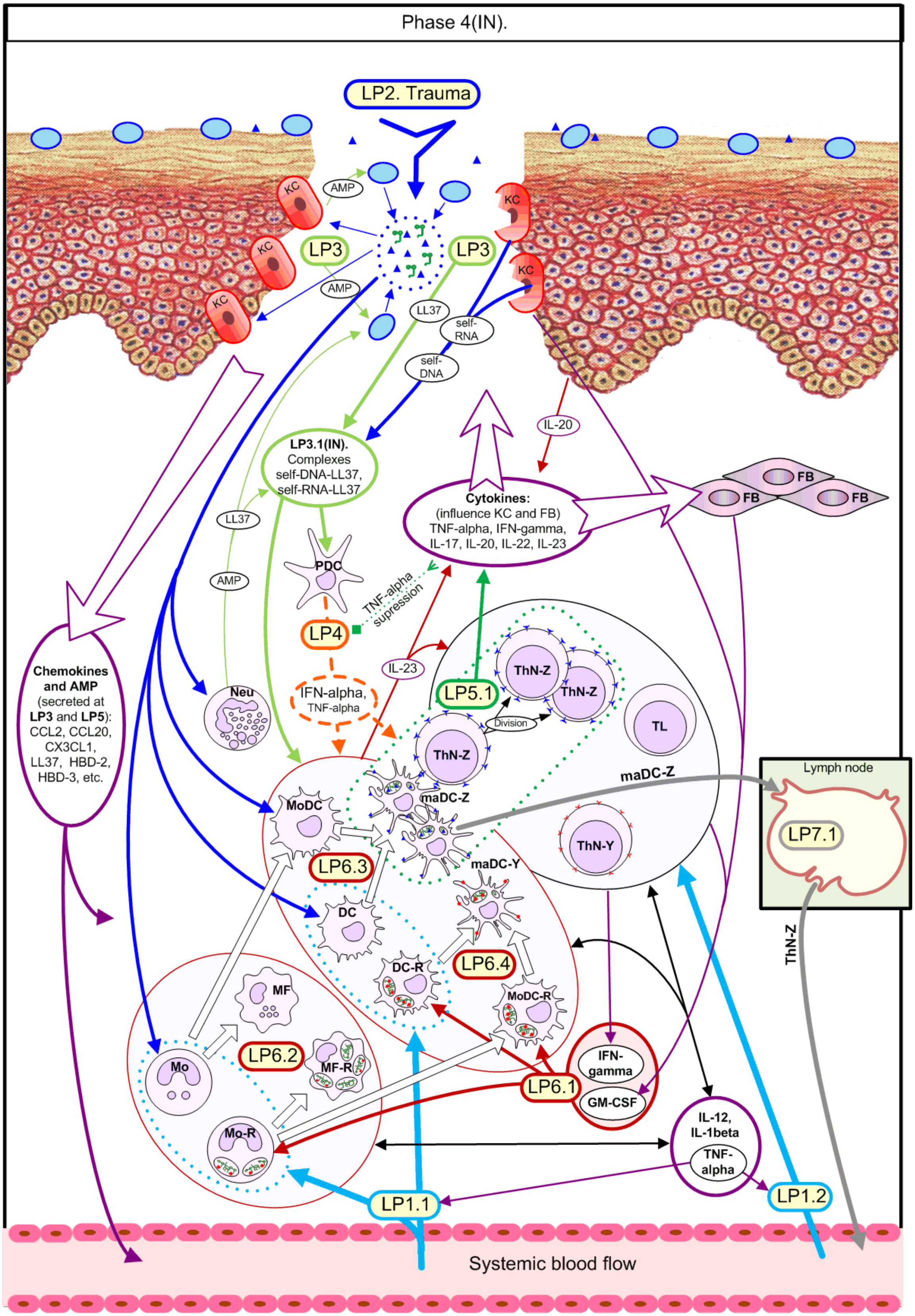

**Fig. 2-15. Prepsoriasis. Phase 4 at LP2(IN) - open trauma of derma.**

Possible transformations of PG-Y(-)Mo-T and PG-Y(-)DC-T are identical to transformations of Mo and DC, therefore their images are absent.

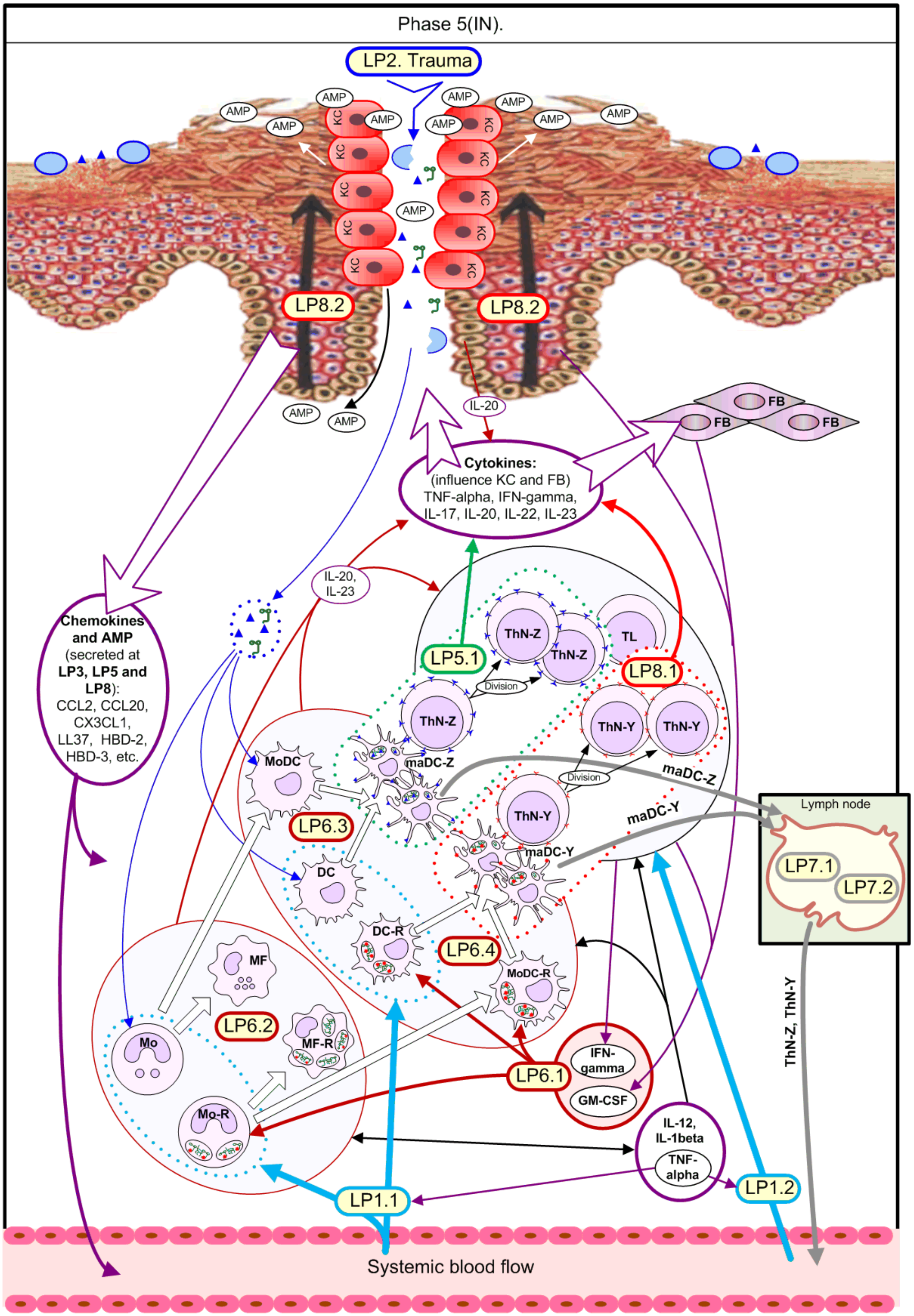

**Fig. 2-16. Psoriasis. Phase 5 at LP2(IN) - open trauma of derma. Trauma healing.**

Plaque initialization occurs after trauma during LP3 - innate and LP5 - adaptive responses against commensals (Koebner effect). Possible transformations of PG-Y(-)Mo-T and PG-Y(-)DC-T are identical to transformations of Mo and DC, therefore their images are absent.

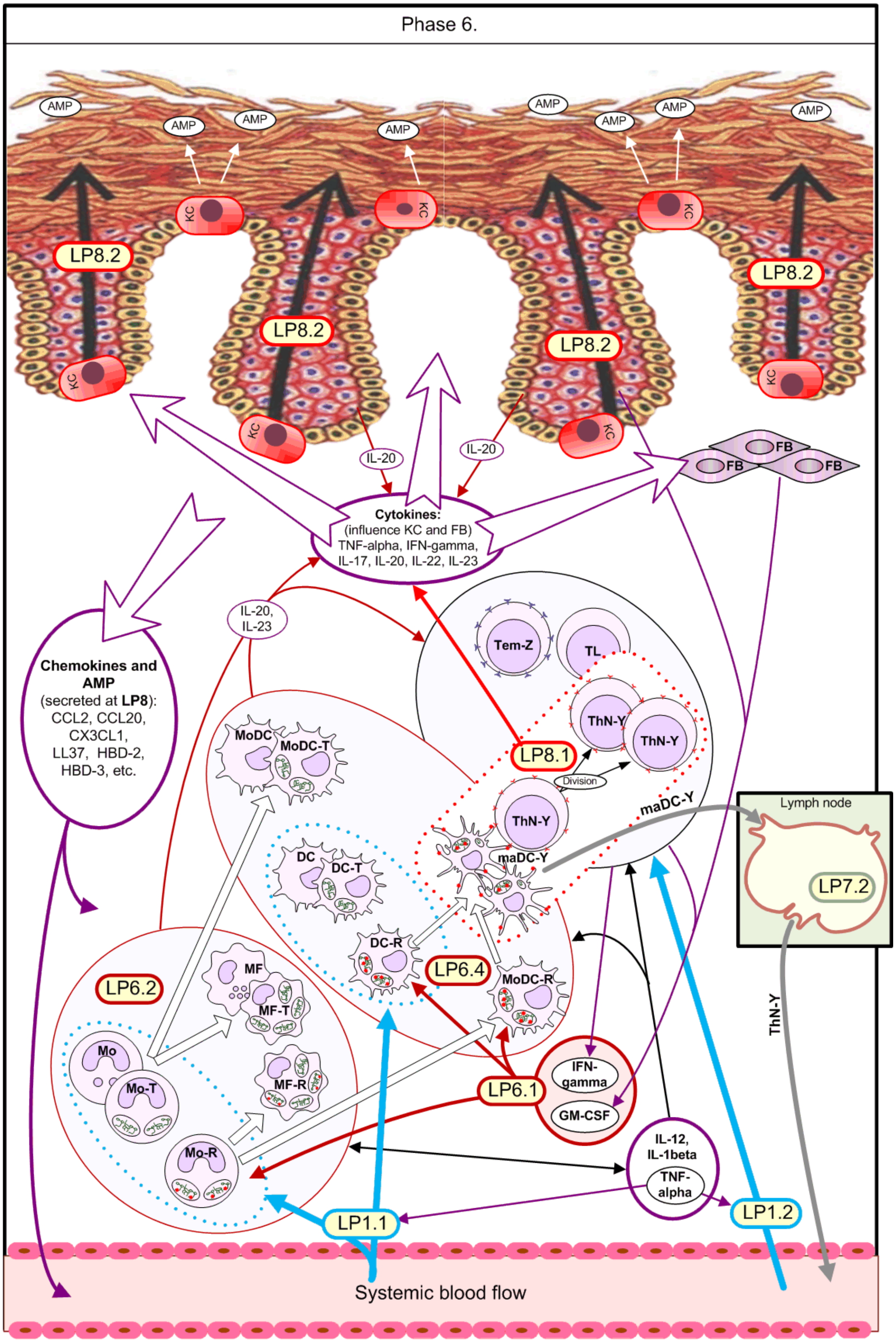

**Fig. 2-17. Psoriasis. Phase 6, common for all LP2 (as LP2 has ended).**
The basic process is self-sufficient LP8.

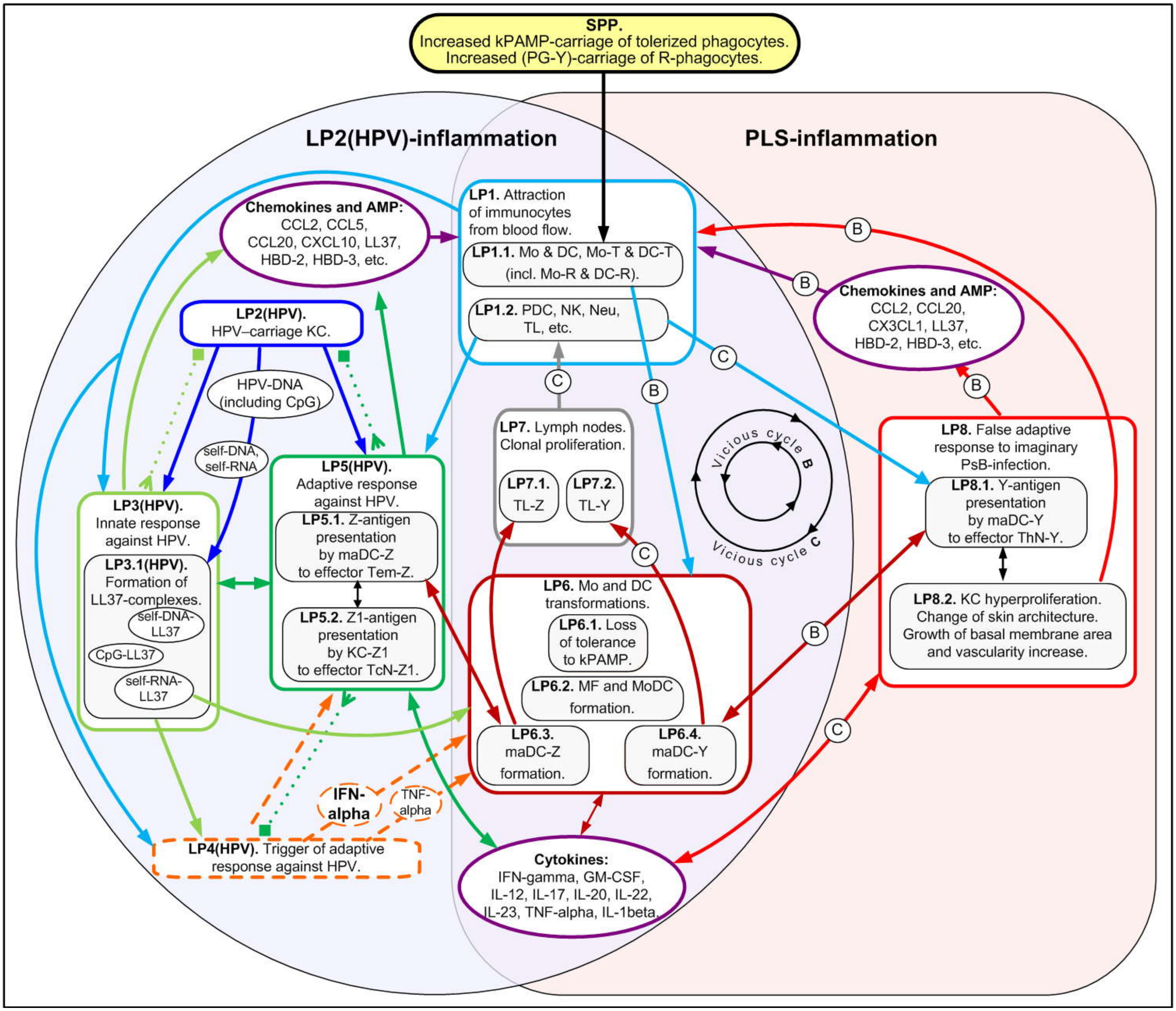

**Fig. 2-18. Y-model of pathogenesis of psoriasis at LP2(HPV) - HPV-carriage of KC.**
Dashed lines - transit process LP4 and influences connected to it.
Dotted arrows with small squares - suppression.

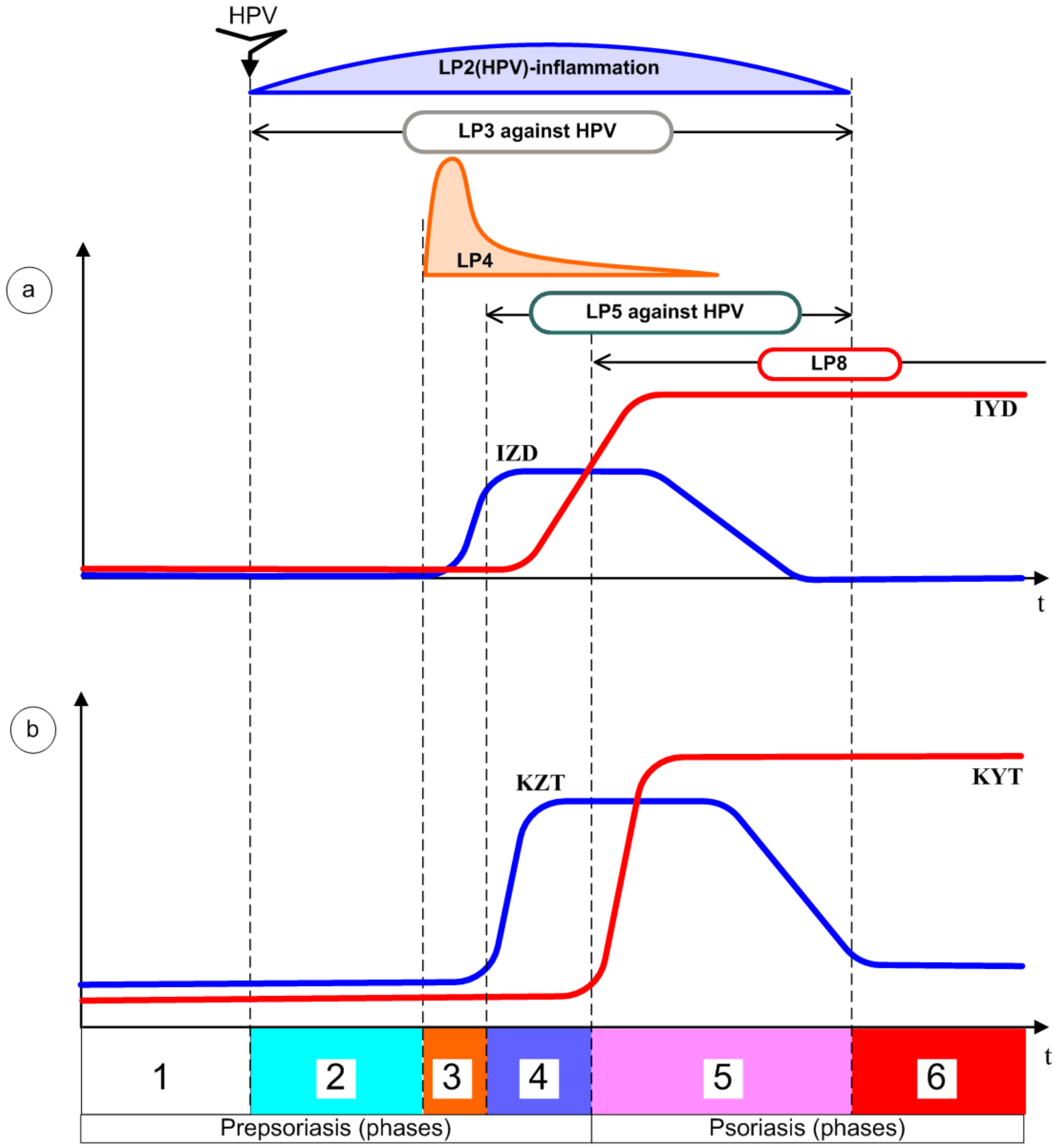

**Fig. 2-19. Phases of psoriatic plaque development at LP2(HPV) and previous Y-priming. Conditional graphs.**

LP4 - trigger of adaptive response; LP3 - innate response, LP5 - adaptive response against Z-antigen, LP8 - false adaptive response against Y-antigen.

a) Unit antigenic presentation by maDC (quantity of antigen presented by maDC in the unit of volume of derma):

IYD - defined by Y-antigen (presented by maDC-Y), IZD - defined by Z1-antigen (presented by maDC-Z1) and Z2-antigen (presented by maDC-Z2).

b) Logarithmic curves of the unit quantity of effector Tem in the unit of volume of derma:

KYT = ln(unit quantity of ThN-Y), KZT = ln (unit quantity of Tem-Z).

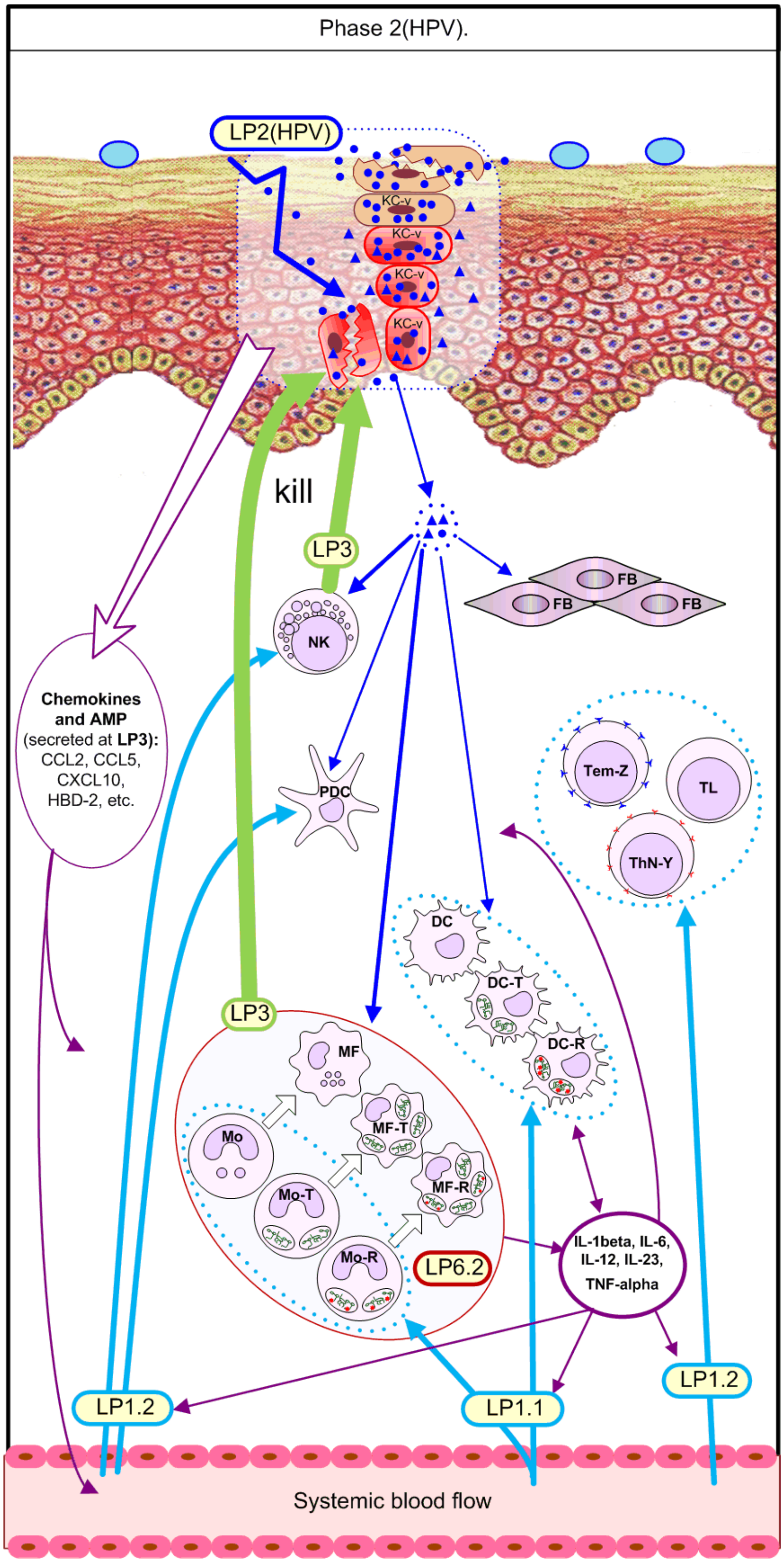

**Fig. 2-20. Prepsoriasis. Phase 2 at LP2(HPV) - HPV-carriage of KC.**
Here (as anywere): Mo-R = PG-Y(+)Mo-T; MF-R = PG-Y(+)MF-T ; DC-R = PG-Y(+)DC-T;

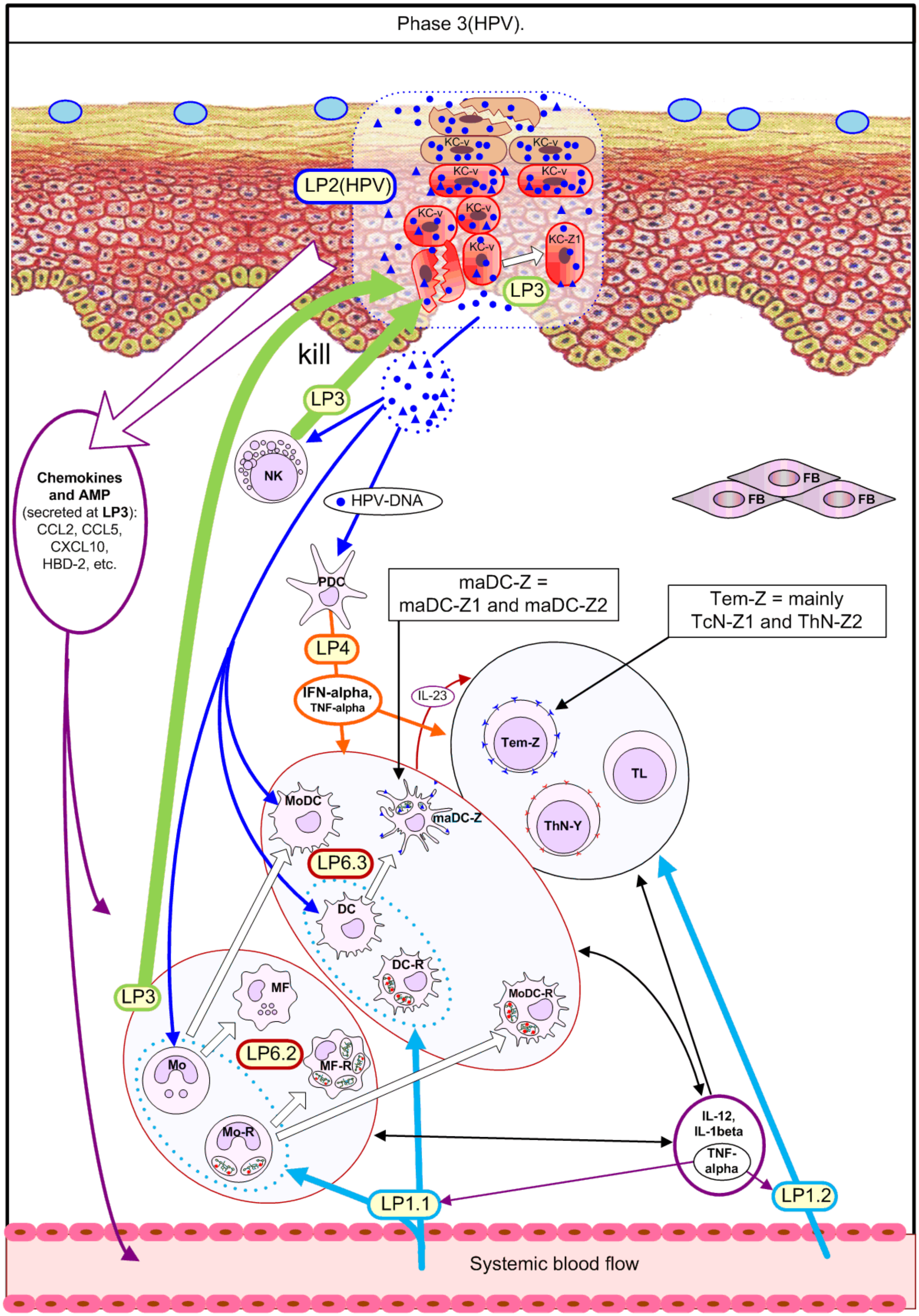

**Fig. 2-21. Prepsoriasis. Phase 3 at LP2(HPV) - HPV-carriage of KC.**
Possible transformations of PG-Y(-)Mo-T and PG-Y(-)DC-T are identical to transformations of Mo and DC, therefore their images are absent.

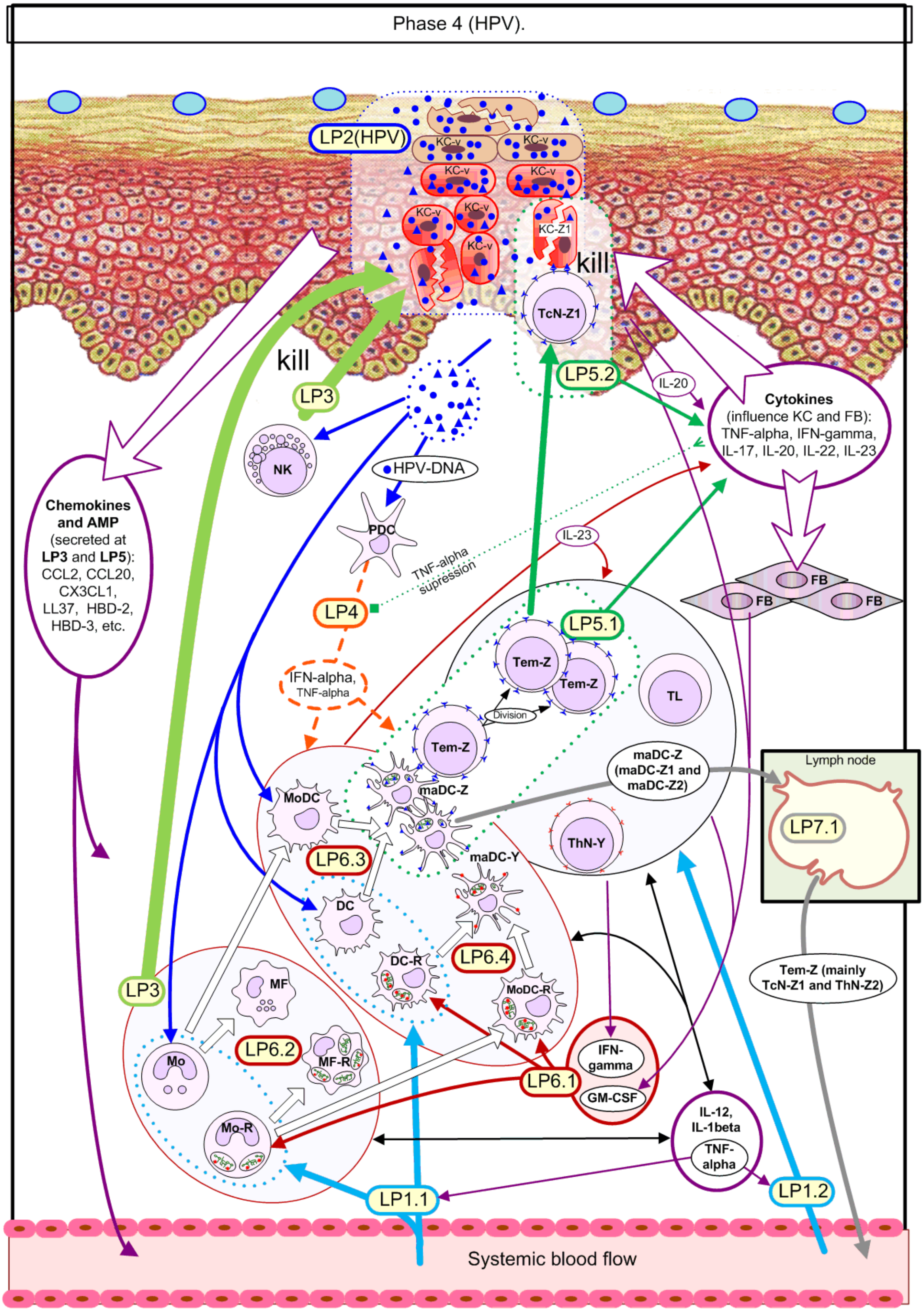

**Fig. 2-22. Prepsoriasis. Phase 4 at LP2(HPV) - HPV-carriage of KC.**
Possible transformations of PG-Y(-)Mo-T and PG-Y(-)DC-T are identical to transformations of Mo and DC, therefore their images are absent.

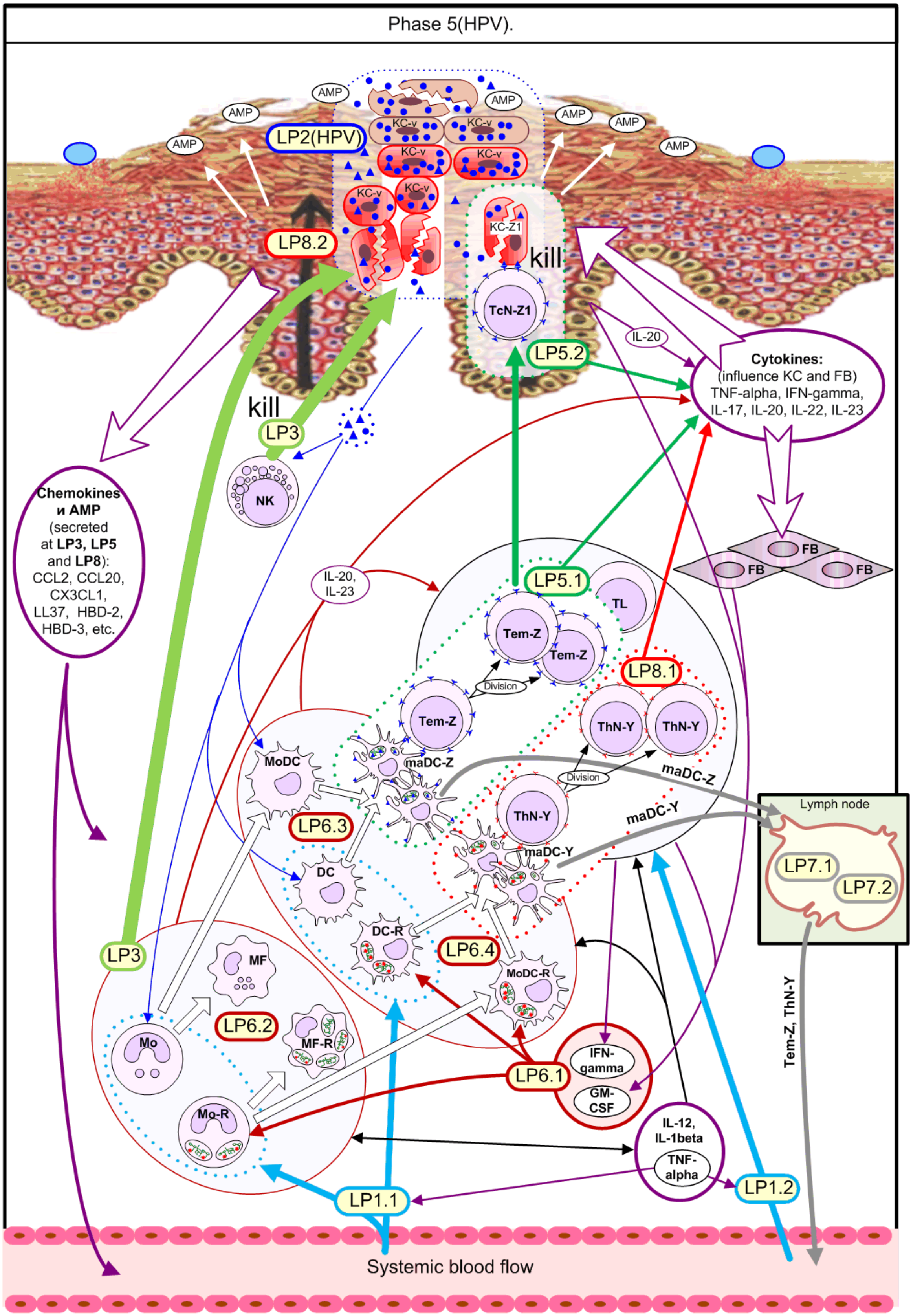

**Fig. 2-23. Psoriasis. Phase 5 at LP2(HPV) - HPV-carriage of KC.**

Possible transformations of PG-Y(-)Mo-T and PG-Y(-)DC-T are identical to transformations of Mo and DC, therefore their images are absent.

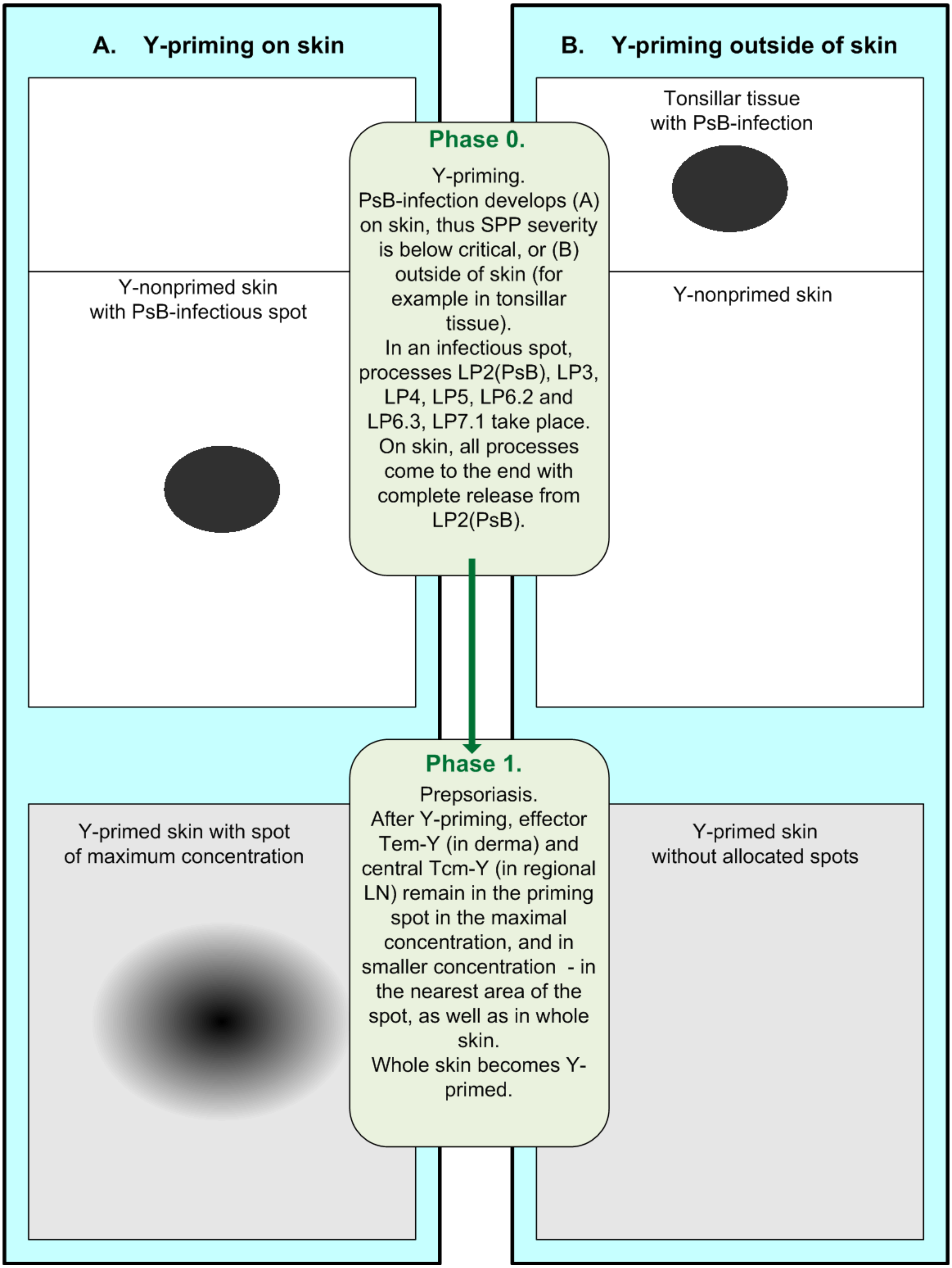

**Fig. 2-24. Y-priming of skin. Transition of whole skin from phase 0 into phase 1.**

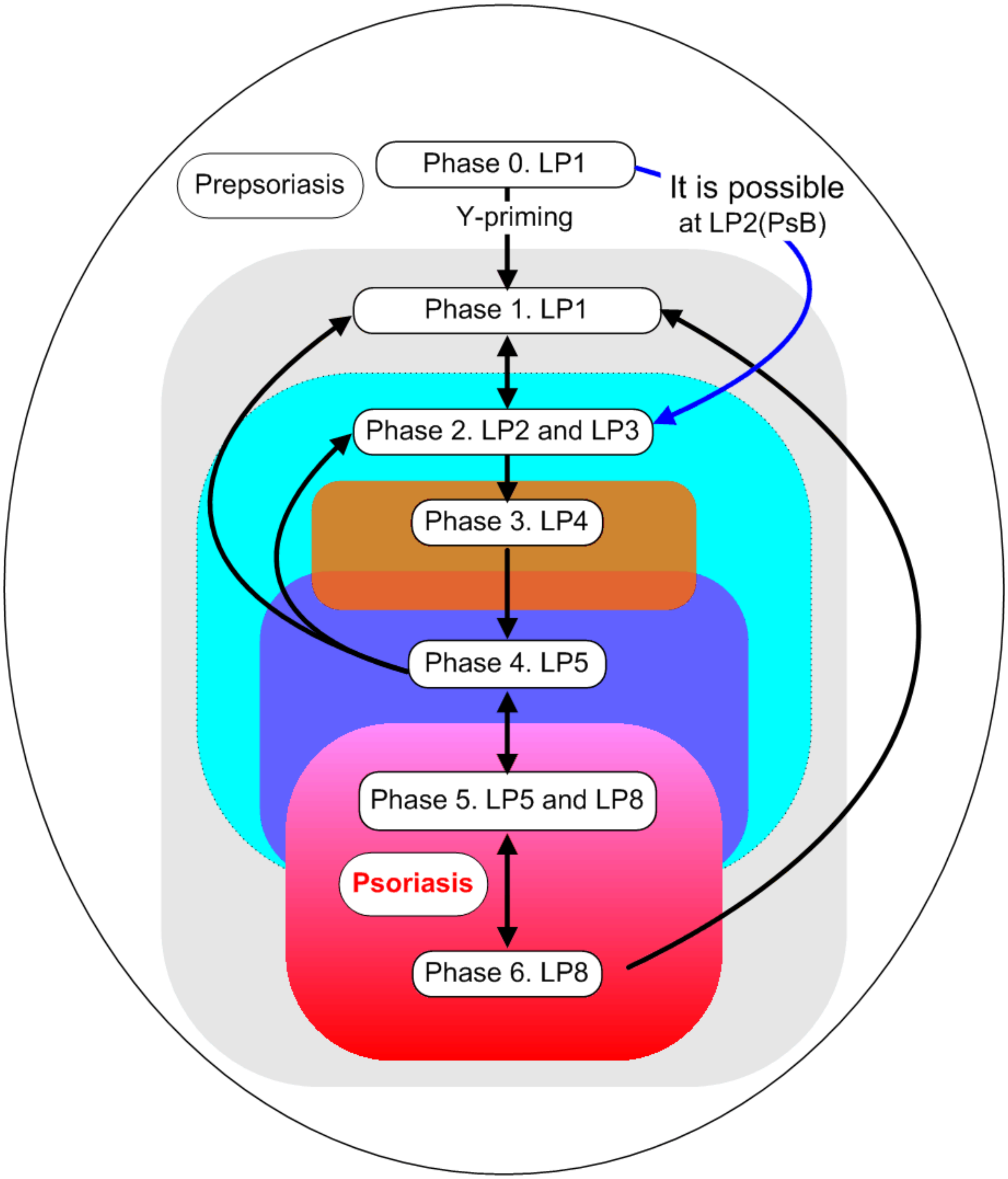

**Fig. 2-25. Phase transitions.**

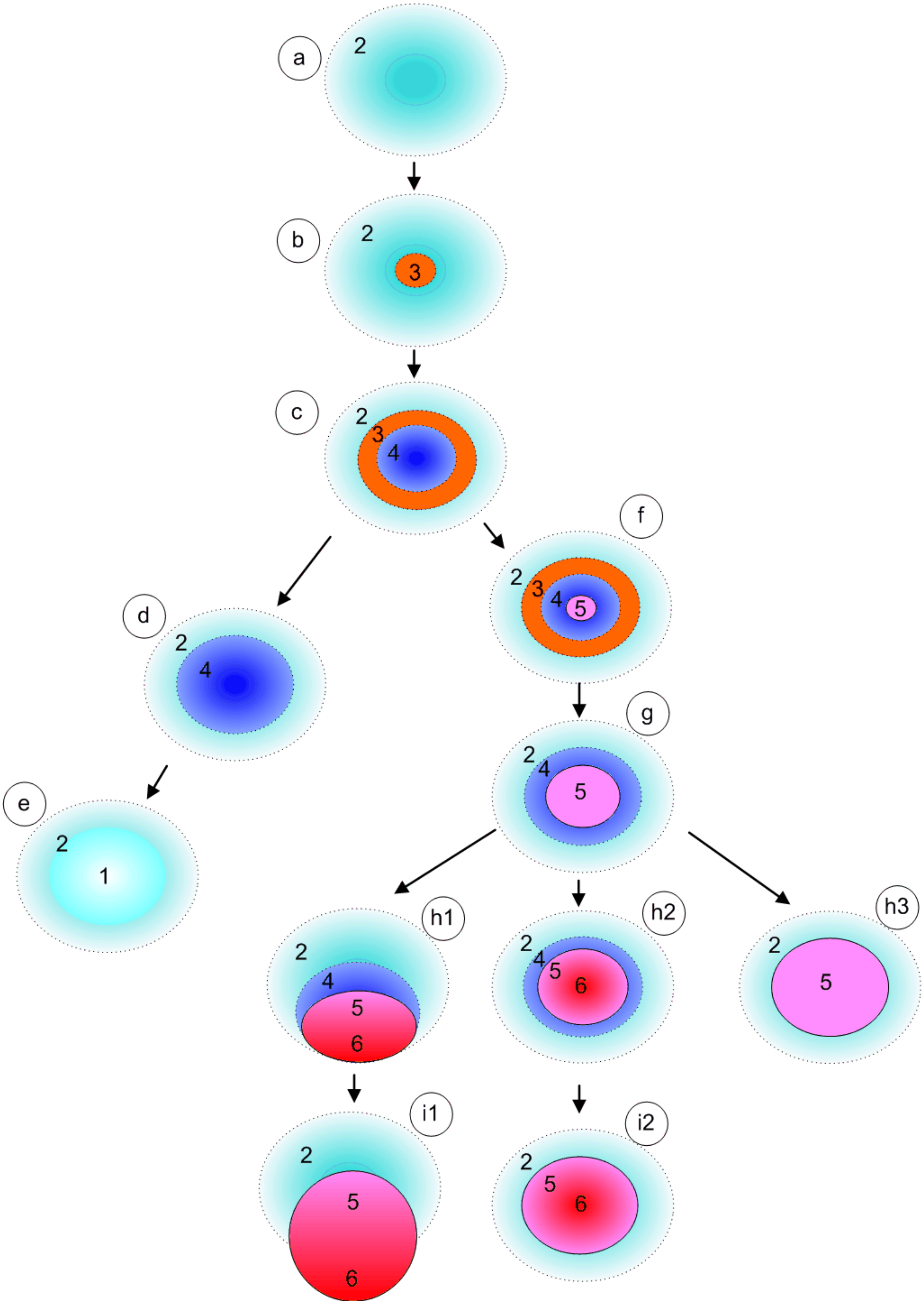

**Fig. 2-26. PLS-plaque development at LP2(HPV) - HPV-carriage of KC.**

**Notes to Fig. 2-26.**

a) Site of HPV-carriage of KC in phase 2; is closer to the edge of this site HPV-carriage of KC smoothly disappears; In the whole site LP2(HPV) and LP3(HPV) develop. In the site center HPV-expansion has begun.

b) In the site of HPV-expansion LP4 has begun and, hence, phase 2 has transferred in phase 3 (orange);

c) Site of LP4 (and phase 3) extend to until LP4 weakening because of decrease of HPV-carriage level of KC. In the middle of the ring, LP5 against HPV is initiated, i.e. there is transition to phase 4 (dark blue);

d) As a result of LP4, LP5 against HPV begins (phase 4). LP8 does not begin because of a low level of IYD and-or KYT (Fig. 2-19)

e) HPV-expansion is suppressed, up to full elimination of HPV-carriage (phase 1). However, where LP5 against HPV was absent (the external ring), skin remains in phase 2.

f) On the site of action of LP5 against HPV, levels of IYD and KYT appear sufficient for initialization of LP8 (phase 5 - pink). The pinpoint PLS-plaque initiates.

g) Simultaneous expansion of sites of phase 4 and phases 5 occurs. It can happen in different ways, but, anyway, the site of phase 4 will transform completely into phase 5.

h1-i1) It is eccentric, so that LP8 extends to NLS site, where is no HPV-carriage of KC and in this site LP8 becomes self-sufficient (phase 6 - red); One part of PLS-plaque is in phase 5, and another part - in phase 6.

h2-i2) It is central, in such a way that in the middle of PLS-plaque HPV-carriage of KC stops completely and LP8 becomes self-sufficient in this site (phase 6 - red); One part of PLS-plaque is in phase 5, and another part - in phase 6.

h3) It is central, but without elimination of HPV-carriage of KC somewhere. PLS-plaque is entirely in phase 5.

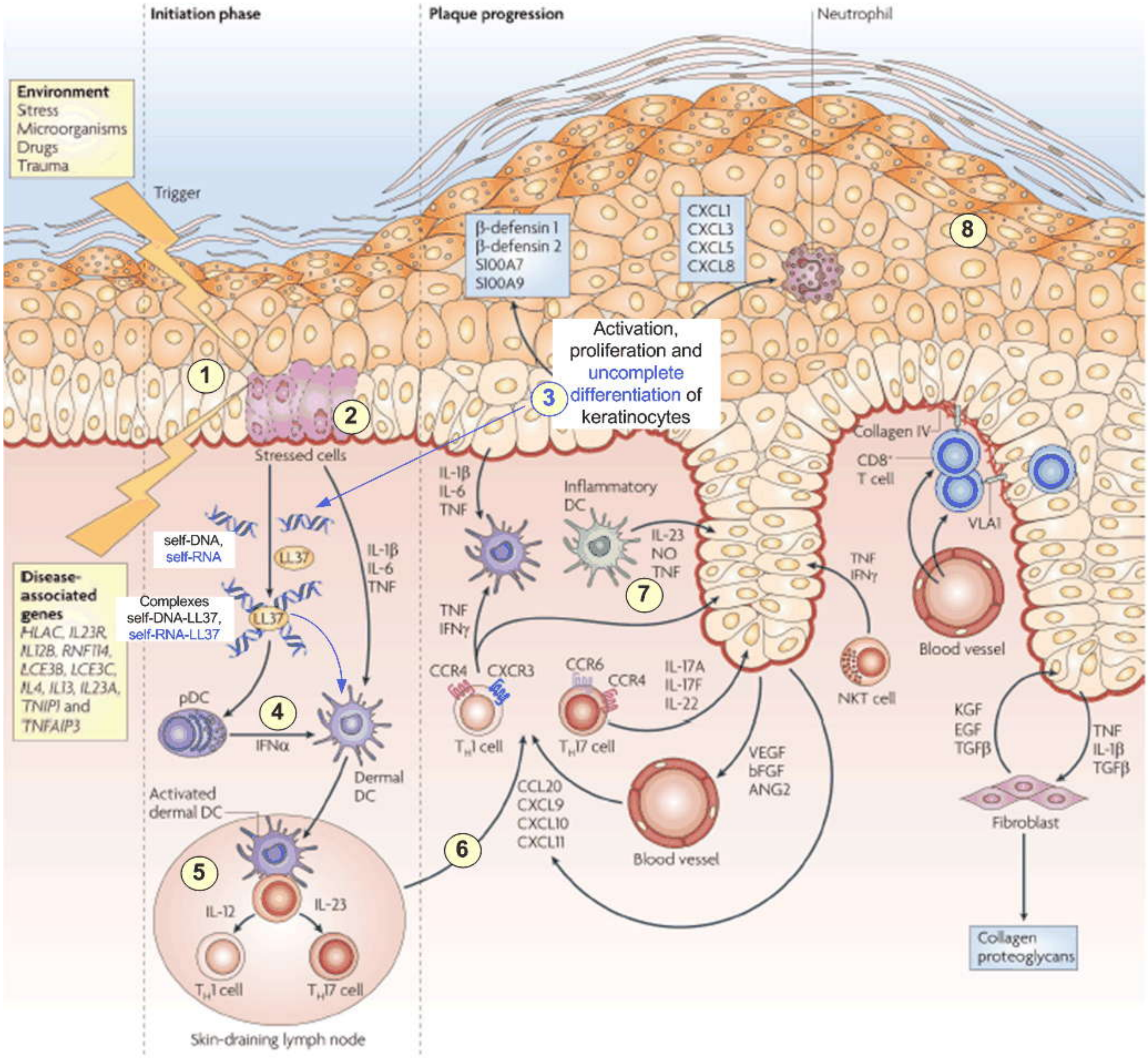

**Fig. 2-27. N-model of pathogenesis of psoriasis.**

*Reproduced from (Nestle 2009a, fig.5) with permission from NATURE PUBLISHING GROUP via Copyright Clearance Center.*

N-model description (Fig. 2-27) according last version (Perera 2012).

(1) A genetically predisposed individual encounters one of many potential environmental triggers, which sets in motion immune responses from both the innate and adaptive arms of the immune system.

(2) Initial triggers such as physical trauma or bacterial products start a cascade of events. Stressed keratinocytes (KC) produce proinflammatory cytokines IL-1alpha/beta, IL-18, TNF-alpha via inflammasome activation, thereby activating other KC, DDC and LC. Stressed KC release self-DNA and self-RNA into extracellular compartment. Self-DNA-LL37 and self-RNA-LL37 complexes are formed.

(3) Psoriatic keratinocytes prematurely enter terminal differentiation, causing the release of self-DNA and self-RNA into the extracellular compartment. Self-DNA-LL37 and self-RNA-LL37 complexes are formed.

(4) PDC induced by chemerin, relocate from blood flow to derma. PDC endocytose viral dsDNA, ssRNA (if trigger is a skin viral infection) or complexes self-DNA-LL37 and self-RNA-LL37 and produce IFN-alpha.

(5) IFN-alpha activates and matures DC. Mature DC circulate to draining lymph nodes (LN), where present putative autoantigen to naive T cells.

(6) TL differentiation and clonal expansion in LN occur; the resulting effector Th1, Th17, Th22 and Tc1, Tc17 and Tc22, bearing skin-homing receptors (CCR4, CCR6, CCR10, CXCR3), circulate to derma under the guidance of selectins and integrin-receptor ligand interactions.

(7) Key processes during disease maintenance are the presentation of putative autoantigen to effector TL and the release of IL-23 by DDC, the production of proinflammatory mediators such as TNF-alpha and NO by TipDC, and the production of IL-17A, IL-17F, and IL-22 by Th17 and Tc17 cells and IFN-gamma and TNF-alpha by Th1 and Tc1. These mediators act on keratinocytes, leading to the activation, proliferation, and production of antimicrobial peptides (e.g., LL-37 and HBD), chemokines (e.g., CXCL1, CXCL9, CXCL10, CXCL11, and CCL20), and S100 proteins (e.g., S100A7-9) by keratinocytes.

DC and effector TL form perivascular clusters and lymphoid-like structures around blood vessels in the presence of chemokines such as CCL19 produced by MF. A key checkpoint of psoriatic plaque initiation is the migration of TL from derma in epidermis; This migration is controlled through the interaction of alpha1beta1integrin (VLA-1) on TL and collagen IV at the basement membrane.
Unconventional TL, including NKT, contribute to the disease process.
Feedback loops involving keratinocytes, fibroblasts, and endothelial cells contribute to tissue reorganization with endothelial-cell activation and proliferation and deposition of extracellular matrix.
Neutrophils in the epidermis are attracted by chemokines, including CXCL8 and CXCL1
.
(8) KC continue to proliferate; there is continued TL recruitment; and the proinflammatory cycle continues unchecked until therapeutic intervention takes place or, very rarely, until spontaneous remission occurs.

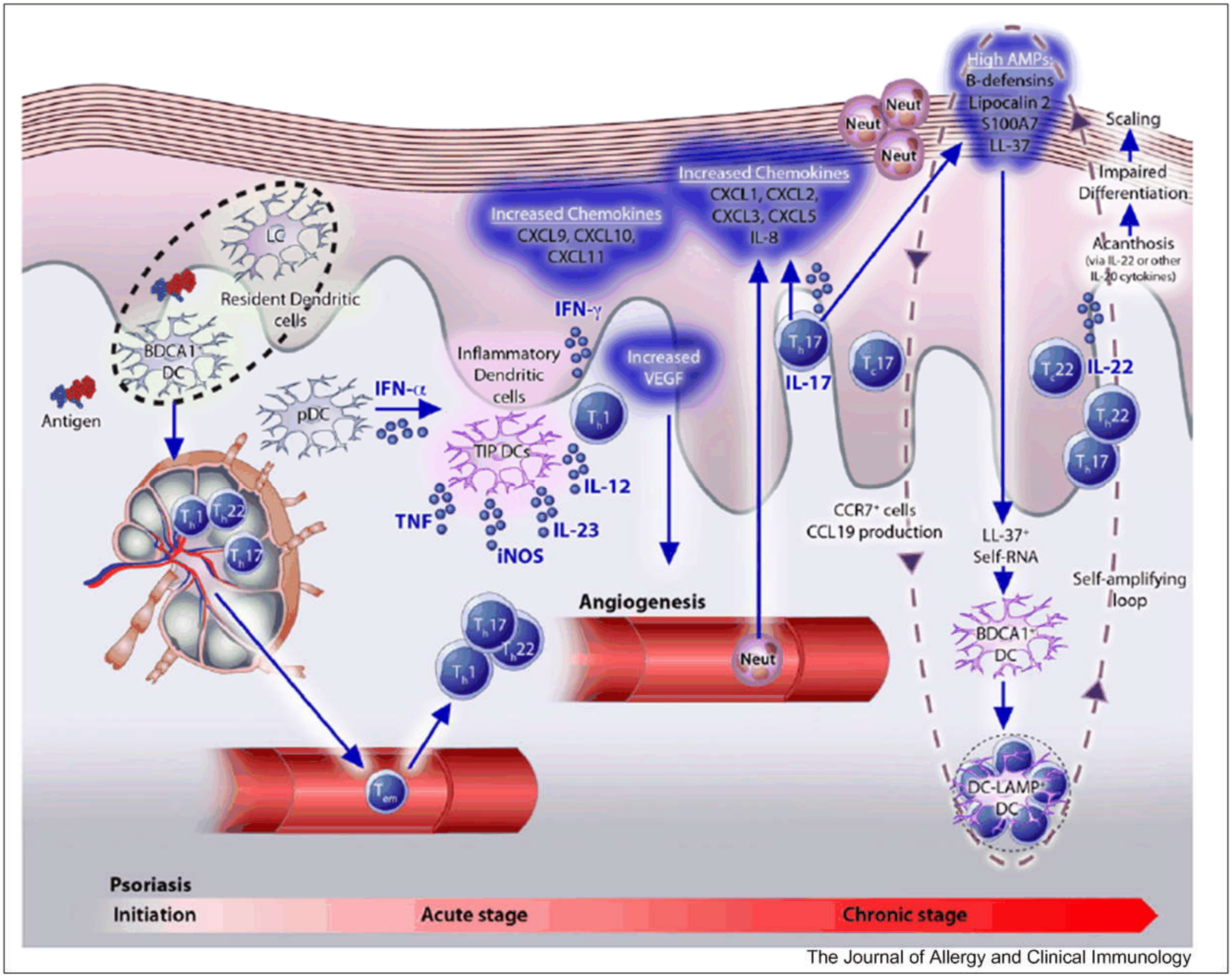

**Fig. 2-28. GK-model of pathogenesis of psoriasis.**
*Reproduced from (Guttman-Yassky 2011, fig.2B) with permission of MOSBY, INC. via Copyright Clearance Center.*

GK-model description (Fig. 2-28) based on (Guttman-Yassky 2011, Nograles 2010).

**Initial, acute and chronic stages of psoriatic plaque.**

**Initial stage.**
Trauma or skin infection lead to occurrence dermal-epidermal X-antigen.

From damaged KC to extracellular space get self-RNA and self-DNA. Neu and KC secrete LL37.
In extracellular space complexes self-DNA-LL37 are formed. Then they endocytosed by PDC and influence through TLR9.

PDC produce IFN-alpha, which induces maturation and differentiation of inflammatory DC.
Both LC and BDCA-1+DDC endocytose X-antigen and carry it to LN. As a result of interaction with maDC in LN are formed effector Th1, Th17 and Th22.

**Acute stage.**
Inflammatory DC produce TNF-alpha, iNOS, IL-20, and IL-23, which induce Th1 and Th17 cell responses. The Th1-cytokine IFN-gamma induces KC to produce proinflammatory chemokines and increase production of VEGF, which promotes angiogenesis. DC secrete IL-23, which stimulates Th17 and Th22 for production of IL-17 and IL-22.
IL-17 induces KC to produce chemoattractants for TL, Neu and Mo. IL-22 and other IL-20 family cytokines promote epidermal acanthosis. IL-17 and IL-22 induce KC for production of AMP that include defensins, lipocalin 2 and LL-37.

**Chronic stage.**
LL-37 upregulation results in a self-amplifying inflammatory loop from self-DNA-LL37 complexes that stimulate PDC production of IFN-alpha and self-RNA-LL37 complexes that stimulate production of TNF-alpha, IL-6, and IL-23 by DC and their maturation into CD208+maDC. CD208+maDC colocalize with TL in lymphoid structures that include CCR7+ cells, which produce CCL19.
Formation of effector Th17, Th22, Tc22.

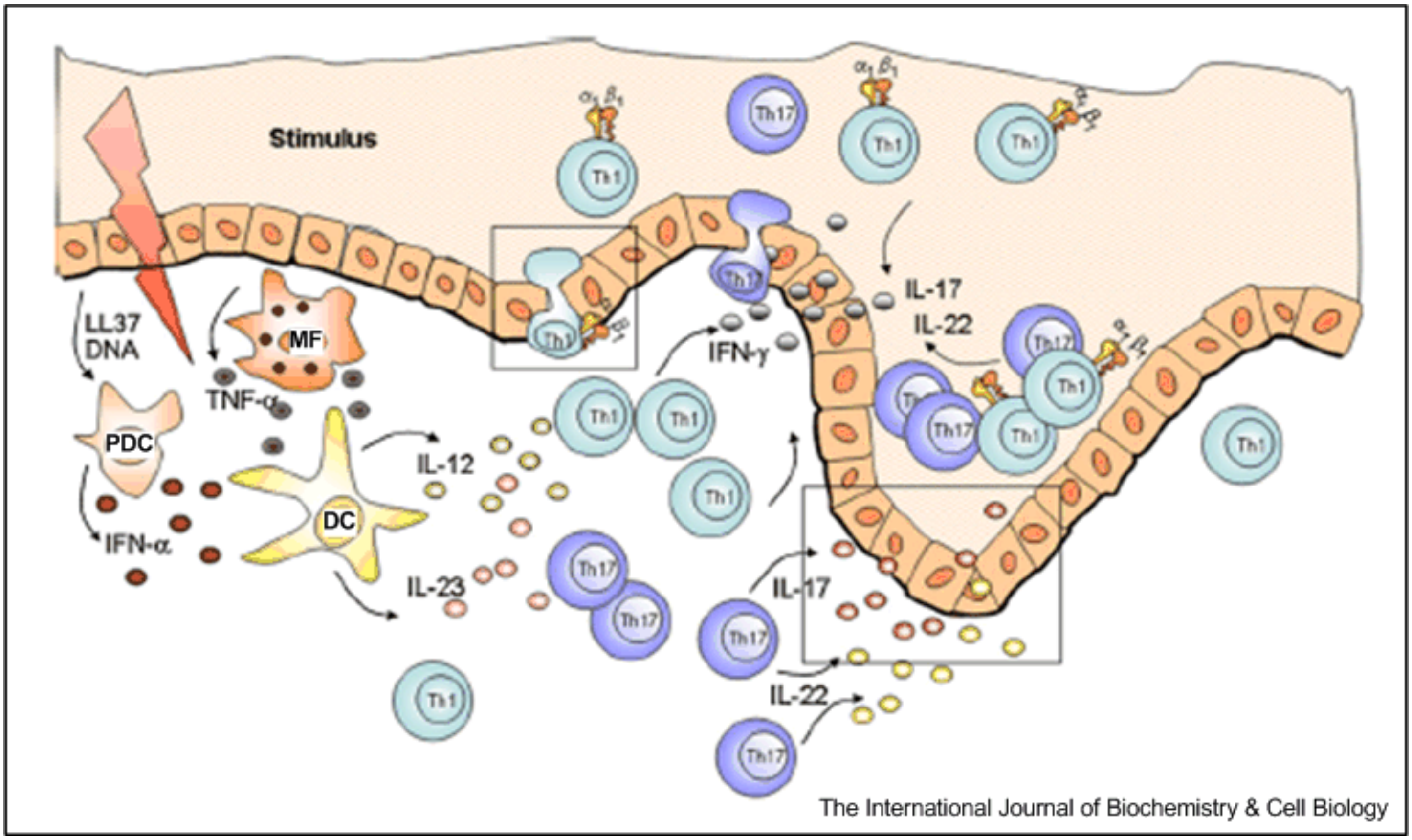

**Fig. 2-29. TC-model of pathogenesis of psoriasis.**

*Reproduced from (Tonel 2009, fig. 1) with permission of PERGAMON via Copyright Clearance Center.*

**Development of psoriatic plaque.**

Various stimuli (such as physical trauma, infections, etc.) can all trigger an initial PLS-plaques in those individuals who already have a genetic predisposition. After the initial trigger, one of the earliest events driving the inflammatory eruption are the secretion of IFN-alpha from plasmacytoid dendritic cells (PDC) and the production of TNF-alpha by cells of the innate and adaptive immune system.

Large amounts of IFN-alpha released by PDC induce activation of the local immune effector cells enabling them to secrete cytokines that further promote the inflammatory cascade.

TNF-alpha is a highly active cytokine of the inflammatory infiltrate and is mainly released by activated macrophages, dermal DC and in a lesser extent by keratinocytes and T cells. The high levels of TNF-alpha lead to the maturation of DC into potent antigen presenting cells (APC), and secondly, in conjunction with other cytokines, TNF-alpha up-regulates the expression of endothelial E-selectin and ICAM-1 attracting further CLA+T cells into the skin.

In addition, the panel of cytokines released by T lymphocytes also contributes to the stimulation of epidermal keratinocytes and is at least partially responsible for typical changes seen in psoriasis. They induce the expression of ICAM-1, CD40 and MHC-II, the release of various cytokines, and trigger keratinocyte hyperproliferation leading to epidermal hyperplasia (acanthosis).

Finally, alpha1beta1 integrin is expressed only on epidermal but not dermal T cells defining key effectors in psoriasis as intraepidermal but not total T cells correlate with onset of the disease.

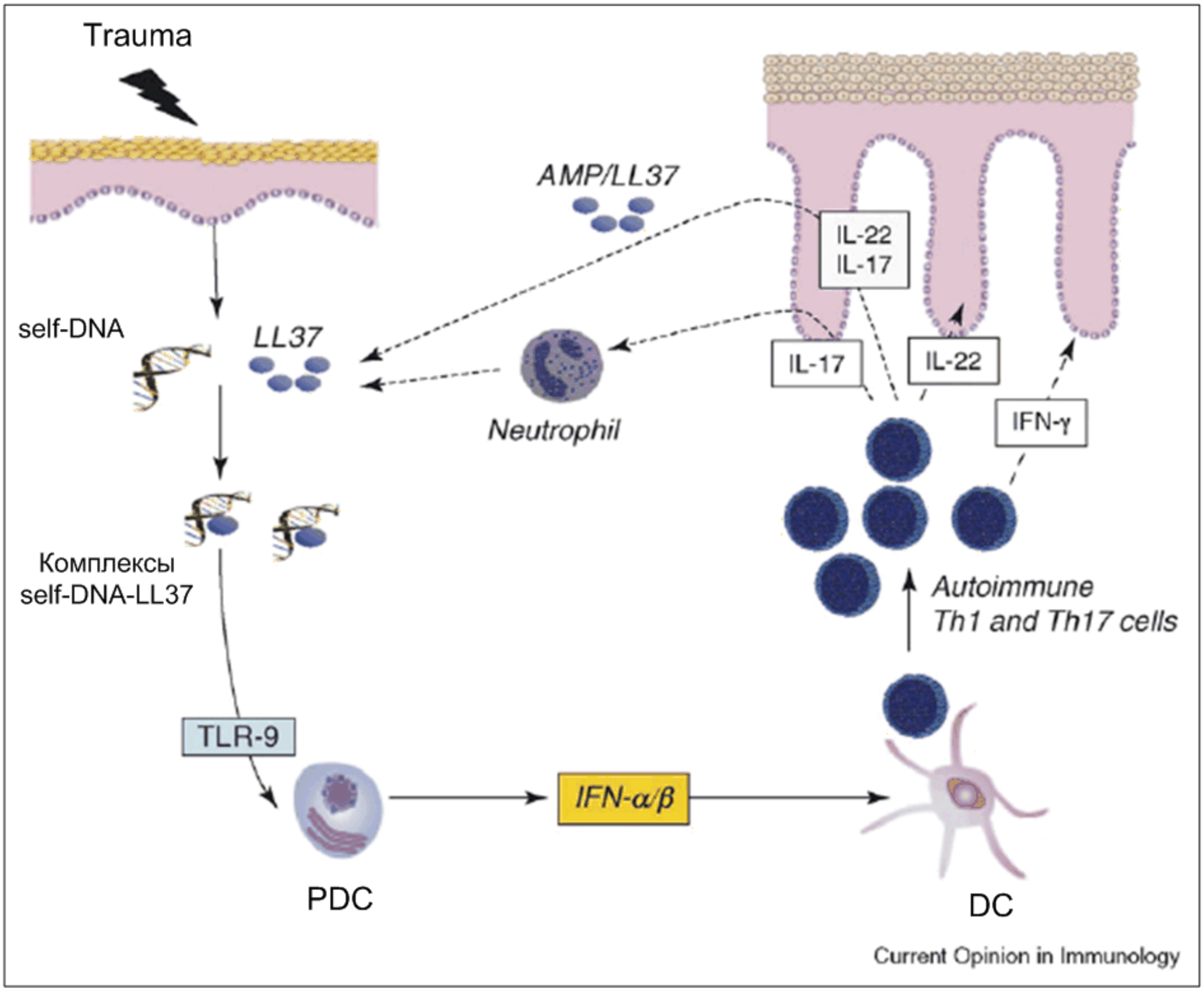

**Fig. 2-30. GL-model of pathogenesis of psoriasis.**

*Reproduced from (Gilliet 2008, fig. 3) with permission of* ELSEVIER LTD. *via Copyright Clearance Center.*

**A model for initiation and maintenance of autoimmune skin inflammation in psoriasis.**

Mechanical injury to prepsoriatic skin induces LL37 production by keratinocytes. LL37 forms a complex with self-DNA released by damaged cells and triggers TLR9-mediated PDC activation to produce type I IFN (IFN-alpha/beta) in prepsoriatic skin.

Type I IFN trigger local maturation of myeloid dendritic cell and activation of autoreactive T cells leading to the formation of psoriatic skin lesions. In this process, activated Th17 cells produce IL-22 and IL-17, which induce epidermal hyperplasia and chemokine (e.g. IL-8)-dependent recruitment of neutrophils, respectively.

IL-17 and IL-22 stimulate psoriatic keratinocytes to sustain the expression of cationic antimicrobial peptides (AMP) including LL37, which further forms complexes with self-DNA that is abundantly released in psoriatic skin lesions by apoptotic cells. As a result, self-DNA-LL37 complexes constantly promote dendritic cell maturation and activation of autoreactive T cells, providing a self-sustaining feedback mechanism that amplifies and maintains autoimmune skin inflammation in psoriasis.